\documentclass[amsmath,amssymb,onecolumn,superscriptaddress,prstper,runinaddress,floatfix]{revtex4}
\usepackage{amsfonts}
\usepackage{amssymb}
\usepackage{amsmath}
\usepackage{epsfig}
\usepackage{tabularx}
\usepackage{array}
\allowdisplaybreaks
\begin{document}

\title{Study of Dynamical Instability of Collapsing Charged Spherically Symmetric Anisotropic Matter Configurations within Non-Minimally Coupled Gravity}

\author{A. Rehman}
\email{atteeq.math@gmail.com}
\affiliation{Department of Mathematics, University of Management and Technology,\\ Johar Town Campus, Lahore-54782, Pakistan.}

\author{M. Yousaf}
\email{myousaf.math@gmail.com}
\affiliation{Department of Mathematics, Virtual University of Pakistan,\\ 54-Lawrence Road, Lahore 54000, Pakistan}
\affiliation{Research Center of Astrophysics and Cosmology, Khazar University,
Baku, AZ1096, 41 Mehseti Street, Azerbaijan}

\author{Mohammed Zakarya}
\email{mzibrahim@kku.edu.sa}
\affiliation{Department of Mathematics, College of Science, King Khalid University, P.O. Box 9004, Abha 61413, Saudi Arabia.}

\author{M. Aslam}
\email{muamin@kku.edu.sa}
\affiliation{Department of Mathematics, College of Science, King Khalid University, P.O. Box 9004, Abha 61413, Saudi Arabia.}

\author{Maram Ali}
\email{Maahmaad@kku.edu.sa}
\affiliation{Department of Mathematics, College of Science, King Khalid University, P.O. Box 9004, Abha 61413, Saudi Arabia.}


\begin{abstract}
We are primarily focused to evaluate the dynamics of stability, limitations for gravitational collapse, and the development of celestial formations in the presence of electric field caused by electric charge within the formalism of $f(R, \mathcal{L}_{m})$ theory. The relationship between anisotropic pressure, nonlinear disturbances, and correction terms associated with the considered gravity theory comprehensively describes the stability for compact structures along with the factors that cause the dynamical instability. These factors considerably emphasize the objectives of higher energy depictions, particularly for modeling the substantial matter configurations  in the existence of high pressure and curvature. In this manuscript, we study the instability constraints related to the charged spherical fluid in $f(R, \mathcal{L}_{m})$ theory and contemplate a certain equation of state (EOS) for establishing the interaction between static and fundamental variables using the adiabatic index. Our study initiates with the description of fundamental formalism of $f(R, \mathcal{L}_{m})$ gravity, the modified form of the hydrostatic equilibrium equation is solved after employing the perturbation approach and we determine the collapse equation to derive the certain findings for adiabatic index in Newtonian and post-Newtonian domains. It is asserted that the equilibrium state between inner gravitating force and the outer pressure establishes the stability of highly dense matter. Furthermore, the consideration of dark source terms associated with $f(R, \mathcal{L}_{m})$ gravity results the increased stability of charged fluid governed by the non-linear function. Our research work advances the interpretation of gravitational collapse
in the context of higher energy density regions and provides important insights for the description of compact systems through the illustration of the fundamental features related to the charged fluid configurations.
\end{abstract}
\maketitle

\section{Introduction}

The development of compact objects influences the formation of planets and stars, whereas gravity brings interstellar gases close to each other. The phenomenon of condensation produces compact objects that expand after consuming their interior matter, resulting in the creation of highly dense components across the universe. Gravitational collapse plays an important role in the life cycle of highly dense objects, specifying the end of their existance, while self-gravitating objects maintain their stability because of the equilibrium between attractive effect of gravity and the pressure acting in the outward direction. However, if the pressure generated by the energy is not sufficient to counteract the attraction induced by inner gravity then the system will undergo the complete collapse. Consequently, the internal mass of the object establish its fate of becoming neutron star, white dwarf, or the black hole. An object with a mass lower than eight times that of the Sun is developed into a white dwarf, while one with more mass is changed into neutron star or a black hole. The behavior of our universe and its components to minor fluctuations is termed as stability whose thorough study implies significant implications in the field of astrophysics such as exceptionally high dense structures, gravitational waves, and the cosmological frameworks. In classical gravity, dynamical stability describes the dynamics of spacetime and corresponding gravitational field in responding to fluctuations and is usually correlated with the solved form of field equations. The simplified form of equations of motion as well as the evaluation of their related eigen values and eigen functions are important in this context. The assessment of stability is performed after perturbing series of small factors and then simplifying the equations after considering only the linear terms in the derived equations. It allows the most basic analysis of celestial objects along with the formation and evolution of compact objects and demonstrates the significance of minor fluctuations in preserving the primary features and stability of such configurations.
The stability evaluation for spherical, cylindrical and axially symmetric compact objects provides the thorough interpretation of the physical composition of celestial formations. The complex interactions between density distributions, mass, and matter dynamics are crucial in describing the resulting instabilities and the comprehensive analysis of such interrelated factors is crucial in determining the stability of highly dense matter.

Chandrasekhar \cite{chandrasekhar1957introduction,chandrasekhar1964dynamical} after analyzing the physical dynamics of celestial objects recommended the particular range of required mass in order to experience the gravitational collapse that is termed as Chandrasekhar limit. Herrera \cite{herrera2009expansion,herrera2010collapsing} evaluated the mechanics of gravitational collapse and dynamical instabilities related to the anisotropic fluid and asserted that the stability restraints may differ from the Chandrasekhar's limit whereas, $\Gamma$ plays a significant role in determining stability as $\Gamma < \frac{4}{3}$ corresponds to the unstable configurations and $\Gamma > \frac{4}{3}$ relates to the stability of the compact structures \cite{herrera2018tilted,herrera2022non,herrera2023expansion}. The stable evolution of compact structures throughout the gravitational collapse is studied in various theories of gravity. Researchers \cite{sharif2014dynamical,sharif2016instability,yousaf2018dynamical} revealed that correction terms related to distinct theories imply considerable impacts on the physical features and value of $\Gamma$ in the consideration of weak field approximations. Skripkin suggested an analytical formalism in order to assess the spherically symmetric matter distributions after taking into account the zero expansion and it describes the formation of singularity, which requires additional analysis. In the context of radial perturbation and specific restraints for the electromagnetic field it was asserted that the range of dynamical instability remains consistent irrespective to the $\Gamma$ parameter  in N and $pN$ domains. However, beyond the $pN$ approximation, the $\Gamma$ parameter implies significant impact in assessing the dynamical instability, whereas additional restraints offer a profound understanding of the issue \cite{torres2005some,mitra2006gravitational,ivanov2010importance}. Yousaf et al. \cite{yousaf2016influence} thoroughly analyzed the relevance of dark source terms in understanding of stable behavior for anisotropic matter. Baffou et al. \cite{baffou2016exploring} studied the stability related to the compact objects after contemplating the de-Sitter solution, and claimed that the corresponding correction terms can provide the interpretation for earlier development of our universe. Further, Wu and Yu \cite{wu2011stability} evaluated the stability of highly dense matter after the consideration of perturbation approach within the context of modified gravity. Izumi and Ong \cite{izumi2014acausality} examined the dynamics of stability against the cosmological disturbances for a specific scalar field and claimed that the more degrees of freedom are identified in the case of non-linear matter. Herrera et al. \cite {herrera1989dynamical} discussed the stability of compact structures after considering the non-adiabatic nature of spherical composition of matter in the context of weak field estimations. Chan et al. \cite{chan1993dynamical} studied the implications of radiating heat on the restraints of stability corresponding to Newtonian and post-Newtonian domains and stability was further analyzed through the perturbation scheme by Herrera et al. \cite{herrera2012dynamical}. Bhatti et al.
\cite{bhatti2020stability,ur2024dynamically}
assessed the fundamental factors having significant relevance in establishing the stability for isotropic and anisotropic compositions of matter.

The existence of electric charge implies significant impacts on the matter configuration and the corresponding electromagnetic forces are regarded as an important factor that reduce the attractive impact of gravitating force. This increased force of repulsion disrupts the position of hydrostatic equilibrium, allowing the system to sustain more matter ahead of the gravitational collapse. Consequently, charge matter distributions possess greater range of mass, larger radii and a wider range of stable equilibrium than uncharged compositions. It can also increase stability in opposition to small perturbations by limiting the probability of cracking and increasing the adiabatic index beyond the established stability criterion. Furthermore, presence of electric charge significantly affects the fundamental physical properties of highly dense structures including rigidity and surface redshift which cause the prevention of gravitational collapse.
As a result, it can be asserted that the appearance of electric field endorses the matter stability through the resistance against gravitational forces. Bekenstein \cite{bekenstein1971hydrostatic} suggested that the force caused by the electric charge implies repulsion that eventually increases the stability of the compact structure, while, Esculpi and Aloma \cite{esculpi2010conformal} continued this study for the anisotropic fluid and concluded that the electromagnetic field along with the positive anisotropy exerts the repulsive force on the considered system. The repulsive force caused by the electric charge is a long-range force that opposes the attractive gravitational force all through the matter. This specifies that the density of charged compact matter is decreased and will possess more radius as compared to uncharged one, that eventually ensures that the stability of matter hinders the phenomenon of gravitational collapse.

General relativity (GR) offers the thorough explanation for the interaction between distribution of matter and the structure of space-time in comparison with the Newtonian gravity and provides feasible description for distinct cosmological phenomenon such as the inflationary era of the cosmos, and the gravitational effects on the dynamics of celestial objects. The empirical results including the type Ia supernova and cosmic microwave background radiations suggest the accelerated development of the universe \cite{riess1998observational,perlmutter1999astrophys,riess2007new}, and reveal the presence of an enigmatic type of energy referred to as dark energy (DE), which accounts for over seventy percent of the universe and plays a crucial role in its accelerated expansion. However, GR does not illustrate the dynamics of rapid cosmic expansion beyond the consideration of DE, as DE in conjunction with the   cosmological constant has significant relevance in this regard \cite{nojiri2007introduction,nojiri2006new,bamba2012dark}. These constraints of GR affirm the significance of different mathematical gravity models in order to evaluate the mysterious nature of DE, in addition, distinct theoretical and interpretive findings suggest the consideration of modified form of GR. Distinct methodologies are  recommended for the explanation of DE \cite{copeland2006dynamics,nojiri2011unified,bamba2015inflationary}, and these approaches are classified in two categories, the first one primarily focuses on the fundamental configuration of matter, that comprises the majority of our cosmos. This technique suggests that the negative essence of pressure leads to the anti-gravitational stress which ultimately drives the accelerated development of the universe and in this case, the cosmological constant $\Lambda$ is inserted in the corresponding field equations.

The contemplation of certain DE models also provide the precise interpretation for the rapid cosmic expansion, however, this approach leads to the discrepancy between the determined value of $\Lambda$ and the approximated value of vacuum density. An alternative approach endorses the modification in the geometrical section of the field equations for the interpretation of accelerated expansion of the universe. Subsequently, these methodologies inferred the formalism of distinct alternative gravity theories and different forms of Ricci scalar $R$ and the variations in Lagrangian in the Einstein-Hilbert action are crucial in determining the formalism for these theories. These alternative theories impart the general scalar field functions, permitting for a thorough explanation of intriguing dark components, rapid cosmic expansion, and the gravitational dynamics. The analysis of distinct astronomical phenomenon in the context of alternative theories of gravity is available in \cite{rehman2026complexity, ditta2026structure,yousaf2025anisotropic,nojiri2017modified,faraoni2005stability,feng2024brief}. The generalized form of curvature functions including Ricci scalar and the Gauss-Bonnet invariant, in addition to the impacts of the trace of energy-momentum tensor play a crucial role in determining the particular models in alternative gravity theories. The $f(R)$ gravity, in the case of which $\mathcal{L}_{GR}=R$ is substituted with a generalized function is thought to be the most basic alteration of GR. The appropriate $f(R)$ frameworks have significant relevance in determining the inconsistent implications of the Riemann scalar in the development of our cosmos \cite{sotiriou2010f,de2010f}. The comprehensive interpretation for the mathematical framework of $f(R)$ gravity is given in \cite{olmo2007limit}, acquiring the exact solutions related to the spherical composition of matter and their consistency were validated after comparing them with the determined results. While, the solution of Einstein field equations has significant relevance in assessing the compact formations through the precise description of their structural characteristics.

The state of equilibrium for compact structures is assessed in $f(R)$ theory by Capozziello et al. \cite{capozziello2011hydrostatic}, in addition to the Newtonian estimation deriving the modified formalism of Lane-Emden and Poisson equations. Further, their conclusions for radial perturbations relating to gravitational potential are compatible with GR, illustrating their significance in the literature. Sharif and Yousaf \cite{sharif2016charged} studied the spherical fluid distribution along with the appearance of the electric charge within the context of $f(R)$ theory and suggested that the electromagnetic field decelerates the physical process of gravitational collapse. Pan et al. \cite{pan2018astronomical} evaluated the non-gravitational interactions of dark energy with dust by considering the Friedmann-Lema\^{\i}tre-Robertson-Walker, in the case of which entire energy density is determined by the the second-order differential equation. Moreover, they considered type Ia supernova findings, conventional cosmic tests, and different values of Hubble parameter in order to approximate the distribution of phantom energy. Whereas, Mustafa \cite{mustafa2020bardeen}  proposed a novel form compact solutions after contemplating the restraint for Bardeen BH geometry, spherically symmetric fluid dispersion, and the electromagnetic field.
The comparison between various inner solutions related to the compact structures with measured radius and mass asserted their consistency with the charged fluid configuration, and graphical illustration confirmed their substantial viability. Bertolami \cite{bertolami2007extra} contributed significantly in interpreting the interaction between fundamental space-time configuration and matter in $f(R)$ gravity.

Another alternative gravity theory named as $f(R, \mathcal{L}_{m})$ theory includes the significant non-minimal relationship between matter composition and geometry, demonstrating the fundamental physical implications \cite{harko2010f}. The consideration of matter and curvature relationships in the formalism of $f(R, \mathcal{L}_{m})$ gravity is regarded as an important addition in the cosmic and astrophysical domains. Furthermore, this specific mathematical formalism specifies the non-zero covariant divergence of energy-momentum tensor, implying the emergence of an additional force which departs from the basic GR model, allowing particles to deviate from their typical trajectory. Additionally, the non-conservation of energy-momentum tensor describes the transfer of energy and momentum between physical configuration and matter fields. This energy transfer is described through the constant creation of matter at the cosmological scale, which suggests that the mysterious dark components of universe should be interpreted without considering the existing particles. Distinct values of  $\mathcal{L}_{m}$ implies important physical consequences and referred as the main feature of $f(R, \mathcal{L}_{m})$ theory. The choice of $\mathcal{L}_{m}=-\rho$, specifies the coupling of gravity with the distribution of energy density and $\mathcal{L}_{m}=p$, describes the interaction between gravity, and the stresses inside the matter. Rehman et al. \cite{rehman2025orthogonal}
analyzed the complexity of anisotropic matter having spherically symmetric distribution in $f(R, \mathcal{L}_{m})$ gravity after determining the structure scalars through the orthogonal splitting of Riemann curvature tensor. Wang and Liao \cite{wang2012energy} established the energy constraints for $f(R, \mathcal{L}_{m})$ gravity in the context of certain models in FRW cosmology and determined specific outcomes through the consideration of observational results. Jaybhaye et al. \cite{jaybhaye2022cosmology} evaluated the late-time acceleration of the universe within the formalism of $f(R, \mathcal{L}_{m})$ gravity and analyzed the stability of the system through the linear perturbations. Naseer et al. \cite{naseer2023constructing} studied the physical viability of the compact structures after the contemplation of karmarker restraint in $f(R, \mathcal{L}_{m})$ gravity. Moreover, the study of different structural features of celestial formations within the context of $f(R, \mathcal{L}_{m})$ gravity can be seen in  \cite{naseer2023constructing, wang2012energy,jaybhaye2022cosmology,jaybhaye2023baryogenesis}.

Our analysis evaluates the dynamical instability of spherically charged fluid configurations within the framework of $f(R, \mathcal{L}_{m})$ gravity having the form $f(R, \mathcal{L}_{m})=R+\zeta R^{2}+\chi \mathcal{L}_{m}^{2}$ \cite{esculpi2010conformal}. The key objective of our work is to establish the criterion for adiabatic index in determining dynamical instability of charged fluid in N and $pN$ domains determined by the model parameters $\zeta$ and $\chi$. The related fundamental physical constraints specify this particular formalism suitable for assessing the unstable dynamics of highly dense charged matter. These results are particularly relevant in extremely dense cosmic scenarios such as compact structures, central collapse supernova, and astrophysical structures with exceptional thermodynamical and physical conditions. The findings reported in this work offer a detailed analysis of matter dynamics in extreme contexts as well as potential instabilities emerged from modified gravity scenarios. Our research article is formatted as described below: The fundamental framework associated with $f(R,\mathcal{L}_{m})$ theory is discussed in the context of spherically symmetric charged fluid along with the revised non-conserved field equations in section $\mathbf{II}$, while, section $\mathbf{III}$ provides the comprehensive description for the perturbation technique corresponding to the collapsed charged fluid through the contemplation of non-conserved perturbed equations. Matching constraints along with the collapse equation is derived in section $\mathbf{IV}$. Section $\mathbf{V}$ addresses the dynamical instability restraints in N and $pN$ regimes having specific dependence on adiabatic index along with the considered gravity model. The acquired results are reviewed as conclusion in section $\mathbf{VI}$.

\section{Spherically symmetric geometry and field equation}

A number of cosmic phenomena, including creation of extremely dense objects, propagation of gravitational waves, and accelerated cosmic expansion are described by general relativity (GR). However, various emipirical findings such as intriguing nature of dark energy and dark matter, interpretation of inflationary cosmic epoch affirm the revised form of GR. In this context, it appears to be a compelling methodology to modify gravitational action by adding dependencies more than Ricci scalar $R$ that results the formalism of distinct modified gravity theories. Within these revised gravity theories, one is $f(R,\mathcal{L}_{m})$ theory that incorporates the explicit dependence on matter lagrangian and was proposed by Harko \cite{harko2010f}. The interaction between Ricci scalar $R$ and matter Lagrabgian $\mathcal{L}_{m}$ is regarded as the substantial characteristic of this theory, which significantly impacts the dynamics of relationship between matter and geometry. In GR and minimal coupling theories, the implicit effects of matter on spatial curvature are described, while,  $f(R,\mathcal{L}_{m})$ theory characterizes the explicit impact of the matter distribution on spatial curvature, leading to significant physical gravitational dynamics. Furthermore, the $R-\mathcal{L}_{m}$ interaction comprehensively describes the gravitational implications that successfully explain the presence of enigmatic dark components. A thorough analysis of gravitational Lagrangian, which is a function of $R$ and $\mathcal{L}_{m}$ in our contemplated theory, has significant relevance in determining the fundamental characteristics associated with highly dense objects within the context of cosmic perturbations. The EHA associated with $f(R, \mathcal{L}_{m})$ theory is described in subsequent form \cite{naseer2023constructing}
\begin{equation}\label{1}
S=\int_{\Sigma} f(R, \mathcal{L}_{m}) \sqrt{-g} d^{4}x,
\end{equation}
here, $\Sigma$ signifies the $4$-dimensional manifold with the $x^{\xi}$ coordinate system, whereas, $g$ represents the determinant of metric tensor $g_{\mu \nu}$ having Greek indices ranging from zero to three. The alternative Friedmann equations related to $f(R, \mathcal{L}_{m})$ gravity demonstrate modifications for $\mathcal{L}_{m}$ term in the framework of homogenous and isotropic fluid dispersion, restrained by the Friedmann-Lema\^{\i}tre-Robertson-Walker metric, that have substantial implications for understanding the rapid cosmic expansion. Furthermore, the particular mathematical framework of $f(R, \mathcal{L}_{m})$ gravity offer a thorough explanation for mysterious essence of dark energy without taking into account the cosmological constant. The inclusion of $\mathcal{L}_{m}$ terms modifies the energy restraints, particularly weak and strong energy restraints, which are important in determining the feasibility of modified gravities. These modifications may lead to the violations of standard energy restraints, which could have implications regarding distribution of exotic matter and accelerated expansion of the universe. The study of gravitational collapse in $f(R, \mathcal{L}_{m})$ theory has received significant consideration in literature, and the modified field equations have substantial implications on rate of collapse, the development of horizons and eventual form of celestial objects. Furthermore, certain forms of $f(R, \mathcal{L}_{m})$ theory might result in delayed collapse, that could avert the development of singularity in highly dense matter distributions. The stability evaluation has significant relevance in establishing the feasibility of modified theories and contemplation of perturbation methodology characterizing physical compositions in $f(R, \mathcal{L}_{m})$ theory enables the interpretation of behavior of cosmological structures in minor fluctuations. Moreover, $N$ and $pN$ approximations illustrate the role of disrupted curvature terms in establishing the cosmic anisotropies and development of compact objects. Additionally, modified Tolman-Oppenheimer Volkoff equations in $f(R, \mathcal{L}_{m})$ theory yield changes in mass-radius relationship and the restraint for maximum mass.

The additional curvature terms associated with $f(R, \mathcal{L}_{m})$ theory are considered for comprehensive study of dynamical instability of compact objects, in order to which modified field equations are determined after varying action Eq. \eqref{1} with respect to $g_{\mu\nu}$,
\begin{equation}\label{3}
\mathcal{G}_{\mu \nu}=T_{\mu \nu}^{(eff)},
\end{equation}
here, $\mathcal{G}_{\mu \nu}$ specifies the Einstein tensor of the form $\mathcal{G}_{\mu \nu}=R_{\mu \nu}-\frac{1}{2}g_{\mu \nu}R$ and $T_{\mu \nu}^{(eff)}$ represents the stress energy-momentum tensor classified as
\begin{equation}\label{3a}
T_{\mu \nu}^{(eff)}=\frac{1}{f_{R}}T^{(m)}_{\mu \nu}+T^{(D)}_{\mu \nu}+\mathbb{E}_{\mu \nu},
\end{equation}
where, $\mathbb{E}_{\mu \nu}$ is written as electromagnetic energy-momentum tensor, $T^{(m)}_{\mu\nu}$ depicts the standard energy-momentum tensor, and $T^{(D)}_{\mu\nu}$ comprises the dark source terms in following form relating to $f(R, \mathcal{L}_{m})$ theory.
\begin{equation}\label{3b}
T^{(D)}_{\mu \nu}=\frac{1}{2}\left(f+\rho f_{\mathcal{L}_{m}}\right)g_{\mu \nu}-R_{\mu \nu}f_{R}-\left(g_{\mu \nu}\Box-\nabla_{\nu}\nabla_{\mu} \right)f_{R}.
\end{equation}
The consideration of spherically symmetric metric is regarded as the simplest and an important formalism in order to describe the intrinsic attributes of the spacetime as these metrics specify no certain directions, implying that the matter composition seems uniform in all directions. Spherical matter distribution has significant relevance in the primary explanation of the self-gravitating compact systems and determines the fundamental and decisive solutions corresponding to the Einstein's field equations, providing the foundation for the description of the static compact structures. The cosmic principle recommends the homogenous and isotropic matter distribution across the universe. The isotropic matter distribution ensures the spherical symmetry across all points and provides the feasible framework for the perturbation technique along with the explanation of various equilibrium states, allowing us to understand the fundamental restraints and the fluctuating behavior of the astrophysical events. Consequently, we can say that the spherical fluid dispersion imparts substantial formalism in order to establish the physically viable models in GR. Moreover, it provides the detailed explanation for the physical composition of celestial formations, paving the way for the illustration of significant and complicated celestial phenomenon.

A three-dimensional sphere like surface $\Sigma^{(e)}$ is assumed that specifies the partition of four-dimensional line element into interior and exterior sections. The inner matter dispersion is characterized by the following spherically symmetric metric
\begin{equation}\label{4}
ds^{2}_{-}=D^{2}(t, r)dt^{2}-E^{2}(t, r)dr^{2}-F^{2}(t, r)d\theta^{2}-F^{2}\sin^{2}\theta d^{2}\phi ,
\end{equation}
While, the exterior metric of $\Sigma^{(e)}$ is described as
\begin{equation}\label{M2}
ds^{2}_{+}=\left(1-\frac{2M}{r}+\frac{Q^{2}}{r^{2}}\right)d \nu^{2}+2d\nu dr -r^{2}d\theta^{2}-r^{2}\sin^{2}\theta d \phi^{2},
\end{equation}
here, $M$, $\nu$ and $Q$ specify the entire mass, retarded time and total enclosed charge within the matter configuration.
Several research papers evaluating the fundamental characteristics of celestial formations with anisotropic distribution of matter can be seen in \cite{thirukkanesh2008charged,nayak1989bianchi,maurya2017relativistic}. Anisotropic matter dispersion allows directional variability in primary features as compared to isotropic fluid which has consistent attributes in each direction. The energy-momentum tensor has a crucial role in cosmology as it incorporates the fundamental features of matter affecting gravity and provides the basis for curvature of spacetime, whereas in the context of alternative theories it establishes direct interface with curvature to offer contemporary interpretations for physical astronomical phenomena. In our manuscript, the following energy-momentum tensor is considered
\begin{equation}\label{2}
T_{\mu \nu}^{(m)}= (\rho+P_{yy})U_{\mu}U_{\nu}+\eta l_{\mu}l_{\nu}-P_{yy}g_{\mu \nu}+h_{\nu}U_{\mu}+(P_{xx}-P_{yy})S_{\mu}S_{\nu}+h_{\mu} U_{\nu},
\end{equation}
here, $\rho$ describes the total energy density which incorporates rest mass and internal energy of the matter composition, and $U_{\mu}$ is written as the time-like unit vector that determines the characteristics of matter after defining local rest frame. While, $S_{\mu}$, and $l_{\mu}$ specify the particular spatial directions being the unit vectors in $x$ and $y$ directions. And, $P_{xx}$ and $P_{yy}$ are pressure components in the directions of $S_{\mu}$ and $l_{\mu}$. Furthermore, $\eta=P_{yy}-P_{\bot}$ determines the deviation of isotropic pressure distribution heading toward $l_{\mu}$, and $h_{\mu}$ is known as four vector heat flux which represents the transmission of energy with respect to the moving observer.
It describes the symmetry of four-velocity $(h_{\mu}U^{\mu}=0)$, and terms $h_{\mu}U^{\nu}+h_{\nu}U^{\mu}$, specifies the dissipative heat transmission in fluid composition. Moreover, the term  $(P_{xx}-P_{yy})S_{\mu}S_{\nu}$ plays a crucial role in assessing the complexity through structure scalars and stability of anisotropic matter.

The electromagnetic energy-momentum tensor has the following form \cite{yousaf2023cylindrical}
\begin{equation}\label{C1}
\mathbb{E}_{\mu \nu}=\frac{1}{4 \pi}\left\{-F^{\delta}_{\mu}F_{\nu \delta}+\frac{1}{4}F^{\delta \tau}F_{ \delta \tau}g_{\mu \nu}\right\},
\end{equation}
here, $F_{\mu \nu}=\psi_{\nu, \mu}-\psi_{\mu, \nu}$ and $\psi_{\mu}$ specify the electromagnetic field tensor and four potential, respectively. The corresponding electromagnetic field equations are
\begin{equation}\label{C2}
F^{\mu \nu}_{; \nu}=\mu_{0}J^{\mu}, \quad F_{[\mu \nu ; \delta]}=0,
\end{equation}
here, $J^{\mu}$ and $\mu_{0}$ are written as four current and magnetic permeability. It is considered that stationary charge distribution implies absence of magnetic field. Therefore,
\begin{equation}\nonumber
\psi_{\mu}=\psi (t, r) \delta ^{0}_{\mu}, \quad J^{\mu}=\varrho(t, r) V^{\mu},
\end{equation}
in which $\varrho$ represents the charge density. The Maxwell field equations result
\begin{equation}\label{C3}
\frac{\partial ^{2}\psi}{\partial r^{2}}-\left(\frac{E^{'}}{E}+\frac{D^{'}}{D}-\frac{2F^{'}}{F}\right)\frac{\partial \psi}{\partial r}=4 \pi \varrho E^{2}D.
\end{equation}
\begin{equation}\label{C4}
\frac{\partial}{\partial t}\left(\frac{\partial \psi}{\partial r}\right)-\left(\frac{\dot{D}}{D}+\frac{\dot{E}}{E}-\frac{2\dot{F}}{F}\right)\frac{\partial \psi}{\partial r}=0.
\end{equation}
here, $.$ and $\prime$ specify the derivative with respect to time and radial coordinate respectively. The integration of Eq. \eqref{C3} concludes
\begin{equation}\nonumber
\psi ^{'}=\frac{q DE}{F^{2}}
\end{equation}
where,
\begin{equation}\label{C5}
q(r)=2 \pi \int _{0}^{r} \varrho E F^{2}dr.
\end{equation}
is the overall quantity of charge within a radius $r$ determined through the conservation law, $J^{\mu}_{; \mu}=0$. The electric field intensity is defined as the charge per unit spherical area and mathematically expressed as
\begin{equation}\label{C6}
\mathbb{E}(t,r)=\frac{q}{4 \pi F^{2}}.
\end{equation}
The non-zero components of Eq. \eqref{3} are derived in subsequent form
\begin{align}\nonumber
\mathcal{G}_{00}&=\frac{2 \ddot{F}}{F}f_{R}+\frac{\ddot{E}}{E}f_{R}-\frac{D D^{''}}{E^{2}}f_{R}-\frac{2F^{'}D^{'}D}{E^{2}F}f_{R}-\frac{2 \dot{D}\dot{F}}{DF}f_{R}-\frac{\dot{E}\dot{D}}{DE}f_{R}+\frac{DD^{'}E^{'}}{E^{3}}f_{R}-\frac{1}{2}f D^{2}-\frac{1}{2} f_{\mathcal{L}_{m}}D^{2}\\\label{5}&-\frac{1}{2} f_{\mathcal{L}_{m}} \rho-\frac{1}{2} f_{\mathcal{L}_{m}} \eta D^{2}-\frac{q^{2}D^{2}}{8 \pi F^{4}}+\varphi_{00},
\\\label{6}
\mathcal{G}_{01}&=\frac{2 \dot{F^{'}}}{F}f_{R}-\frac{2 \dot{E}F^{'}}{E F}f_{R}-\frac{2 D^{'}\dot{F}}{D F}f_{R}+\frac{1}{2}\eta DEf_{\mathcal{L}_{m}}-\frac{1}{2}h D Ef_{\mathcal{L}_{m}} +\varphi_{01},
\\\nonumber
\mathcal{G}_{11}&=\frac{2F^{''}}{F}f_{R}-\frac{E \ddot{E}}{D^{2}}f_{R}+\frac{D^{''}}{D}f_{R}-\frac{2 E^{'}F^{'}}{E F}f_{R}-\frac{2 E \dot{E}\dot{F}}{D^{2}F}f_{R}+\frac{\dot{D}\dot{E}E}{D^{3}}f_{R}-\frac{D^{'}E^{'}}{D E}f_{R}+\frac{1}{2}E^{2}f+\frac{1}{2}E^{2}f_{\mathcal{L}_{m}}\\\label{7}&+\frac{q^{2}E^{2}}{8 \pi F^{4}}-\frac{1}{2}E^{2}\eta f_{\mathcal{L}_{m}}-\frac{1}{2}E^{2}P_{xx}f_{\mathcal{L}_{m}}+\varphi_{11},
\\\nonumber
\mathcal{G}_{22}&=\frac{F F^{'}}{E^{2}}f_{R}-\frac{F \ddot{F}}{D^{2}}f_{R}+\frac{F^{'2}}{E^{2}}f_{R}-\frac{F F^{'}E^{'}}{E^{3}}f_{R}+\frac{F D^{'}F^{'}}{D E^{2}}f_{R}-\frac{\dot{F}^{2}}{D^{2}}f_{R}-\frac{\dot{F}\dot{E}F}{D^{2}E}f_{R}+\frac{\dot{F}\dot{D}F}{D^{3}}f_{R}-f_{R}
+\\\label{8}&\frac{1}{2}f F^{2}-\frac{q^{2}}{8 \pi F^{2}}+\frac{1}{2}f_{\mathcal{L}_{m}}F^{2}-\frac{1}{2}f_{\mathcal{L}_{m}}F^{2}P_{yy}+\varphi_{22},
\end{align}
the values of $\varphi_{00}$, $\varphi_{01}$, $\varphi_{11}$, and $\varphi_{22}$ are written in Appendix. The non-conserved configuration of energy-momentum tensor is regarded as the most conspicuous aspect of $f(R, \mathcal{L}_{m})$ theory which is emerged from the direct interaction between matter and geometry. These relationships ought to be described as the violation of the fundamental conservation law as well as the energy transmission through matter and gravitational field. In our case study, we aim to analyze the dynamical instability related to the spherically symmetric fluid dispersions in $f(R, \mathcal{L}_{m})$ gravity, despite the fact that the non-conservation of energy-momentum tensor is endemic to $f(R, \mathcal{L}_{m})$ gravity and the methodology we contemplated, encompassing perturbation and stability restraints; is relevant to the particular form of considered $f(R, \mathcal{L}_{m})$  theory. In the context of GR, the conservation law is applied to energy momentum tensor specified as \cite{herrera2010collapsing,herrera2012dynamical}
\begin{equation}\label{3c}
\nabla ^{\mu}T_{\mu \nu}=0
\end{equation}
However, in $f(R, \mathcal{L}_{m})$ gravity the conservation equation is impacted by the relationship between matter and geometry as the Lagrangian is directly based on matter Lagrangian. As a result, the covariant divergence persists \cite{wang2012energy}, resulting the subsequent non-conserved form
\begin{equation}
\nabla ^{\mu}T_{\mu \nu}=2\nabla ^{\mu} ln \left\{f_{\mathcal{L}_{m}}\right\}\frac{\partial \mathcal{L}_{m}}{\partial g^{\mu \nu}}.
\end{equation}
This significant interaction is explicitly presented before Eq. (\ref{10}), demonstrating the deviation from standard energy-momentum conservation. For the analysis of anisotropic fluid distribution in  $f(R, \mathcal{L}_{m})$ theory, non-conserved equations are determined by allocating $\nu$ value of zero and one by giving variation to $\mu$ and obtained two non-zero components. The derivative of radial coordinate is represented by prime  $(')$, whereas dot $(.)$ specifies the time coordinate derivative in these equations. The following are the resultant equations
\begin{align}\nonumber
&\frac{2 \ddot{F}}{D^{4}F}\dot{f}_{R}+\frac{2 \dddot{F}}{D^{4}F}f_{R}-\frac{8 \dot{D}\ddot{F}}{D^{5}F}f_{R}+\frac{ \dot{F}\ddot{F}}{D^{4}F^{2}}f_{R}+\frac{\ddot{E}}{D^{4}E}\dot{f}_{R}+\frac{\dddot{E}}{D^{4}E}f_{R}
-\frac{2\dot{D}\ddot{E}}{D^{5}E}f_{R}
-\frac{\dot{E}\ddot{E}}{D^{4}E^{2}}f_{R}-\frac{D^{''}}{D^{3}E^{2}}\dot{f}_{R}\\\nonumber&+\frac{\dot{D}}{D^{4}}
 \frac{D^{''}}{E^{2}}f_{R}-\frac{\dot{D}^{''}}{D^{3}E^{2}}f_{R}
-\frac{2 D^{'}F^{'}}{D^{3}E^{2}F}\dot{f}_{R}-\frac{2\dot{D}^{'}F^{'}}{D^{3}E^{'}F}f_{R}-\frac{4 D^{'}\dot{F}^{'}}{D^{3}E^{2}F}f_{R}
+\frac{2D^{'}F^{'}\dot{D}}{D^{4}E^{2}F}f_{R}+\frac{4 D^{'}F^{'}\dot{E}}{D^{3}E^{3}F}f_{R}+\frac{3}{D}\\\nonumber& \times \frac{D^{2}D^{'}F^{'}\dot{F}}{E^{2}F^{2}}f_{R}-\frac{2 \dot{D}\dot{F}}{D^{5}F}\dot{f_{R}}-\frac{2 \dot{D}^{'}\dot{F}}{D^{5}F}f_{R}
-\frac{2 \dot{D}\dot{F}^{'}}{D^{5}F}f_{R}-\frac{2\dot{D}^{2}\dot{F}}{D^{6}F}f_{R}-\frac{\dot{D}\dot{E}}{D^{5}E}\dot{f_{R}}
-\frac{2\ddot{D}\dot{E}}{D^{5}E}f_{R}+\frac{3\dot{D}^{2}\dot{E}}{D^{6}E}f_{R}\\\nonumber&+\frac{\dot{D}\dot{E}^{2}}{D^{5}E^{2}} f_{R}+\frac{D^{'}E^{'}}{D^{3}E^{3}}\dot{f_{R}}+\frac{q^{2}\dot{D}}{4 \pi D^{3}F^{4}}+\frac{\dot{D^{'}}E^{'}}{D^{3}E^{3}}f_{R}+\frac{\dot{E^{'}}D^{'}}{D^{3}E^{3}}f_{R}
-\frac{D^{'}E^{'}\dot{D}}{D^{4}E^{3}}f_{R}-\frac{2D^{'}E^{'}\dot{E}}{D^{3}E^{4}}f_{R}-\frac{\dot{f}}{2D^{2}}+\frac{f }{D}\\\nonumber&\times \frac{\dot{D}}{D^{2}} -\frac{1}{2}\frac{\dot{f}_{_{\mathcal{L}_{m}}}}{D^{2}}+\frac{f_{\mathcal{L}_{m}}}{D}\frac{\dot{D}}{D^{2}} +\frac{\dot{\varphi_{00}}}{D^{4}}
-\frac{4 \dot{D}}{D^{5}}\varphi_{00}-\frac{\rho}{2 D^{4}}\dot{f}_{\mathcal{L}_{m}}-\frac{\eta}{2 D^{4}}\dot{f_{\mathcal{L}_{m}}}-\frac{\dot{\rho}}{2 D^{4}}f_{\mathcal{L}_{m}}+\frac{2 \dot{D}\rho}{ D^{5}}f_{\mathcal{L}_{m}}+\frac{ \eta \dot{D}}{ D^{3}}f_{\mathcal{L}_{m}}\\\nonumber&-\frac{2\dot{F^{'}}}{D^{2}E^{2}F}f_{R}^{'}-\frac{2}{D^{2}} \frac{\dot{F^{''}}}{E^{2}F}f_{R}
-\frac{6 \dot{F}^{'}E^{'}}{E^{3}D^{2}} \frac{f_{R}}{F} -\frac{4 \dot{F}^{'}F^{'}}{D^{2}E^{2}F^{2}}f_{R}+\frac{2 \dot{E}F^{'}}{D^{2}E^{3}F}
f_{R}^{'}+\frac{2 \dot{E^{'}}F^{'}}{D^{2}E^{3}F}f_{R}+\frac{2\dot{E}F^{''}}{D^{2}E^{3}F}f_{R}-\\\nonumber&\frac{6\dot{E}F^{'}E^{'}}{D^{2}E^{4}F}f_{R}
+\frac{2}{D^{3}} \frac{\dot{F}D^{'}}{E^{2}F}f_{R}^{'}+\frac{2\dot{F}}{D^{3}} \frac{D^{''}}{E^{2}F}f_{R}
-\frac{2\dot{F}}{D^{4}E^{2}}\frac{D^{'2}}{F}f_{R} -\frac{2\dot{F}D^{'}F^{'}}{D^{3}E^{2}F}f_{R}-\frac{q^{2}\dot{D}}{4 \pi D^{3}F^{4}}-\frac{\varphi_{01}^{'}}{D^{2}E^{2}}
-\frac{2D^{'}}{D^{3}E^{2}}\\\nonumber& \times \varphi_{01}-\frac{2E^{'}}{D^{2}E^{3}}\varphi_{01}-\frac{\eta}{D^{3}E^{3}}f_{\mathcal{L}_{m}}
+\frac{3\eta D^{'}}{D^{4}E^{3}}f_{\mathcal{L}_{m}}+\frac{q^{2}\dot{F}}{2 \pi F^{5}D^{2}}+\frac{3E^{'}\eta}{D^{3}E^{4}}f_{\mathcal{L}_{m}}+\frac{h}{D^{3}} \frac{f_{\mathcal{L}_{m}}^{'}}{E^{3}}
-\frac{3D^{'}h}{D^{4}E^{3}}f_{\mathcal{L}_{m}}-\frac{3}{D^{3}}\\\nonumber& \times \frac{E^{'}h}{E^{3}}f_{\mathcal{L}_{m}}+\frac{q^{2}\dot{F}}{4 \pi F^{5}D^{2}}+\frac{4\ddot{F}\dot{D}}{D^{5}F}f_{R}-\frac{\dot{D}}{D^{3}} f
-\frac{\dot{D}}{D^{3}}f_{\mathcal{L}_{m}}-\frac{\dot{D}}{D} \frac{\rho}{D^{4}}f_{\mathcal{L}_{m}}-\frac{\dot{D}\eta}{D^{3}}f_{\mathcal{L}_{m}}
-\frac{2\dot{D}}{D^{5}}\varphi_{00}+\frac{2\ddot{F}\dot{E}}{D^{4}E} \frac{f_{R}}{F}-\\\nonumber& \times \frac{4\dot{F}\dot{E}\dot{D}}{D^{5}EF}f_{R}
-\frac{\dot{E}}{2D^{2}E}f-\frac{\dot{E}}{2D^{2}E}f_{\mathcal{L}_{m}}-\frac{\dot{E}}{2D^{2}} \frac{f_{\mathcal{L}_{m}}}{E}-\frac{\dot{E}\rho}{2D^{4}E}  f_{\mathcal{L}_{m}} -\frac{\eta \dot{E}}{2D^{2}E}f_{\mathcal{L}_{m}}+\frac{\dot{E}}{D^{4}E}\varphi_{00}+\frac{\dot{F}}{D^{4}} \frac{\ddot{E}}{EF}f_{R}\\\nonumber&-\frac{q^{2}\dot{E}}{8 \pi D^{2}EF^{4}}
-\frac{2D^{''}\dot{F}}{D^{3}E^{2}F}f_{R}-\frac{\dot{F}}{D^{2}F}f-\frac{\dot{F}f_{\mathcal{L}_{m}}}{D^{2}F}-\frac{\rho \dot{F}}{D^{4}F}f_{\mathcal{L}_{m}}-\frac{\eta \dot{F}}{D^{2}F}f_{\mathcal{L}_{m}}+\frac{2\dot{F}}{D^{4}F}\varphi_{00}-\frac{D^{'}}{D^{3}E^{2}}
 \varphi_{01}-\frac{2}{D^{4}}\\\nonumber& \times \frac{D^{'}\eta}{E^{3}} f_{\mathcal{L}_{m}}+\frac{2D^{'}h}{D^{4}E^{3}}f_{\mathcal{L}_{m}}-\frac{E^{'}}{D^{2}E^{3}}\varphi_{01}
-\frac{E^{'}\eta}{D^{3}E^{4}}f_{\mathcal{L}_{m}}+\frac{hE^{'}}{D^{3}} \frac{f_{\mathcal{L}_{m}}}{E^{4}} -\frac{2F^{'}}{D^{2}E^{2}F}\varphi_{01}+\frac{q^{2}\dot{E}}{8 \pi D^{2}EF^{4}}-\frac{2\eta F^{'}}{D^{3}E^{3}}\\\nonumber& \times \frac{f_{\mathcal{L}_{m}}}{F}+\frac{2h F^{'}}{D^{3}E^{3}F}f_{\mathcal{L}_{m}}+\frac{2F^{''}\dot{E}}{D^{2}E^{3}F}f_{R}
-\frac{2\dot{E}^{2}\dot{F}}{D^{4}E^{2}F}f_{R}+\frac{\dot{E}}{2D^{2}E} f-\frac{q^{2}}{8 \pi} \frac{\dot{F}}{ D^{2}F^{5}}+\frac{\dot{E}}{2D^{2}} \frac{f_{\mathcal{L}_{m}}}{E}
+\frac{\dot{E}}{D^{2}E^{3}}\varphi_{11}-\\\nonumber&\frac{\eta \dot{E}}{2D^{2}E}f_{\mathcal{L}_{m}}-\frac{\dot{E} P_{xx}}{2D^{2}E}f_{\mathcal{L}_{m}}
+\frac{\dot{F}F^{''}}{D^{2}E^{2}F^{2}}f_{R}-\frac{\dot{F}F^{'}E^{'}}{D^{2}E^{3}F^{2}}f_{R}
-\frac{\dot{F}^{3}}{D^{4}} \frac{f_{R}}{F^{3}}
-\frac{\dot{F}^{2}\dot{E}}{D^{4}F^{2}E} f_{R}+\frac{\dot{F}^{2}\dot{D}}{D^{5}F^{2}}f_{R}-\frac{\dot{F}}{D^{2}F^{3}}f_{R}\\\label{10}&
+\frac{\dot{F}}{2D^{2}F}f_{R}+\frac{\dot{F}}{2D^{2}F}f_{\mathcal{L}_{m}}
-\frac{\dot{F}P_{yy}}{2D^{2}}f_{\mathcal{L}_{m}}
+\frac{\dot{F}}{D^{2}F^{3}}
\varphi_{22}-\mathcal{Z}_{1}=0,
\\\nonumber
&\frac{-2\dot{F}^{'}}{D^{2}E^{2}F}\dot{f_{R}}-\frac{2\ddot{F}^{'}}{D^{2}E^{2}F}f_{R}+\frac{2\dot{F}^{'}\dot{D}}{D^{3}E^{2}F}f_{R}
+\frac{2\dot{F}^{'}\dot{E}}{D^{2}E^{3}F}f_{R}-\frac{2\dot{F}^{'}\dot{F}}{D^{2}E^{2}F^{2}}f_{R}-\frac{4\dot{E}^{2}F^{'}}
{D^{2}E^{4}F}f_{R}+\frac{\dot{E}\dot{F}F^{'}}{D^{2}E^{3}F^{2}}f_{R}+\frac{\dot{F}}{D^{3}}\\\nonumber& \times \frac{D^{'}}{E^{2}} \frac{\dot{f_{R}}}{F}  +\frac{2\dot{D}^{'}\dot{F}}{D^{3}E^{2}F}f_{R}+\frac{4D^{'}\ddot{F}}{D^{3}E^{2}F}f_{R}-\frac{6D^{'2}\dot{F}}
{D^{4}E^{2}F}f_{R}
+\frac{2D^{'}\dot{F}^{2}}{D^{3}E^{2}F^{2}}f_{R}+\frac{q q^{'}}{4 \pi E^{2}F^{4}}-\frac{D^{2} q^{'}}{4 \pi E^{3}F^{5}}-\frac{q^{2} F^{'}}{ E^{2}F^{5}}-\\\nonumber&\frac{\dot{\varphi}_{01}}{D^{2}E^{2}}-\frac{2\dot{D}}{D^{3}E^{2}}\varphi_{01}
-\frac{2\dot{E}}{D^{2}E^{3}}\varphi_{01}
+\frac{2 F^{''}}{E^{4}} \frac{f_{R}^{'}}{F} +\frac{2 F^{'''}}{E^{4}F}f_{R}-\frac{8 F^{''}E^{'}}{E^{5}F}f_{R}+\frac{ F^{''}F^{'}}{E^{4}F^{2}}f_{R}-\frac{\ddot{E}}{D^{2}E^{3}}f_{R}^{'}-\frac{\ddot{E}^{'}}{D^{2}E^{3}}f_{R}\\\nonumber&
+\frac{2\ddot{E}D^{'}}{D^{3}E^{3}} f_{R}+\frac{2\ddot{E}E^{'}}{D^{2}E^{4}}f_{R}+\frac{D^{''}}{DE^{4}}f_{R}^{'}
+\frac{D^{'''}}{D} \frac{f_{R}}{E^{3}} -\frac{D^{'}D^{''}}{D^{2}E^{4}}f_{R}-\frac{4D^{'}D^{''}E^{'}}{D^{2}E^{5}}f_{R}-\frac{2E^{'}F^{'}}{E^{5}}f_{R}^{'}
-\frac{2E^{''}F^{'}}{E^{5}F}f_{R}\\\nonumber&+\frac{6E^{'2}F^{'}}{E^{6}F}f_{R}+\frac{3E^{'}F^{'2}}{E^{5}F^{2}}f_{R}
-\frac{2\dot{E}\dot{F}}{D^{2}E^{3}F}
f_{R}- \frac{2}{D^{2}} \frac{\dot{E}^{'}\dot{F}}{E^{3}F}f_{R}  -\frac{q^{2} D^{'}}{8 \pi DE^{2}F^{4}} +\frac{\dot{E}\dot{D}}{D^{3}E^{3}}f_{R}^{'}+\frac{\dot{E}\dot{D}^{'}}{D^{3}E^{3}}f_{R}
+\frac{\dot{D}}{D^{3}}\\\nonumber& \times\frac{\dot{E}^{'}}{E^{3}}f_{R}-\frac{3\dot{E}\dot{D}D^{'}}{D^{4}E^{3}}
 f_{R}+\frac{3q^{2} F^{'}}{8 \pi E^{2}F^{5}}+\frac{2\dot{E}\dot{F}E^{'}}{D^{2}FE^{4}}f_{R}
-\frac{\dot{E}\dot{D}E^{'}}{D^{3}E^{4}}f_{R}-\frac{D^{'}E^{'}}{DE^{5}}f_{R}^{'}+\frac{D^{''}}
{D} \frac{E^{'}}{E^{5}} f_{R}-\frac{E^{''}D^{'}}{DE^{5}}
 f_{R}\\\nonumber&+\frac{3D^{'}E^{'2}}{DE^{6}}f_{R}+\frac{f^{'}}{2E^{2}}-\frac{f E^{'}}{E^{3}}+\frac{f_{\mathcal{L}_{m}}^{'}}{2 E^{2}}
-\frac{E^{'}}{E^{3}}f_{\mathcal{L}_{m}}-\frac{\eta}{2 E^{2}}f_{\mathcal{L}_{m}}^{'}-\frac{P_{xx}}{2 E^{2}}f_{\mathcal{L}_{m}}+\frac{\eta E^{'}}{E^{3}}f_{\mathcal{L}_{m}}-\frac{P_{xx}^{'}}{2E^{2}} f_{\mathcal{L}_{m}}+\frac{q^{2}}{8 \pi}\\\nonumber& \times\frac{ D^{'}}{ DE^{2}F^{4}} +\frac{E^{'}P_{xx}^{'}}{E^{3}} f_{\mathcal{L}_{m}}-\frac{2D^{'2}F^{'}}{D^{2}E^{4}F}f_{R} -\frac{D^{'}f}{2E^{2}D}-\frac{D^{'}}{2E^{2}D}f_{\mathcal{L}_{m}}-\frac{\rho D^{'}}{2E^{2}D^{3}}f_{\mathcal{L}_{m}}-\frac{\eta D^{'}}{2E^{2}D}f_{\mathcal{L}_{m}}+\frac{2q^{2}}{8 \pi}\\\nonumber& \times \frac{ E^{'}}{ D^{3}F^{4}} +\frac{D^{'}}{E^{2}D^{3}}\varphi_{00}\frac{\ddot{F}F^{'}}{F^{2}D^{2}E^{2}}f_{R} -\frac{F^{'3}}{E^{4}F^{3}}
 f_{R}- \frac{F^{'2}D^{'}}{DE^{4}F^{2}}f_{R}+\frac{F^{'}\dot{F}^{2}}{D^{2}E^{2}F^{3}}f_{R}
-\frac{\dot{F}F^{'}\dot{D}}{E^{2}D^{3}F^{2}} f_{R}+\frac{F^{'}}{E^{2}F^{3}}\\\nonumber& \times f_{R}+\frac{F^{'}}{2E^{2}F}f+\frac{F^{'}}
{2E^{2}F}f_{\mathcal{L}_{m}}
+\frac{P_{yy}F^{'}}{E^{2}F} f_{\mathcal{L}_{m}}-\frac{F^{'}}{E^{2}F^{3}} \varphi_{22}-\frac{2\dot{D}\dot{F}^{'}}{D^{3}E^{2}F}f_{R}
+\frac{2\dot{D}\dot{E}}{D^{3}E^{3}}f_{R}-\frac{\dot{D}}{D^{3}E^{2}}\varphi_{01}-\\\nonumber&\frac{2}{D^{2}} \frac{\dot{E}}{E^{3}}
\varphi_{01}-\frac{2\dot{F}\varphi_{01}}{D^{2}E^{2}F}
+\frac{2F^{''}D^{'}}{DE^{4}F}f_{R} -\frac{2D^{'}}{D} \frac{E^{'}F^{'}}{E^{5}F}f_{R}+\frac{D^{'}}{2DE^{2}}f +\frac{D^{'}}{2DE^{2}}f_{\mathcal{L}_{m}}+\frac{D^{'}}{DE^{4}}\varphi_{11}-\frac{\eta D^{'}}{2DE^{2}}f_{\mathcal{L}_{m}}\\\nonumber&-\frac{D^{'}P_{xx}}{2DE^{2}}f_{\mathcal{L}_{m}}+\frac{E^{'}}{E^{3}}f
+\frac{E^{'}}{E^{3}}f_{\mathcal{L}_{m}}
+\frac{2E^{'}}{E^{5}} \varphi_{11}
-\frac{\eta E^{'}}{E^{3}}f_{\mathcal{L}_{m}}-\frac{\eta E^{'}P_{xx}}{E^{3}} +\frac{2D^{''}F^{'}}{DE^{4}F}f_{R}-\frac{4E^{'}F^{'2}}{E^{5}F^{2}}f_{R}+\frac{2F^{'}}{E^{4}F}\\\label{11}& \times \varphi_{11}-\frac{\eta F^{'}}{E^{2}F}f_{\mathcal{L}_{m}}-\frac{F^{'}P_{xx}}{E^{2}F}f_{\mathcal{L}_{m}}-\mathcal{Z}_{2}=0.
\end{align}
The direct interaction between matter and geometry related to $f(R, \mathcal{L}_{m})$ theory in above equations particularly demonstrates the non-conserved form of energy-momentum tensor and these modified continuity equations are crucial in our perturbed analysis of the collapsing anisotropic composition. Additionally, the non-zero divergence refute the conserved form of GR and implies the exchange of energy and momentum inside the matter and geometric sectors that is an inherent feature of $f(R, \mathcal{L}_{m})$ theory. Therefore, the specific components of non-conserved interaction that correlate the contemplated spherically symmetric fluid configuration are stated in Eqs. (\ref{10}) and (\ref{11}). They define the radial and time components of $\nabla^{\mu}T_{\mu \nu} \neq 0$, featuring terms $\mathcal{Z}_{1}(t,r)$ and $\mathcal{Z}_{2}(t,r)$, including the dark source terms which emerged from $f(R, \mathcal{L}_{m})$ theory in the subsequent form:
\begin{align}\nonumber
\mathcal{Z}_{1}(t,r)&=-\frac{D^{2}\rho \dot{f}_{\mathcal{L}_{m}}}{f_{\mathcal{L}_{m}}}-\frac{\rho \dot{f}_{\mathcal{L}_{m}}}{f_{\mathcal{L}_{m}}}-\frac{\eta D^{2}\dot{f}_{\mathcal{L}_{m}}}{f_{\mathcal{L}_{m}}},
\\\nonumber
\mathcal{Z}_{2}(t,r)&=\frac{E^{2}\rho f_{\mathcal{L}_{m}}^{'}}{f_{\mathcal{L}_{m}}}-\frac{\eta E^{2}f_{\mathcal{L}_{m}}^{'}}{f_{\mathcal{L}_{m}}}-\frac{E^{2}P_{xx}f_{\mathcal{L}_{m}}^{'}}{f_{\mathcal{L}_{m}}}.
\end{align}
We placed $\mathcal{Z}_{1}(t, r, )$ and $\mathcal{Z}_{2}(t, r, )$ to left side of non-conserved equations to streamline our computations ensuring the proper representation of modified equations. The Ricci scalar is derived relating to constrained spherical charged composition and is revised according to the basic structural constraints and symmetries. Ricci scalar within the context of specified constraints is expressed in following:
\begin{equation}\label{12}
R=\frac{2D^{''}}{DE^{2}}-\frac{2E^{'}D^{'}}{D^{2}E^{3}}-\frac{4E^{'}F^{'}}{DE^{3}F}-\frac{2}{E^{2}}-\frac{2F^{''}}{EF^{2}}
+\frac{2\dot{D}\dot{F}}{DE^{2}F}-\frac{D\dot{F}^{2}}{D^{2}F^{2}}-\frac{2\ddot{E}}{D^{2}E}+\frac{4\ddot{F}}{D^{2}F}
+\frac{2\dot{D}\dot{E}}
{D^{3}E}-\frac{4\dot{E}\dot{F}}{D^{2}EF}.
\end{equation}

\section{Dynamics with $f(R, \mathcal{L}_{m})$ functions and Perturbation approach}

Different $f(R, \mathcal{L}_{m})$ gravity models with significant implications are studied in literature \cite{harko2010f,jaybhaye2023baryogenesis,jaybhaye2022constraints}. Usually, considered gravity models are : i) $f(R, \mathcal{L}_{m})= f(R) +\mathcal{L}_{m}$, determining the minimal coupling between matter and geometry and can easily be simplified to $f(R)$ theory framework. ; ii) $f(R, \mathcal{\mathcal{L}}_{m})= R +\alpha R \mathcal{L}_{m}$ illustrates the non conserved form of energy-momentum tensor and non-geodesic motion of test particles through the explicit interaction between curvature and matter. This gravity model plays a crucial role in studying the divergence from equivalence principle and particular cosmic constraints.
The subsequent form of $f(R, \mathcal{L}_{m})$ gravity model is contemplated for our study \cite{devi2024constraining}
\begin{equation}\nonumber
f(R, \mathcal{L}_{m})=R+\zeta R^{2}+\chi \mathcal{L}_{m}^{2},
\end{equation}
where, $R$ represents the Ricci scalar that describes the relationship between presence of matter and the curvature of spacetime, $\mathcal{L}_{m}$ refers to the Lagrangian density associated with various matter fields, and $\zeta$ and $\chi$ are the coupling constants. The addition of the simplest quadratic correction term $\zeta R^{2}$ in the Einstein-Hilbert action implies important results for the higher gravitating domains and appears to be the interacting term relating to the gravitational field influencing its behavior against the presence of high energies. The constant $\zeta$ regulates the magnitude of curvature terms and sign of $\zeta$ has significant relevance in establishing the fundamental significance of highly dense matter. The positive $\zeta$ ensures the absence of ghost-like instabilities in the context of higher correction terms, moreover, it recommends the emergence of massive scalaron particles and specifies the mass for these particles. While, negative $\zeta$ corresponds to the ghost instabilities in the gravitational region, which causes the catastrophic vacuum depletion due to the energy of associated degrees of freedom and does not imply compatibility with the observable results. While, the term $\chi \mathcal{L}_{m}^{2}$ incorporates quadratic non-minimal interaction between matter and gravity and suggests that the relationship between different matter fields is profoundly impacted by corresponding Lagrangian density. The mathematical form of GR proposes that the test particles acquire uniform acceleration in the gravitational field. But, the term $\chi$ describes the specific interior matter distribution and implies that various test particles accelerate separately in a gravitational field, which is a considerable violation from the  basic assumptions of GR. The particular interpretation of sign of $\chi$ depends on the Lagrangian density relating to the matter fields. The consideration of Lagrangian density as the negative pressure along with the positive value of $\chi$ serves as the constant pressure for the stress-energy dispersion. In the case of high-density compositions, this positivity specifies the stiffer EOS, which corresponds to increased matter resilience towards pressure and resists the attractive force of gravity, which may influence the dynamics of gravitational collapse. While, the negative $\chi$ specifies the relatively `` softer '' matter which is readily compressed and negative pressure increases the rate of gravitational collapse that eventually impacts the dynamics of supernova or contribute in the development of compact structures.

Now, we address the importance of perturbing field equations as well as the dynamical equations in evaluating the stability of highly dense fluid composition. Perturbations are small-scaled variations in physical system caused by the gravitational interactions with adjacent compact formations. The study of stability associated with the highly dense matter against the oscillatory perturbations is the topic of research for last few years \cite{semelin2001self,yousaf2017role,herrera2011physical}. In this manuscript, subscript $0$ specifies that the quantity is dependent to radius $r$ and label them as static and we consider the technique of linear perturbation after contemplating the small perturbation parameter $\epsilon$ which allows us to neglect its higher order derivatives. The perturbation methodology leads to the particular form of dynamical equations that are essential for assessing the unstable zones in $f(R, \mathcal{L}_{m})$ theory. The considered system is supposed to be at balanced state but with time it experiences fluctuations. The perturbation technique corresponding to the metric coefficients and matter parameters is specified as follows \cite{herrera2012dynamical}.
\begin{align}\label{P1}
D(t,r)&=D_{0}(r)+\epsilon \mathcal{E}(t)a(r),
\\\label{P2}
E(t,r)&=E_{0}(r)+\epsilon \mathcal{E}(t)b(r),
\\\label{P3}
F(t,r)&=F_{0}(r)+\epsilon \mathcal{E}(t)c(r),
\\\label{P4}
R(t,r)&=R_{0}(r)+\epsilon \mathcal{E}(t)e(r),
\\\label{PE}
\mathbb{E}(t,r)&=\mathbb{E}_{0}(r)+\epsilon \mathcal{E}(t)i(r),
\\\label{P5}
P_{xx}(t,r)&=P_{xx_{0}}(r)+\epsilon \bar{P}_{xx}(t,r),
\\\label{P6}
P_{yy}(t,r)&=P_{yy_{0}}(r)+\epsilon \bar{P}_{yy}(t,r),
\\\label{P8}
\rho(t,r)&=\rho_{{0}}(r)+\epsilon \bar{\rho}(t,r),
\\\label{P9}
T(t,r)&=T_{{0}}(r)+\epsilon \bar{T}(t,r).
\end{align}
This methodology substantially simplifies the perturbation technique and the perturbed formalism corresponding to $f(R, \mathcal{L}_{m})$ function is as follows
\begin{equation}\label{13a2}
f(R,\mathcal{L}_{m})=R_{0}+\zeta R_{0}^{2}+\chi \mathcal{L}_{m_{0}}^{2}+\epsilon \left(e\mathcal{E}+2\zeta e\mathcal{E}R_{0}+2\chi \mathcal{L}_{m_{0}}\bar{\mathcal{L}}_{m}\right).
\end{equation}

Our considered perturbation approach adheres to a standard framework in order to evaluate highly dense matter compositions opposed to the slight variations from the equilibrium condition and this methodology is successfully established and used in \cite{herrera2012dynamical,herrera1989dynamical,bhatti2026dynamics}. This approach develops metric coefficients and fundamental parameters exceeding their equilibrium values through incorporating the perturbation parameter $\epsilon$. The subscript $0$ denotes the equilibrium state corresponding to these parameters, while the perturbed values are defined as the product of spatial and temporal components. The complex partial differential equations are simplified into standard differential equations through the contemplation of spatial and temporal parts. This methodology ensures mathematical compliance with basic Einstein's field equations and their modified variants. The perturbation approach described in Eqs. (\ref{P1}-\ref{P9}), distinguishes between spatial and temporal components through the inclusion of revised metric and matter functions into field equations while preserving the linearized appearance of $\epsilon$; which validates the compliance of linear equations with basic dynamics of the system. The separation of time-related components is important in assessing the stability of highly dense matter by thoroughly describing the emergence of matter composition with time.

The conventional approach for evaluating the behavior of celestial formations and anisotropic fluid configurations is splitting the temporal and spatial parts as described in \cite{herrera2012dynamical,herrera2010cavity,ur2024dynamically}. These manuscripts emphasize the significance of separation technique in assessing the behavior of compact objects in the formalism of certain restraints, additionally, this methodology is crucial in analyzing isotropic and anisotropic fluid configurations in the existence of electric field and different gravitational fields. Furthermore, the consideration of $\mathcal{E}(t)$ in Eqs. (\ref{P1}-\ref{PE}), permits linear perturbation scheme after separating spatial and temporal parts of the field equations and its inclusion enables the analysis of temporal evolution of compact objects after maintaining the spatial formalism of perturbation. This assumption is particularly pertinent in evaluating minor variations from state of equilibrium, given that the temporal part prevails in establishing fundamental characteristics of matter composition. Moreover, solving reduced field equations in the presence of minor variations requires the generalized time dependent function. The absence of greater order $\epsilon$ terms in the process of obtaining simplifying terms yields only the time-related perturbations.
The corresponding temporal representation resulted from interaction between perturbed metric and matter parts in field equations ensures the consistency of these equations with our considered perturbation approach. The static components of modified field equations are derived after considering the most basic perturbation approach along with the usage of Eqs. (\ref{P1}-\ref{P9}), whereas associated non-static components are of the form:
\begin{align}\nonumber
\bar{\rho}&=\frac{4c\zeta R_{0}\ddot{\mathcal{\mathcal{E}}}}{\chi \mathcal{\mathcal{L}}_{m_{0}}F_{0}}+\frac{2\zeta R_{0}b\ddot{\mathcal{\mathcal{E}}}}{\chi \mathcal{\mathcal{L}}_{m_{0}}E_{0}}-\frac{2\zeta R_{0}D_{0}\mathcal{\mathcal{E}}a^{''}}{\chi \mathcal{\mathcal{L}}_{m_{0}}E_{0}^{2}}+\frac{2\zeta R_{0}D_{0}^{''}\mathcal{\mathcal{E}}a}{\chi \mathcal{\mathcal{L}}_{m_{0}}E_{0}^{2}}-\frac{2\zeta D_{0}D_{0}^{''}\mathcal{\mathcal{E}}e}{\chi \mathcal{\mathcal{L}}_{m_{0}}E_{0}^{2}}+\frac{4\zeta R_{0}D_{0}D_{0}^{''}\mathcal{\mathcal{E}}b}{\chi \mathcal{\mathcal{L}}_{m_{0}}E_{0}^{3}}
-\frac{4\zeta R_{0}}{\chi \mathcal{\mathcal{L}}_{m_{0}}}\\\nonumber&
\times \frac{D_{0}D_{0}^{'}\mathcal{\mathcal{E}}c^{'}}{E_{0}^{2}F_{0}}+\frac{2 \zeta }{\chi \mathcal{\mathcal{L}}_{m_{0}}E_{0}}
\frac{R_{0}D_{0}\mathcal{\mathcal{E}}a^{'}}{\chi \mathcal{\mathcal{L}}_{m_{0}}E_{0}}-\frac{4\zeta R_{0}D_{0}D_{0}^{'}\mathcal{\mathcal{E}}c^{'}}{\chi \mathcal{\mathcal{L}}_{m_{0}}E_{0}^{2}F_{0}}
-\frac{4\zeta R_{0} F_{0}^{'}D_{0}^{'}\mathcal{\mathcal{E}}a}{\chi \mathcal{\mathcal{L}}_{m_{0}}E_{0}^{2}F_{0}}-\frac{4\zeta F_{0}^{'}D_{0}^{'}D_{0}\mathcal{\mathcal{E}}e}{\chi \mathcal{\mathcal{L}}_{m_{0}}E_{0}^{2}F_{0}}+\frac{4\zeta R_{0}}{\chi \mathcal{\mathcal{L}}_{m_{0}}}\\\nonumber&
\times \frac{ F_{0}^{'}D_{0}^{'}D_{0}\mathcal{\mathcal{E}}c}{E_{0}^{2}F_{0}^{2}}+\frac{8\zeta R_{0} F_{0}^{'}D_{0}^{'}D_{0}\mathcal{\mathcal{E}}b}{\chi \mathcal{\mathcal{L}}_{m_{0}}E_{0}^{3}F_{0}^{2}}
+\frac{2\zeta R_{0} \mathcal{\mathcal{E}}aD_{0}^{'}E_{0}^{'}}{\chi \mathcal{\mathcal{L}}_{m_{0}}E_{0}^{3}}+\frac{2\zeta R_{0} \mathcal{E}b^{'}D_{0}^{'}D_{0}^{'}}{\chi \mathcal{\mathcal{L}}_{m_{0}}E_{0}^{3}}+
 \frac{2\zeta R_{0} E_{0}^{'}D_{0}^{'}D_{0} \mathcal{\mathcal{E}}e}{\chi \mathcal{\mathcal{L}}_{m_{0}}E_{0}^{3}}-\frac{6\zeta}{\chi}\\\nonumber& \times \frac{ R_{0}\mathcal{\mathcal{E}}bE_{0}^{'}D_{0}^{'}D_{0}}{ \mathcal{\mathcal{L}}_{m_{0}}E_{0}^{4}}-\frac{\mathcal{\mathcal{E}}eD_{0}^{2}}{2\chi \mathcal{\mathcal{L}}_{m_{0}}}-\frac{\zeta \mathcal{\mathcal{E}}}{2\chi \mathcal{\mathcal{L}}_{m_{0}}eD_{0}^{2}R_{0}} +\frac{2D_{0}^{2}\mathcal{\mathcal{E}}c^{''}}{\chi \mathcal{\mathcal{L}}_{m_{0}}F_{0}E_{0}^{2}}+\frac{4\mathcal{\mathcal{E}}aD_{0}F_{0}^{''}}{\chi \mathcal{\mathcal{L}}_{m_{0}}F_{0}E_{0}^{2}}
-\frac{4\mathcal{E}bD_{0}^{2}F_{0}^{''}}{\chi \mathcal{\mathcal{L}}_{m_{0}}E_{0}^{3}F_{0}}-\frac{2\mathcal{E}}{\chi}\\\nonumber&
 \times \frac{cD_{0}^{2}F_{0}^{''}}{F_{0}^{2}E_{0}^{2} \mathcal{\mathcal{L}}_{m_{0}}}+\frac{2\mathcal{E}c^{'}D_{0}^{'}F_{0}^{'}}{\chi \mathcal{\mathcal{L}}_{m_{0}}F_{0}^{2}E_{0}^{2}}
+\frac{\mathcal{E}a^{'}F_{0}^{'2}}{\chi \mathcal{\mathcal{L}}_{m_{0}}F_{0}^{2}E_{0}^{2}}-\frac{2\mathcal{E}cD_{0}^{'}F_{0}^{'2}}{\chi \mathcal{\mathcal{L}}_{m_{0}}F_{0}^{3}E_{0}^{2}}
-\frac{2\mathcal{E}bD_{0}^{'}F_{0}^{'2}}{\chi \mathcal{\mathcal{L}}_{m_{0}}F_{0}^{2}E_{0}^{3}}
-\frac{2\mathcal{E}b^{'}F_{0}^{'}D_{0}}{\chi \mathcal{\mathcal{L}}_{m_{0}}E_{0}^{3}}-\frac{2\mathcal{E}c^{'}}{\chi \mathcal{\mathcal{L}}_{m_{0}}}\\\nonumber& \times \frac{E_{0}^{'}D_{0}}{E_{0}^{3}}-\frac{4\mathcal{E}aF_{0}^{'}E_{0}^{'}}{\chi \mathcal{\mathcal{L}}_{m_{0}}E_{0}^{3}}
+\frac{6F_{0}^{'}E_{0}^{'}D_{0}}{\chi \mathcal{\mathcal{L}}_{m_{0}}E_{0}^{4}}+\frac{2\mathcal{E}aF_{0}^{'}E_{0}^{'}}{\chi \mathcal{\mathcal{L}}_{m_{0}}E_{0}^{3}}-  \frac{2a\mathcal{E}D_{0}}{\chi \mathcal{\mathcal{L}}_{m_{0}}F_{0}^{2}}+
 \frac{2\mathcal{E}cD_{0}^{2}}{\chi \mathcal{\mathcal{L}}_{m_{0}}F_{0}^{3}}
+2 \mathcal{E}a \eta D_{0}-\frac{4 \pi}{\chi \mathcal{L}_{m_{0}}}\\\label{16}&-\frac{4 \pi a\mathcal{E} \mathbb{E}^{2}D_{0}}{\chi \mathcal{L}_{m_{0}}} -\frac{4 \pi i \mathcal{E} \mathbb{E}D_{0}^{2}}{\chi \mathcal{L}_{m_{0}}}-\frac{2 \chi \eta D_{0}^{2}\bar{\mathcal{L}_{m}}}{\chi \mathcal{L}_{m_{0}}}-\frac{2 \rho_{0}\bar{\mathcal{\mathcal{L}}}_{m}}{ \mathcal{L}_{m_{0}}}
-\frac{\bar{\mathcal{L}_{m}}D_{0}^{2}}{\chi \mathcal{L}_{m_{0}}}-\frac{D_{0}R_{0}  \mathcal{E}a}{\chi \mathcal{L}_{m_{0}}}-\frac{ \bar{\mathcal{L}_{m}}D_{0}^{2}}{ \mathcal{L}_{m_{0}}}-2\mathcal{E}aD_{0},
\\\nonumber
\varphi_{01}&=-2 \dot{\mathcal{E}}c-\eta \chi \mathcal{L}_{m_{0}} \mathcal{E}bD_{0}-\eta a \mathcal{E} \chi \mathcal{L}_{m_{0}}E_{0}-\eta \chi D_{0}E_{0}\bar{\mathcal{L}_{m}}
+h \chi \mathcal{L}_{m_{0}}\mathcal{E} b D_{0}+h \chi \mathcal{L}_{m_{0}}\mathcal{E} a E_{0}+h\chi  \bar{\mathcal{L}_{m}} \\\label{17}& \times E_{0} D_{0}+\frac{2 bF_{0}^{'} \dot{\mathcal{E}}}{E_{0}F_{0}} -\frac{4 \zeta R_{0}\dot{\mathcal{E}}c^{'}}{F_{0}}+\frac{4 \zeta R_{0}\dot{\mathcal{E}}bF_{0}^{'}}{E_{0}F_{0}}+\frac{4 \zeta R_{0}\dot{\mathcal{E}}cD_{0}^{'}}{D_{0}F_{0}}+\frac{2 \dot{\mathcal{E}}cD_{0}^{'}}{D_{0}F_{0}},
\\\nonumber
\varphi_{11}&=\frac{4 \zeta e \mathcal{E} F_{0}^{''}}{F_{0}}+\frac{4 \zeta R_{0}  \mathcal{E} c^{''}}{F_{0}}-\frac{4 \zeta c \mathcal{E} R_{0} F_{0}^{''}}{F_{0}^{2}}+\frac{2 \zeta e \mathcal{E} D_{0}^{''}}{D_{0}}+\frac{2 \zeta a^{''} \mathcal{E} R_{0}}{D_{0}}
-\frac{2 \zeta b \ddot{\mathcal{E}} R_{0}E_{0}}{D_{0}^{2}}- \frac{2\zeta a \mathcal{E} R_{0} D_{0}^{''}}{D_{0}^{2}}
\\\nonumber& 
+\frac{2 \zeta b \mathcal{E} R_{0}E_{0}^{'}F_{0}^{'}}{E_{0}^{2}F_{0}}+\frac{2 \zeta c \mathcal{E} R_{0}E_{0}^{'}F_{0}^{'}}{F_{0}^{2}E_{0}}
+\frac{2 \zeta c^{'} \mathcal{E} R_{0}E_{0}^{'}}{E_{0}F_{0}}+\frac{2 \zeta b^{'} \mathcal{E} R_{0}F_{0}^{'}}{E_{0}F_{0}}+\frac{2 \zeta e \mathcal{E}E_{0}^{'}F_{0}^{'} }{E_{0}F_{0}} +\frac{2 \zeta b \mathcal{E} R_{0}E_{0}^{'}D_{0}^{'}}{E_{0}^{2}D_{0}}\\\nonumber&
+\frac{2 \zeta a \mathcal{E} R_{0}E_{0}^{'}D_{0}^{'}}{D_{0}^{2}E_{0}}+\frac{2 \zeta a^{'} \mathcal{E} R_{0}E_{0}^{'}}{D_{0}E_{0}}+\frac{2 \zeta b^{'} \mathcal{E} R_{0}D_{0}^{'}}{E_{0}D_{0}}+\frac{e \mathcal{E}E_{0}^{2}}{2}+\zeta R_{0}e \mathcal{E} E_{0}^{2}+\chi E_{0}^{2}\mathcal{L}_{m_{0}}\bar{\mathcal{L}}_{m}+b\mathcal{E}E_{0}R_{0}\\\nonumber&+b\mathcal{E}E_{0}\zeta R_{0}^{2}+b\mathcal{E}E_{0}\chi \mathcal{L}_{m_{0}}^{2}+\chi E_{0}^{2}\bar{\mathcal{L}}_{m}
+2b\mathcal{E}E_{0}  \chi \mathcal{L}_{m_{0}}- \eta \chi E_{0}^{2}\bar{\mathcal{L}}_{m}+4\pi f \mathbb{E}_{0}
\mathbb{F}E_{0}^{2}-\eta b \mathcal{E} \chi E_{0}^{2}\mathcal{L}_{m_{0}}\\\nonumber&-2b\mathcal{E}\chi E_{0}
\mathcal{L}_{m_{0}}P_{xx_{0}}-E_{0}^{2}P_{xx_{0}}\chi \bar{\mathcal{L}}_{m}+\frac{2cE_{0}^{2}\ddot{\mathcal{E}}}{D_{0}^{2}F_{0}} +\frac{2c\mathcal{E}D_{0}^{'}F_{0}^{'}}
{D_{0}^{2}F_{0}^{2}}-E_{0}^{2}\chi \mathcal{L}_{m_{0}}\bar{P}_{xx}+4\pi b \mathcal{E}E_{0}\mathbb{E}_{0}^{2}-\frac{2c^{'}\mathcal{E}}{F_{0}}\\\label{M8}& \times \frac{F_{0}^{'}}{F_{0}}+\frac{2a\mathcal{E}D_{0}^{'}F_{0}^{'}}
{D_{0}^{2}F_{0}}+\frac{2c\mathcal{E}F_{0}^{'}}
{F_{0}^{3}}+\frac{2b\mathcal{E}
E_{0}}{F_{0}^{2}}-\frac{2c\mathcal{E}E_{0}^{2}}
{F_{0}^{3}},
\\\nonumber
\varphi_{22}&=\chi \mathcal{L}_{m_{0}}F_{0}^{2}\bar{P}_{yy}-\left(2 \chi \mathcal{L}_{m_{0}}c \mathcal{E} F_{0}+\chi \bar{\mathcal{L}_{m}}F_{0}^{2}\right)P_{yy_{0}}+\frac{2 \zeta c^{'} R_{0}\mathcal{E} F_{0}}{E_{0}^{2}}+\frac{2 \zeta c R_{0}\mathcal{E} F_{0}^{'}}{E_{0}^{2}}+\frac{2 \zeta e R_{0} \mathcal{E}F_{0}F_{0}^{'} }{E_{0}^{2}}\\\nonumber&  -\frac{4 \zeta b R_{0}\mathcal{E} F_{0}F_{0}^{'}}{E_{0}^{3}}-\frac{2 \zeta c R_{0}F_{0}\ddot{\mathcal{E}}}{D_{0}^{2}}
+\frac{2 \zeta e \mathcal{E} F_{0}^{'}}{E_{0}^{2}}+\frac{4 \zeta c^{'} R_{0}\mathcal{E} F_{0}^{'}}{E_{0}^{2}}-\frac{4 \zeta b R_{0}\mathcal{E} F_{0}^{'}}{E_{0}^{3}}-\frac{2 \zeta c^{'}R_{0}\mathcal{E} E_{0}^{'} }{E_{0}^{3}}-\frac{2 \zeta c}{E_{0}}\\\nonumber& \times \frac{ R_{0}\mathcal{E} E_{0}^{'}F_{0}^{'}}{E_{0}^{2}}
-\frac{2 \zeta R_{0}\mathcal{E} b^{'} F_{0}^{'}F_{0}}{E_{0}^{3}}-4 \pi F_{0}^{2}f \mathbb{E}\mathbb{F}-4\pi c\mathcal{E}F_{0}\mathbb{E}_{0}^{2}-\frac{2 \zeta e R_{0}\mathcal{E} E_{0}^{'}F_{0}F_{0}^{'}}{E_{0}^{3}} 
+\frac{6 \zeta b R_{0}\mathcal{E} E_{0}^{'}F_{0}^{'}}{E_{0}^{3}}\\\nonumber& \times \frac{F_{0}}{E_{0}}+ \frac{2c\zeta  R_{0}\mathcal{E} D_{0}^{'}F_{0}^{'}}{D_{0}E_{0}^{2}}+\frac{2 \zeta a^{'} R_{0}\mathcal{E} F_{0}^{'}}{D_{0}E_{0}^{2}}+\frac{2 \zeta c^{'} R_{0}\mathcal{E} D_{0}^{'}F_{0}}{D_{0}E_{0}^{2}}+\frac{2 \zeta e \mathcal{E} F_{0}F_{0}^{'}D_{0}^{'}}{D_{0}E_{0}^{2}} -\frac{4 \zeta b R_{0}\mathcal{E} D_{0}^{'}F_{0}^{'}F_{0}}{D_{0}E_{0}^{3}}\\\nonumber& - \frac{ 2a \zeta  R_{0}\mathcal{E} D_{0}^{'}F_{0}^{'}F_{0}}{D_{0}^{2}E_{0}^{2}}-\frac{2 \zeta}{D_{0}}
 \frac{ e \ddot{\mathcal{E}}F_{0}^{2}}{D_{0}}+\frac{2 \zeta R_{0}c \mathcal{E} F_{0}D_{0}^{'}}{D_{0}E_{0}^{2}}+\frac{2 \zeta R_{0}a^{'}\mathcal{E} F_{0}^{2}}{E_{0}^{2}D_{0}} +\frac{2 \zeta e \mathcal{E} F_{0}^{2}D_{0}^{'}}{D_{0}E_{0}^{2}}-\frac{4 \zeta R_{0}b \mathcal{E} F_{0}^{2} }{D_{0}E_{0}^{2}} \\\nonumber& \times \frac{D_{0}^{'}}{E_{0}}-\frac{2 \zeta R_{0} a \mathcal{E} F_{0}^{2}D_{0}^{'}}{D_{0}^{2}E_{0}^{2}}+\frac{2 \zeta e^{''}\mathcal{E} F_{0}^{2}}{E_{0}^{2}}+\frac{4 \zeta c \mathcal{E} F_{0}R_{0}^{''}}{E_{0}^{2}} -\frac{ 4\zeta b \mathcal{E} F_{0}^{2}R_{0}^{''}}{E_{0}^{3}}
-\frac{2 \zeta c^{'}\mathcal{E} F_{0}R_{0}^{'}}{E_{0}^{2}}+\frac{2 \zeta c\mathcal{E} F_{0}^{'}R_{0}^{'}}{E_{0}^{2}}\\\nonumber&-\frac{2 \zeta e^{'}\mathcal{E} F_{0}^{'}F_{0}}{E_{0}^{2}}+\frac{4 \zeta b\mathcal{E} F_{0}^{'}R_{0}^{'}F_{0}}{E_{0}^{3}}-\frac{2 \zeta b^{'}\mathcal{E} F_{0}^{2}R_{0}^{'}}{E_{0}^{3}}-2 \zeta   \mathcal{E} e -\frac{4 \zeta c \mathcal{E} E_{0}^{'}R_{0}^{'}F_{0}}{E_{0}^{4}}-\frac{2 \zeta e^{'} F_{0}^{2} \mathcal{E} E_{0}^{'}}{E_{0}^{3}}+\mathcal{E}eF_{0}^{2}\\\nonumber&+2 \zeta R_{0}e\mathcal{E}F_{0}^{2}+\frac{6 \zeta b F_{0}^{2} \mathcal{E} E_{0}^{'}R_{0}^{'}}{E_{0}^{4}}+2 \chi \mathcal{L}_{m_{0}} \bar{\mathcal{L}_{m}} F_{0}^{2}+2cF_{0} \mathcal{E}R_{0}+2 \zeta R_{0}^{2}\mathcal{E}c F_{0}+2 \chi \mathcal{L}_{m_{0}}^{2}\mathcal{E}cF_{0}+\chi \bar{\mathcal{L}}_{m}\\\label{18}& \times F_{0}^{2}+2 c \chi \mathcal{E} F_{0}\mathcal{L}_{m_{0}}.
\end{align}
The contemplation of Eqs. (\ref{P1}-\ref{P9}), concludes the static parts of non-conservation equation in the form given below:
\begin{align}\nonumber
&-\frac{\psi_{01}^{'}}{D_{0}^{2}E_{0}^{2}}-\frac{2 D_{0}^{'}}{D_{0}^{3}E_{0}^{2}}\psi_{01}-\frac{2E_{0}^{'}}{D_{0}^{2}E_{0}^{3}}\psi_{01}-\frac{D_{0}^{'}}{D_{0}^{3}E_{0}^{2}}\psi_{01}
-\frac{ E_{0}^{'}}{D_{0}^{2}E_{0}^{3}}\psi_{01}-\frac{2 F_{0}^{'}}{D_{0}^{2}}\psi_{01}-\frac{2\eta b \chi \mathcal{E}\mathcal{L}_{m_{0}}}{D_{0}^{2}E_{0}}-\frac{ b \chi \dot{\mathcal{E}}\mathcal{L}_{m_{0}}\rho_{0}}{D_{0}^{4}E_{0}}\\\label{20}& -\frac{2\eta  \chi \mathcal{L}_{m_{0}}E_{0}^{'}}{D_{0}^{3}E_{0}^{2}}+\frac{2 h  \chi \mathcal{L}_{m_{0}}E_{0}^{'}}{D_{0}^{3}E_{0}^{2}}-\frac{\eta}{D_{0}^{2}}
-\frac{4 D_{0}^{'}\chi \mathcal{L}_{m_{0}}}{D_{0}^{4}E_{0}^{3}}-\frac{4 D_{0}^{'}\chi \mathcal{L}_{m_{0}}\eta}{D_{0}^{4}E_{0}^{3}}
+\frac{4 D_{0}^{'}\chi \mathcal{L}_{m_{0}}h}{D_{0}^{4}E_{0}^{3}},
\\\nonumber
&\frac{4\zeta R_{0}^{'}F_{0}^{''}}{E_{0}^{2}F_{0}}+\frac{4\zeta R_{0}F_{0}^{'''}}{E_{0}^{4}F_{0}}+\frac{2\zeta R_{0}F_{0}^{'}F_{0}^{''}}{E_{0}^{4}F_{0}^{2}}+\frac{4 \pi \mathbb{E}\mathbb{E}_{0}^{'}}{E_{0}^{2}}+ \frac{8 \pi \mathbb{E}^{2}F_{0}^{'}}{E_{0}^{2}F_{0}^{2}}
+\frac{2\zeta R_{0}^{'}D_{0}^{''}}{E_{0}^{4}D_{0}}-\frac{2\zeta R_{0}D_{0}^{'}D_{0}^{''}}{E_{0}^{4}D_{0}^{2}}-\frac{8\zeta R_{0}D_{0}^{'}D_{0}^{''}E_{0}^{'}}{E_{0}^{5}D_{0}^{2}}\\\nonumber&
-\frac{4\zeta R_{0}^{'}E_{0}^{'}F_{0}^{'}}{E_{0}^{5}F_{0}}-\frac{4\zeta R_{0}F_{0}^{'}E_{0}^{''}}{E_{0}^{5}F_{0}}
+\frac{12\zeta R_{0}F_{0}E_{0}^{'2}}{E_{0}^{6}F_{0}}+\frac{6\zeta R_{0}F_{0}^{'2}E_{0}^{'}}{E_{0}^{5}F_{0}^{2}}
-\frac{2\zeta R_{0}^{'}D_{0}^{'}E_{0}^{'}}{E_{0}^{5}D_{0}}+\frac{2\zeta R_{0}D_{0}^{''}E_{0}^{'}}{E_{0}^{5}D_{0}}
-\frac{2\zeta R_{0}D_{0}^{'}E_{0}^{''}}{E_{0}^{5}D_{0}}\\\nonumber&+\frac{R_{0}^{'}}{2 E_{0}^{2}}+\frac{6\zeta R_{0}D_{0}^{'}E_{0}^{'2}}{E_{0}^{6}D_{0}}+\frac{\zeta R_{0}R_{0}^{'}}{E_{0}^{2}}+\frac{\chi \mathcal{L}_{m_{0}}\mathcal{L}_{m_{0}^{'}}}{E_{0}^{2}}-\frac{4 \pi E_{0}^{'}\mathbb{E}_{0}^{2}}{E_{0}^{3}F_{0}}
-\frac{16\pi^{2}F_{0}\mathbb{E}_{0}^{2}}{E_{0}^{2}F_{0}}
-\frac{R_{0}E_{0}^{'}}{E_{0}^{2}}+\frac{\zeta R_{0}^{2}E_{0}^{'}}{E_{0}^{2}}-\frac{\chi E_{0}^{'}\mathcal{L}_{m_{0}}^{2}}{E_{0}^{2}}\\\nonumber&
+\frac{\chi \mathcal{L}_{m_{0}}^{'}}{E_{0}^{2}}-\frac{4 \zeta R_{0}D_{0}^{'}F_{0}^{'}}{D_{0}^{2}E_{0}^{4}F_{0}}-\frac{D_{0}^{'}}{2 E_{0}^{2}} \frac{R_{0}}{D_{0}}-\frac{\zeta R_{0}^{2}D_{0}^{'}}{2 E_{0}^{2}D_{0}}-\frac{\zeta e\mathcal{E} R_{0}D_{0}^{'}}{2 E_{0}^{2}D_{0}}-\frac{\chi \mathcal{L}_{m_{0}}^{2}D_{0}^{'}}{2E_{0}^{2}D_{0}}+\frac{D_{0}^{'}}{E_{0}^{2}D_{0}^{3}}
+\frac{2 \zeta R_{0}F_{0}^{'3}}{E_{0}^{4}F_{O}^{3}}-\frac{2 \zeta R_{0}F_{0}^{'}}{E_{0}^{4}F_{O}^{3}}\\\nonumber&-\frac{2 \zeta R_{0}F_{0}^{'2}D_{0}^{'}}{D_{0}E_{0}^{2}F_{0}^{2}}+\frac{2 \zeta R_{0}F_{0}^{'}}{E_{0}^{2}F_{O}^{3}}+\frac{R_{0}F_{0}^{'}}{2E_{0}^{2}F_{O}}+\frac{\zeta R_{0}^{2}F_{0}^{'}}{2E_{0}^{2}F_{0}}+\frac{\chi \mathcal{L}_{m_{0}}^{2}F_{0}^{'}}{2E_{0}^{2}F_{0}}+\frac{\chi \mathcal{L}_{m_{0}}F_{0}^{'}}{E_{0}^{2}F_{0}}-\frac{F_{0}^{'}}{E_{0}^{2}F_{0}^{3}}\psi_{22}+\frac{4 \zeta R_{0}F_{0}^{''}D_{0}^{'}}{D_{0}E_{0}^{4}F_{0}}\\\nonumber&-\frac{2\pi D_{0}^{'}\mathbb{E}_{0}^{2}}{D_{0}E_{0}^{2}}-\frac{4 \zeta R_{0}F_{0}^{'}D_{0}^{'}E_{0}^{'}}{D_{0}E_{0}^{5}F_{0}}+\frac{D_{0}^{'}R_{0}}{2 D_{0}E_{0}^{2}} +\frac{\zeta D_{0}^{'} R_{0}^{2}}{2 D_{0}E_{0}^{2}}+\frac{6 \pi \mathbb{E}_{0}^{2}}{E_{0}^{2}F_{0}}+\frac{\chi \mathcal{L}_{m_{0}}^{2}D_{0}^{'}}{2 D_{0}E_{0}^{2}}-\frac{\chi \mathcal{L}_{m_{0}}D_{0}^{'}}{2 D_{0}E_{0}^{2}}+\frac{D_{0}^{'}}{D_{0}E_{0}^{2}}+\frac{2 \pi \mathbb{E}_{0}^{2}D_{0}^{'}}{D_{0}E_{0}^{2}}\\\nonumber&+\frac{R_{0}E_{0}^{'}}{E_{0}^{3}}+\frac{\zeta R_{0}^{2}E_{0}^{'}}{E_{0}^{3}}
+\frac{\chi E_{0}^{'}\mathcal{L}_{m_{0}}^{2}}{E_{0}^{3}}+\frac{2 \chi E_{0}^{'}\mathcal{L}_{m_{0}}}{E_{0}^{3}}-\frac{2 E_{0}^{'}}{E_{0}^{5}}\psi_{11}
 +\frac{4 \zeta R_{0}D_{0}^{''}F_{0}^{'}}{D_{0}E_{0}^{4}F_{0}}+\frac{8 \zeta R_{0}E_{0}^{'}F_{0}^{'}}{E_{0}^{5}F_{0}}
-\frac{4 \zeta R_{0}D_{0}^{'}E_{0}^{'}F_{0}^{'}}{D_{0}E_{0}^{5}F_{0}}\\\nonumber&+\frac{2 F_{0}^{'}\psi_{11}}{E_{0}^{4}F_{0}}+\frac{4 \pi \mathbb{E}_{0}^{2}E_{0}^{'}}{E_{0}^{3}}-\eta \chi \mathcal{L}_{m_{0}}-\chi \mathcal{L}_{m_{0}}P_{xx_{0}}
-\frac{\chi D_{0}D_{0}^{'}\mathcal{L}_{m_{0}}\mu_{0}}{D_{0}^{4}E_{0}^{2}}-\frac{\chi D_{0}D_{0}^{'}\mathcal{L}_{m_{0}}\eta}{D_{0}^{2}E_{0}^{2}}+\frac{2\chi F_{0}F_{0}^{'}P_{yy_{0}}}{E_{0}^{2}F_{0}^{2}}-\frac{\chi \eta }{D_{0}}\\\label{21}&\times \frac{\mathcal{L}_{m_{0}}D_{0}^{'}}{E_{0}^{2}}-\frac{\chi D_{0}^{'}\mathcal{L}_{m_{0}}P_{xx_{0}}}{D_{0}E_{0}^{2}}-\frac{2\chi \eta E_{0}^{'}\mathcal{L}_{m_{0}}}{E_{0}^{3}}-\frac{2\chi E_{0}^{'}\mathcal{L}_{m_{0}}P_{xx_{0}}}{E_{0}^{3}}.
\end{align}
The consideration of non-static parts corresponding to non-conservation equation in addition to the modified field equations, Ricci scalar and Eqs. (\ref{P1}-\ref{P9}), in Eqs. (\ref{5}-\ref{8}), concludes the following
\begin{align}\nonumber
&\dot{\bar{\rho}}+\left\{\left(\frac{2cE_{0}}{bF_{0}}-\frac{4a\mathcal{E}}{D_{0}}+\frac{2aE_{0}}{bD_{0}}
+\frac{\bar{\mathcal{L}_{m}}E_{0}}{b\mathcal{L}_{m_{0}}}
\right)\rho_{0}+D_{0}^{2}P_{xx_{0}}+\frac{cD_{0}^{2}E_{0}}{bF_{0}}P_{yy_{0}}
+\frac{2 \pi c\mathbb{E}_{0}^{2}}{D_{0}^{2}F_{0}}+\frac{4 \pi a\mathbb{E}_{0}^{2}}{D_{0}^{3}}+\mathcal{Y}_{1}\right.\\\label{A}&\left.+\frac{2 \pi b\mathbb{E}_{0}^{2}}{D_{0}^{2}E_{0}}
\right\}\dot{\mathcal{E}}=0,
\\\nonumber
&\bar{P}_{xx}^{'}+\left\{-\chi \mathcal{L}_{m_{0}}-\frac{\chi \mathcal{L}_{m_{0}}D_{0}^{'}}{D_{0}E_{0}^{2}}-\frac{2\chi \mathcal{L}_{m_{0}}F_{0}^{'}}{F_{0}^{3}}-\frac{2\chi \mathcal{L}_{m_{0}}E_{0}^{'}}{E_{0}^{3}}\right\}\bar{P_{xx}}+\left\{-\chi \bar{\mathcal{L}_{m}}-
\frac{\chi \mathcal{L}_{m_{0}}\mathcal{E}a^{'}}{D_{0}E_{0}^{2}}-\frac{\chi D_{0}^{'}\bar{\mathcal{L}_{m}}}{D_{0}E_{0}^{2}}\right.\\\nonumber&\left.
+\frac{2b \mathcal{E}\chi \mathcal{L}_{m_{0}}D_{0}^{'}}{D_{0}E_{0}^{3}}+\frac{a\mathcal{E}\chi \mathcal{L}_{m_{0}}D_{0}^{'}}{D_{0}^{2}E_{0}^{2}}
-\frac{2c^{'} \mathcal{E}\chi \mathcal{L}_{m_{0}}}{F_{0}^{3}}-\frac{2 \chi F_{0}^{'}\bar{\mathcal{L}_{m}}}{F_{0}^{3}}+\frac{6c\chi \mathcal{E} \mathcal{L}_{m_{0}}F_{0}^{'}}{F_{0}^{4}}-\frac{2\chi \mathcal{E} b^{'} \mathcal{L}_{m_{0}}}{E_{0}^{3}}-\frac{2 \chi E_{0}^{'}\bar{\mathcal{L}_{m}}}{E_{0}^{3}}
\right.\\\nonumber&\left.+\frac{6b\chi \mathcal{E} \mathcal{L}_{m_{0}}E_{0}^{'}}{E_{0}^{4}}\right\}P_{xx_{0}}+\left\{-\frac{a^{'}\mathcal{E} \chi D_{0}\mathcal{L}_{m_{0}}}{D_{0}^{4}E_{0}^{2}}-\frac{a\mathcal{E} \chi D_{0}^{'}\mathcal{L}_{m_{0}}}{D_{0}^{4}E_{0}^{2}}-\frac{\chi \bar{\mathcal{L}_{m}}D_{0}D_{0}^{'}}{D_{0}^{4}E_{0}^{2}}+\frac{2b\mathcal{E} \chi D_{0}D_{0}^{'}\mathcal{L}_{m_{0}}}{D_{0}^{4}E_{0}^{3}}
+\frac{4a\mathcal{E} \chi }{E_{0}^{2}}\right.\\\nonumber& \left. \times \frac{D_{0}D_{0}^{'}\mathcal{L}_{m_{0}}}{D_{0}^{5}}\right\}\rho_{0}
-\frac{\chi \mathcal{L}_{m_{0}}D_{0}^{'}}{D_{0}^{3}E_{0}^{2}}\bar{\rho}+\left\{\frac{2\chi c^{'}\mathcal{E}\mathcal{L}_{m_{0}}F_{0}}{E_{0}^{2}F_{0}^{2}}
+\frac{2\chi c\mathcal{E}\mathcal{L}_{m_{0}}F_{0}^{'}}{E_{0}^{2}F_{0}^{2}}+\frac{2\chi \bar{\mathcal{L}_{m}}F_{0}F_{0}^{'}}{E_{0}^{2}F_{0}^{2}}
-\frac{4\chi c F_{0}^{'}\mathcal{E}\mathcal{L}_{m_{0}}F_{0}}{E_{0}^{2}F_{0}^{3}}\right.\\\nonumber&\left.-\frac{4\chi F_{0}^{'}b\mathcal{E}\mathcal{L}_{m_{0}}F_{0}}{E_{0}^{3}F_{0}^{2}}\right\}P_{yy_{0}}+\frac{2\chi F_{0}^{'}\mathcal{L}_{m_{0}}F_{0}}{E_{0}^{2}F_{0}^{2}}\bar{P_{yy}}+\frac{4 \pi \mathbb{E}\mathcal{E}i^{'}}{E_{0}^{2}}+\frac{2 \pi c^{'}\mathcal{E}\mathbb{E}_{0}^{2}}{D_{0}^{2}E_{0}^{2}}-\frac{16\pi b^{'}\mathcal{E}\mathbb{E}_{0}^{2}}{E_{0}^{3}F_{0}}-\frac{8 \pi i E_{0}^{'}\mathbb{E}\mathcal{E}}{E_{0}^{3}F_{0}}\\\nonumber&+\frac{4 \pi c\mathcal{E}E_{0}^{'}\mathbb{E}_{0}^{2}}{E_{0}^{3}F_{0}}+\frac{12 \pi b\mathcal{E}E_{0}^{'}\mathbb{E}_{0}^{2}}{E_{0}^{4}F_{0}}-\frac{20 \pi^{2}iF_{0}^{'}\mathbb{E}_{0}\mathcal{E}}{E_{0}^{2}F_{0}}+\frac{16 \pi^{2}c^{'}\mathcal{E}\mathbb{E}_{0}^{2}}{E_{0}^{2}F_{0}}+\frac{16 \pi^{2}c\mathcal{E}F_{0}^{'}\mathbb{E}_{0}^{2} }{E_{0}^{2}F_{0}^{2}}+\frac{20 \pi^{2}b\mathcal{E}F_{0}^{'}\mathbb{E}_{0}^{2}}{E_{0}^{3}F_{0}}-\frac{2 \pi }{D_{0}}\\\label{23}&\times \frac{a^{'}\mathcal{E}\mathbb{E}_{0}^{2}}{E_{0}^{2}}-\frac{4 \pi f D_{0}^{'}\mathcal{E}\mathbb{E}_{0}}{D_{0}E_{0}^{2}}+\frac{4 \pi b\mathcal{E}D_{0}^{'}\mathbb{E}_{0}^{2}}{A_{0}B_{0}^{3}}+\frac{2 \pi a^{'}\mathcal{E}\mathbb{E}_{0}^{2}}{D_{0}E_{0}^{2}}+\mathcal{Y}_{2},
\end{align}
The values of dark source terms ( $\mathcal{Y}_{1}$ and $\mathcal{Y}_{2}$) corresponding to $f(R, \mathcal{L}_{m})$ gravity can be seen in Appendix. Harrison et al. \cite{harrison1965gravitation} proposed the EOS, a renowned interrelation describing the dynamics of anisotropic fluid in the presence of highly dense gravitational fields. The adiabatic index $\Gamma$ establishes the interaction between perturbed pressure components with perturbed energy density parts in subsequent form:
\begin{equation}\label{24}
\bar{P_{k}}=\Gamma \frac{\bar{\rho}P_{k_{0}}}{\rho_{0}+P_{k_{0}}}.
\end{equation}
The aforementioned expression can be specified in other ways by substituting different values, such as $k=xx,xy,yy$. This enables the representation of perturbed pressure parts in the form of $\bar{\rho}$, that is frequently employed to characterize the fundamental attributes of extremely dense anisotropic fluid configurations. The EOS, described in Eq. \eqref{24} that was established in GR, is useful in evaluating the dynamical instability and development of perturbations in alternative theories of gravity. The contemplation of $f(R, \mathcal{L}_{m})$ gravity results the non-conserved effects of gravity and modified form of gravitational dynamics after the insertion of additional structural terms. However, the EOS proposed by Harrison et al. \cite{harrison1965gravitation} is used to provide the fundamental description of substantial characteristics of the anisotropic matter that allows the suitable perturbation technique in order to assess the physical stability. Our contemplated perturbation approach is determined by the linearized energy density and pressure through adiabatic index, which is frequently employed in studying the stability of compact objects within the formalism of GR and modified gravity theories. Moreover, in Harrison's EOS the incorporation of anisotropy through unidirectional and non-diagnol pressure components makes it effective in studying the dynamics of spherical fluid configurations. In addition, the relevance of this EOS in evaluating the stability constraints through adiabatic index and perturbation approach is detailed in \cite{ur2024dynamically,sharif2015instability}. The integration of Eq. \eqref{A} yields
\begin{align}\label{25}
&\bar{\rho}=\left[\left(\frac{2cE_{0}}{bF_{0}}-\frac{4a\mathcal{E}}{D_{0}}+\frac{2aE_{0}}{bD_{0}}+\frac{\bar{L}_{m}E_{0}}{bL_{m_{0}}}
\right)\rho_{0}+D_{0}^{2}P_{xx_{0}}+\frac{cD_{0}^{2}E_{0}}{bF_{0}}P_{yy_{0}}+\frac{2\pi c\mathbb{E}_{0}^{2}}{D_{0}^{2}F_{0}}+\frac{4 \pi a\mathbb{E}_{0}^{2}}{D_{0}^{3}}+\frac{2 \pi b\mathbb{E}_{0}^{2}}{D_{0}^{2}E_{0}}+\mathcal{Y}_{1}\right]\mathcal{E}.
\end{align}

\section{Matching constraints and the Collpase equation}

The discontinuity of energy-momentum tensor is avoided by following Darmois matching conditions that correspond to the matching between inner spherical matter composition and Schwarzschild spacetime. The continuity of first and second basic forms throughout the surface provides a consistent connection between two different sections of spacetime. This assumption concludes the subsequent mathematical results
\begin{equation}\label{M3}
\left(1-\frac{2M}{r}+\frac{Q^{2}}{r^{2}}\right)d \nu=^{\Sigma^{(e)}}Adt, \quad M =^{\Sigma^{(e)}} m(t,r) \Leftrightarrow q(r)=^{\Sigma^{(e)}} Q .
\end{equation}
The above-mentioned equations are determined from the matching between internal and external metrics. The hypersurface $\Sigma ^{(e)} $ corresponds to the separation of fluid from spherically symmetric solution, while the boundary $\Sigma^{(i)}$ specifies the division between Minkoeskian internal vacuum cavity and matter, which yields the following \cite{sharif2013dynamical}
\begin{equation}\label{M4}
m(t,r)=^{\Sigma ^{(i)}} 0, \quad P_{xx} =^{\Sigma ^{(i)}} -\frac{T_{01}^{(eff)}}{DE}-\frac{T_{11}^{(eff)}}{E^{2}}.
\end{equation}

Further, the contemplated perturbation technique concludes the following non-static form of second matching condition
\begin{align}\nonumber
\bar{P}_{xx}&=-\frac{4 \zeta c^{'}R_{0}\dot{\mathcal{E}}}{D_{0}E_{0}F_{0}}-\frac{4 \zeta bR_{0}\dot{\mathcal{E}}E_{0}^{'}}{D_{0}E_{0}^{2}F_{0}}
-\frac{4 \zeta D_{0}^{'}R_{0}c\dot{\mathcal{E}}}{D_{0}E_{0}F_{0}^{2}}+\frac{a\mathcal{E}}{D_{0}^{2}E_{0}}\psi_{01}+\frac{b\mathcal{E}}
{E_{0}^{2}D_{0}}\psi_{01}
+h\chi \bar{\mathcal{L}}_{m}-\frac{4 \zeta c\mathcal{E} F_{0}^{''}}{E_{0}^{2}F_{0}}-\frac{4 \zeta c^{''} }{F_{0}}\\\nonumber& \times
\frac{R_{0}\mathcal{E}}{E_{0}^{2}}+\frac{2 \zeta R_{0}b \ddot{\mathcal{E}}}{D_{0}^{2}E_{0}}-\frac{2 \zeta a\mathcal{E}R_{0}D_{0}^{''}}{D_{0}^{2}E_{0}^{2}}
-\frac{4 \zeta b\mathcal{E}R_{0}D_{0}^{''}}{D_{0}E_{0}^{3}}-\frac{2 \zeta e \mathcal{E}D_{0}^{''}}{D_{0}E_{0}^{2}}-4 \pi f \mathbb{E}_{0}\mathbb{F}-\frac{2 \zeta a^{''}\mathcal{E}R_{0}}{D_{0}E_{0}^{2}}+\frac{4 \zeta c^{'}\mathcal{E}R_{0}E_{0}^{'}}{F_{0}E_{0}^{3}}\\\label{NM2}&+\frac{4 \zeta b^{'}\mathcal{E}R_{0}F_{0}^{'}}{E_{0}^{3}F_{0}} +\frac{4 \zeta c^{'}\mathcal{E}R_{0}E_{0}^{'}}{F_{0}E_{0}^{3}}.
\end{align}
The consideration of matching restraints in addition to the Eqs. \eqref{16} and \eqref{M8} results
\begin{equation}\label{M10}
u\ddot{\mathcal{E}}+v\dot{\mathcal{E}}+w\mathcal{E}=0,
\end{equation}
where, the values of $u$, $v$ and $w$ can be seen in Appendix. It is vital to note that Eq. \eqref{M10}, leads to stable or unstable solutions and $\mathcal{E}(t)$ provides a description for physical dynamics of collapsed spherical matter. We are mainly interested in establishing the range of dynamical instability corresponding to the spherical charged matter distribution; therefore only unstable solution is considered. The subsequent solution is determined from Eq. \eqref{M10}
\begin{equation}\label{M11}
\mathcal{E}(t)=-exp (\tau_{\Sigma^{(e)}}t),\quad where \quad \tau_{\Sigma^{(e)}}=\frac{-v+\sqrt{v^{2}-4uw}}{2u}.
\end{equation}
in addition to $\mathcal{E}(- \infty)=0$, and the fluid distribution is in stationary state, this function suggests that the considered system collapses at $t=-\infty$, and firstly it is supposed that $\mathcal{E}(- \infty)=0$. In astrophysics, a static, extremely dense self-gravitating fluid distribution with zero size  along with the decreasing radius of matter with time is considered for the simplification of derived results.

\subsection{Collapse equation}

In alternative gravity theories `` unstable eras'' are identified as the periods in accelerated universe when significant matter compositions illustrate instability and various cosmic events or fluctuations that cause instability across different celestial objects describe these periods. Moreover, instability resulted by distinct events that includes phase variations in cosmic evolution potentially disturb the basic interaction between matter and forces. In certain cosmic frameworks, variations in the distribution of dark energy may impact the physical dynamics of compact structures and cause instability in cosmic evolution. Instabilities emerged from fluctuations in matter density can play a role in the development of extremely dense objects including black holes. Detectable remnants are caused by instabilities that were developed due to the CMB radiations during initial inflationary cosmic epoch. The identification of physical instabilities by mathematical frameworks may provide comprehensive interpretation for the complex interactions between fluid distribution and gravitational forces impacting the accelerated expansion of the universe. The non-linear framework of dynamical equations in addition to the non-homogenous matter distribution makes it difficult to analyze the compact structures, while, the usage of approximation approach in determining viable solutions has significant relevance in confronting these challenges. The simplified gravity methodology offers feasible results relating to weak gravitational fields, rendering it distinctive within conventional approaches. It simplifies non-linear field equations to generalized form through their description in linear formalism. Certain constraints are implemented in N and $pN$ regimes to determine the plausible solutions within the formalism of such approximation strategies. The physical instability associated with spherical anisotropic distribution of fluid in N and $pN$ regimes is analyzed through the contemplation of non-conservation equation in the framework of collapse equation. The substitution of Eq. \eqref{24} in Eq. \eqref{23} results the subsequent form of collapse equation
\begin{align}\nonumber
&\Gamma \left(\frac{P_{xx_{0}}}{\rho_{0}+P_{xx_{0}}}\right)^{'}\bar{\rho}^{'}-\Gamma\left\{\chi \mathcal{L}_{m_{0}}+\frac{\chi \mathcal{L}_{m_{0}}D_{0}^{'}}{D_{0}E_{0}^{2}}+\frac{2\chi \mathcal{L}_{m_{0}}F_{0}^{'}}{F_{0}^{3}}-\frac{2\chi \mathcal{L}_{m_{0}}E_{0}^{'}}{E_{0}^{3}}\right\}\frac{\bar{\rho} P_{xx_{0}}}{\rho_{0}+P_{xx_{0}}}-\left\{-\chi \bar{\mathcal{L}_{m}}
-\frac{\chi a^{'}}{D_{0}}\right.\\\nonumber&\left.\times \frac{\mathcal{L}_{m_{0}}\mathcal{E}}{E_{0}^{2}}-\frac{\chi \bar{\mathcal{L}_{m}}D_{0}^{'}}{D_{0}E_{0}^{2}}+\frac{2\chi \mathcal{L}_{m_{0}}b\mathcal{E}D_{0}^{'}}{D_{0}E_{0}^{3}}+\frac{\chi \mathcal{L}_{m_{0}}a\mathcal{E}D_{0}^{'}}{D_{0}^{2}E_{0}^{2}}-\frac{2\chi \mathcal{L}_{m_{0}}c^{'}\mathcal{E}}{F_{0}^{3}}-\frac{2\chi \bar{\mathcal{L}_{m}}F_{0}^{'}}{F_{0}^{3}}+\frac{6 \chi c\mathcal{E}\mathcal{L}_{m_{0}}F_{0}^{'}}{F_{0}^{4}}-\frac{2\chi }{E_{0}^{3}}\right.\\\nonumber&\left. \times \frac{b^{'}\mathcal{E}\mathcal{L}_{m_{0}}}{E_{0}}
-\frac{2\chi E_{0}^{'}\mathcal{L}_{m_{0}}}{E_{0}^{3}}+\frac{6\chi b\mathcal{E}\mathcal{L}_{m_{0}}E_{0}^{'}}{E_{0}^{4}}\right\}P_{xx_{0}}
+\left\{\frac{a^{'}\mathcal{E}\chi \mathcal{L}_{m_{0}}D_{0}}{D_{0}^{4}E_{0}^{2}}+\frac{a\mathcal{E}\chi \mathcal{L}_{m_{0}}D_{0}^{'}}{D_{0}^{4}E_{0}^{2}}
-\frac{\chi \bar{\mathcal{L}_{m}}D_{0}D_{0}^{'}}{D_{0}^{4}E_{0}^{2}}-\frac{2b}{E_{0}^{3}}\right.\\\nonumber&\left.\times \frac{\mathcal{E}\chi \mathcal{L}_{m_{0}}D_{0}D_{0}^{'}}{D_{0}^{4}}-
\frac{4a\mathcal{E}\chi \mathcal{L}_{m_{0}}D_{0}D_{0}^{'}}{D_{0}^{5}E_{0}^{2}}\right\}\rho_{0}-\frac{\chi \mathcal{L}_{m_{0}}D_{0}^{'}\bar{\rho}}{D_{0}^{3}E_{0}^{2}}-\left\{\frac{2\chi c^{'}\mathcal{E}\mathcal{L}_{m_{0}}}{E_{0}^{2}F_{0}}+\frac{2\chi c\mathcal{E}F_{0}^{'}\mathcal{L}_{m_{0}}}{E_{0}^{2}F_{0}^{2}}+\frac{2\chi \bar{\mathcal{L}_{m}}F_{0}^{'}}{E_{0}^{2}F_{0}}\right.\\\nonumber&\left.
-\frac{4 \chi c\mathcal{E} \mathcal{L}_{m_{0}}F_{0}^{'}}{E_{0}^{2}F_{0}}-\frac{4 \chi b\mathcal{E} \mathcal{L}_{m_{0}}F_{0}^{'}}{E_{0}^{3}F_{0}}\right\}P_{yy_{0}}
+\frac{\Gamma\bar{\rho}P_{yy_{0}}}{\rho_{0}+P_{yy_{0}}}\left(\frac{2\chi \mathcal{L}_{m_{0}}F_{0}^{'}}{E_{0}^{2}F_{0}}\right)-\frac{4 \pi i^{'}\mathbb{E}\mathcal{E}}{E_{0}^{2}}-\frac{2 \pi c^{'}\mathcal{E}\mathbb{E}_{0}^{2}}{E_{0}^{2}F_{0}^{2}}+\frac{16 \pi b^{'}\mathcal{E}\mathbb{E}_{0}^{2}}{E_{0}^{3}F_{0}}\\\nonumber&+\frac{8 \pi i E_{0}^{'}\mathcal{E}\mathbb{E}}{E_{0}^{3}F_{0}}-\frac{4 \pi E_{0}^{'}c\mathcal{E}\mathbb{E}_{0}^{2}}{E_{0}^{3}F_{0}}-\frac{12 \pi E_{0}^{'}b\mathcal{E}\mathbb{E}_{0}^{2}}{E_{0}^{4}F_{0}}+\frac{20 \pi^{2}i\mathbb{E}\mathbb{E}_{0}F_{0}^{'}}{E_{0}^{2}F_{0}}-\frac{16 \pi^{2}c^{'}\mathcal{E}\mathbb{E}_{0}^{2}}{E_{0}^{2}F_{0}}-\frac{16 \pi^{2}c\mathcal{E}\mathbb{E}_{0}^{2}F_{0}^{'}}{E_{0}^{2}F_{0}^{2}}-\frac{20 \pi^{2}b\mathcal{E}}{E_{0}^{3}}\\\label{26}& \times \frac{\mathbb{E}_{0}^{2}F_{0}^{'}}{F_{0}}+\frac{2 \pi a^{'}\mathcal{E}\mathbb{E}_{0}^{2}}{D_{0}E_{0}^{2}}+\frac{4 \pi iD_{0}^{'}\mathbb{E}_{0}\mathcal{E}}{D_{0}E_{0}^{2}}-\frac{4 \pi bD_{0}^{'}\mathbb{E}_{0}^{2}\mathcal{E}}{D_{0}E_{0}^{3}}-\frac{2 \pi a^{'}\mathcal{E}\mathbb{E}_{0}^{2}}{D_{0}E_{0}^{2}}-\mathcal{Y}_{2}.
\end{align}
The collapse equation includes energy density fluctuations, pressure components, geometrical variations, and the curvature corrections related to $f(R, \mathcal{L}_{m})$ theory.

\section{Interpretation of instability zones in Newtonian
and post-Newtonian domains}

Further, we assess the physical instability associated with restrained spherical non-static charged fluid configuration in N and $pN$ regimes within homogenous and curved contexts. The consideration of metric coefficients in addition to the restricted fundamental components yields an unstable limitation implying that energy density dominates surface pressure across the relevant axes. Our analysis establishes the certain constraints that lead to instability in non-static matter composition. The instability extent in these regimes is influenced by adiabatic index, anisotropic matter distribution, rotating energy density, electric charge and gravity related curvature variables, all of which are thought to possess substantial limitations.

\subsection{Constraint for adiabatic index in Newtonian epoch with
quadratic corrections of $f (R, \mathcal{L}_{m})$ gravity}

We develop the interaction representing the collapsed period  in N and $pN$ regimes through allocating certain values as $D_{0}=E_{0}=1$ and $F_{0}=r$. We establish the relation for adiabatic index for analyzing stable and unstable regions with specific limitations. Establishing a precise interaction between static parts including dark source terms, anisotropic pressure, and energy density is crucial for satisfying $\Gamma$ constraints. The assumption of $P_{k0} << \rho_{0}$, where, $\rho_{0}$ represents energy density and $P_{k0}$ corresponds to pressure parts in certain directions $k=xx, xy, yy$ corresponding to static fluid distribution, implies the assertion that $\frac{P_{k0}}{\rho_{0}} \rightarrow 0$. The precise description of related terms and limitations is crucial in identifying instabilities in N regime. The inclusion of these restraints in determined collapse equation relating to N domain results subsequent mathematical form:
\begin{align}\nonumber
&\Gamma P_{xx_{0}}^{'}\left(-4a\mathcal{E}+\frac{2c}{br}+\frac{2a}{b}+\frac{\bar{\mathcal{L}_{m}}}{\mathcal{L}_{m_{0}}}\right)^{'}+\left(a^{'}\chi \mathcal{L}_{m_{0}}+2b^{'}\chi \mathcal{L}_{m_{0}}+\frac{\chi \bar{\mathcal{L}_{m}}}{\mathcal{E}}+\frac{2c^{'}\chi \mathcal{L}_{m_{0}}}{r^{3}\mathcal{E}}+\frac{2 \chi \bar{\mathcal{L}_{m}}}{r^{3}\mathcal{E}}\right.\\\nonumber&\left.-\frac{6 \chi c\mathcal{L}_{m_{0}}}{r^{4}}\right)P_{xx_{0}}-\Gamma\left(-4a\mathcal{E}+\frac{2c}{br}+\frac{2a}{b}+\frac{\bar{\mathcal{L}_{m}}}{\mathcal{L}_{m_{0}}}\right)\left\{\left(\chi \mathcal{L}_{m_{0}}+\frac{2 \chi \mathcal{L}_{m_{0}}}{r^{3}}\right)P_{xx_{0}}-P_{yy_{0}}\right\}\\\nonumber&+a^{'}\chi \mu_{0}\mathcal{L}_{m_{0}}-\left(\frac{2\chi c^{'}\mathcal{L}_{m_{0}}}{r}-\frac{2\chi c\mathcal{L}_{m_{0}}}{r^{2}}-\frac{4\chi b\mathcal{L}_{m_{0}}}{r}+
\frac{2\chi \bar{\mathcal{L}_{m}}}{r\mathcal{E}}\right)P_{yy_{0}}+\frac{4 \pi i^{'}\mathcal{E}\mathbb{E}}{\mathcal{E}}+\frac{2 \pi c^{'}\mathbb{E}_{0}^{2}}{r^{2}}\\\label{27}&+\frac{16 \pi b^{'}\mathbb{E}_{0}^{2}}{r}+\frac{16 \pi^{2}c^{'}\mathbb{E}_{0}^{2}}{r}+\frac{16 \pi^{2}c\mathbb{E}_{0}^{2}}{r^{2}}-\frac{20 \pi^{2}i\mathbb{E}\mathbb{E}_{0}}{r\mathcal{E}}+\frac{20 \pi^{2}b\mathbb{E}_{0}^{2}}{r}+\mathcal{Y}_{2N}=0,
\end{align}
The mathematical formalism of $\mathcal{Y}_{2N}$, is generally specified as dark source term in N approximation and is written in Appendix. The preceding collapse equation can be written as:
\begin{align}\nonumber
\Gamma& < \left(\chi a^{'}\mathcal{L}_{m_{0}}+2b^{'}\chi \mathcal{L}_{m_{0}}+\mathcal{F}_{1}\right)P_{xx_{0}}+a^{'}\chi \mathcal{L}_{m_{0}}\rho_{0}
\\\label{28}& \times \frac{+\left(\frac{-2\chi c^{'}\mathcal{L}_{m_{0}}}{r}-\frac{2\chi c\mathcal{L}_{m_{0}}}{r^{2}}-\frac{2\chi \bar{\mathcal{L}_{m}}}{r\mathcal{E}}
+\frac{4\chi c \mathcal{L}_{m_{0}}}{r^{2}}+\frac{4\chi b \mathcal{L}_{m_{0}}}{r}\right)P_{yy_{0}}+\mathcal{F}_{2}-\mathcal{Y}_{2N}}{-P_{xx_{0}}^{'}\Omega^{'}+\left\{\left(\chi \mathcal{L}_{m_{0}}+\frac{2 \chi \mathcal{L}_{m_{0}}}{r^{3}}\right)P_{xx_{0}}-P_{yy_{0}}\right\}\Omega}
\end{align}
The constraint Eq.\eqref{28} implies the unstable behavior of spherically symmetric charged anisotropic fluid. The instability range is determined by parameters such as anti-gravitating force, energy density, pressure, and extra curvature resulted by the framework of $f(R, \mathcal{L}_{m})$  theory. The above-mentioned inequality assures the instability of proposed system and subsequent conclusions are determined: the inequality Eq.\eqref{28} reveals that effective pressure, gravitational and anti gravitational forces should be perfectly balanced if the  $|\left(\chi a^{'}\mathcal{L}_{m_{0}}+2b^{'}\chi \mathcal{L}_{m_{0}}+\mathcal{F}_{1}\right)P_{xx_{0}}+a^{'}\chi \mathcal{L}_{m_{0}}\rho_{0}
+\left(\frac{-2\chi c^{'}\mathcal{L}_{m_{0}}}{r}-\frac{2\chi c\mathcal{L}_{m_{0}}}{r^{2}}-\frac{2\chi \bar{\mathcal{L}_{m}}}{r\mathcal{E}}
+\frac{4\chi c \mathcal{L}_{m_{0}}}{r^{2}}+\frac{4\chi b \mathcal{L}_{m_{0}}}{r}\right)P_{yy_{0}}+\mathcal{F}_{2}-\mathcal{Y}_{2N}|$ balances $| -P_{xx_{0}}^{'}\Omega^{'}+\left\{\left(\chi \mathcal{L}_{m_{0}}+\frac{2 \chi \mathcal{L}_{m_{0}}}{r^{3}}\right)P_{xx_{0}}-P_{yy_{0}}\right\}\Omega |$, and thus fluid configuration acquires the position of equilibrium. However, if the revised gravitational forces described in the numerator of Eq.\eqref{28} surpass $| -P_{xx_{0}}^{'}\Omega^{'}+\left\{\left(\chi \mathcal{L}_{m_{0}}+\frac{2 \chi \mathcal{L}_{m_{0}}}{r^{3}}\right)P_{xx_{0}}-P_{yy_{0}}\right\}\Omega |$, system is stable by preventing gravitational collapse, consequently, it can be asserted that the instability constraints are fulfilled by the relationship of effective pressure and gravitating forces.  Whereas, the dominance of $| -P_{xx_{0}}^{'}\Omega^{'}+\left\{\left(\chi \mathcal{L}_{m_{0}}+\frac{2 \chi \mathcal{L}_{m_{0}}}{r^{3}}\right)P_{xx_{0}}-P_{yy_{0}}\right\}\Omega |$, over $|\left(\chi a^{'}\mathcal{L}_{m_{0}}+2b^{'}\chi \mathcal{L}_{m_{0}}+\mathcal{F}_{1}\right)P_{xx_{0}}+a^{'}\chi \mathcal{L}_{m_{0}}\rho_{0}
+\left(\frac{-2\chi c^{'}\mathcal{L}_{m_{0}}}{r}-\frac{2\chi c\mathcal{L}_{m_{0}}}{r^{2}}-\frac{2\chi \bar{\mathcal{L}_{m}}}{r\mathcal{E}}
+\frac{4\chi c \mathcal{L}_{m_{0}}}{r^{2}}+\frac{4\chi b \mathcal{L}_{m_{0}}}{r}\right)P_{yy_{0}}+\mathcal{F}_{2}-\mathcal{Y}_{2N}|$ specifies the unstable dynamics of the system in correspondence to which $\Gamma$ ranges between $0$ and $1$.

The derived form of inequality interrelated to adiabatic index incorporates matter functions and dark source terms and specifies the significance of a key parameter in self-gravitating systems. These scenarios incorporate dark source terms relating to $f(R, \mathcal{L}_{m})$ theory, impacts of electric charge, energy density, and anisotropy resulted form pressure components.
The fluctuations in energy density relating to fluid configuration result pressure disparities in diverse directions which are described in mathematical expression of $\Gamma$, further simplicity of inequality in the context of considered system assures the endurance of the inequality. Specific restraints corresponding to pressure parts i.e., $P_{yy_{0}} < P_{xx_{0}}$, $\frac{4 \chi \mathcal{L}_{m_{0}}}{r}\left\{\frac{2c}{r}P_{yy_{0}}+bP_{yy_{0}}\right\} > 0$, $\frac{-2\chi c^{'}\mathcal{L}_{m_{0}}}{r}P_{yy_{0}} > \frac{-2 \chi}{r}\left( \frac{c\mathcal{L}_{m_{0}}}{r}+\frac{\bar{\mathcal{L}}_{m}}{\mathcal{E}}\right)P_{yy_{0}}$ and $\chi a^{'}\mathcal{L}_{m_{0}}P_{xx_{0}}+\mathcal{F}_{1}P_{xx_{0}}+a^{'}\chi \mathcal{L}_{m_{0}}\rho_{0}+\mathcal{F}_{2}-\mathcal{Y}_{2N} > 0$. should be fulfilled in Eq. \eqref{28} to attain stability.

It is important to mention that the inclusion of correction terms leads to the development of instability in N regime, reducing the overall stability of considered system. Our computations are reduced through the consideration of abbreviated terms $\mathcal{F}_{1}$, $\mathcal{F}_{2}$ and $\Omega$ having the following form
\begin{align}\nonumber
\mathcal{F}_{1}&=\frac{\chi \bar{\mathcal{L}}_{m}}{\mathcal{E}}+\frac{2c^{'}\chi \mathcal{L}_{m_{0}}}{r^{3}\mathcal{E}}+\frac{2\chi \bar{\mathcal{L}}_{m}}{r^{3}\mathcal{E}}-\frac{6c \chi \mathcal{L}_{m_{0}}}{r^{4}}, \\\nonumber
\mathcal{F}_{2}&=-\frac{4 \pi f^{'}\mathbb{E}\mathbb{F}}{\mathcal{E}}-\frac{2 \pi c^{'}\mathbb{E}_{0}^{2}}{r^{2}} -\frac{16 \pi b^{'}\mathbb{E}_{0}^{2}}{r}-\frac{16 \pi^{2}c^{'}\mathbb{E}_{0}^{2}}{r}-\frac{16 \pi^{2}c\mathbb{E}_{0}^{2}}{r^{2}}+\frac{20 \pi^{2}f\mathbb{E}\mathbb{E}_{0}}{r\mathcal{E}}-\frac{20 \pi^{2}b\mathbb{E}_{0}^{2}}{r},
\\\nonumber
\Omega&=-4a\mathcal{E}+\frac{2c}{br}+\frac{2a}{b}+\frac{\bar{\mathcal{L}}_{m}}{\mathcal{L}_{m_{0}}}.
\end{align}

\subsection{Constraint for adiabatic index in post-Newtonian epoch}

Our analysis is important in interpreting the behavior of charged anisotropic matter distribution within the context of $f(R, \mathcal{L}_{m})$ theory. We contemplate essential constraints for assessing and modifying collapse equation to determine the parameters relating to which mass radius ratio $\frac{m_{0}}{r}$ is that of the first order, while, greater order terms are avoided for the sake of simplicity. Several significant assumptions are $D_{0}=1+\frac{m_{0}}{r}$, $E_{0}=1-\frac{m_{0}}{r}$ and $F_{0}=r$, that reduce equations and demonstrate the relevance of related features of compact objects, in the case of which anisotropic composition determined by $f(R, \mathcal{L}_{m})$ theory in $pN$ regime adds the matter component. The contemplation of this approach determines the basic variables influencing the instability of compact structures and these factors incorporating curvature parameters, matter and metric coefficients impact the dynamics of extremely dense matter distribution.
\begin{itemize}
\item We are primarily intended to study the spherically symmetric highly dense charged fluid configurations in correspondence with the additional curvature terms resulting from alternative gravity implications that could effect the gravity-related interactions and change the limitation for physical instability.
\item Fluctuations in pressure and energy density in related dimensions of anisotropic fluid are regarded as the crucial
factors in determining the unstable behavior of the structure.
\item Furthermore, metric functions describe geometrical features of matter, which eventually influence the instability of the system and related fluctuations may have extra impacts on gravitational field and geometrical instability.
\end{itemize}
The absence of greater order terms ensure the concentration on basic properties of the matter. The following from of collapse equation offers the detailed illustration of instability restraints and their interaction with associated parameters. Our complete approach is intended to demonstrate the unstable behavior of non-static spherical compositions, establishing the room for subsequent research and evaluations of such structures in modified gravity theories. Higher-order ratio portions are ignored when computing the collapse equation, leaving the terms of the type $\frac{m_{0}}{r}$ having a ratio of ordering one. Consequently, collapse equation is derived in following form:
\begin{align}\nonumber
&\Gamma \eta_{1}^{'}\mathcal{K}^{'}+\left\{-\chi \mathcal{L}_{m_{0}}-\chi \mathcal{L}_{m_{0}}\mathcal{N}_{6}\mathcal{N}_{2}^{'}
-\frac{2\chi \mathcal{L}_{m_{0}}}{r^{3}}-2\chi \mathcal{L}_{m_{0}}\mathcal{N}_{6}\mathcal{N}_{1}^{'}\right\}\eta_{1}\Gamma\mathcal{K}
+\left\{\chi \bar{\mathcal{L}_{m}}+\chi \mathcal{L}_{m_{0}}a^{'}\mathcal{E}\mathcal{N}_{2}\right.\\\nonumber&\left.+\chi \bar{\mathcal{L}_{m}}\mathcal{N}_{2}\mathcal{N}_{2}^{'}-2 \chi b\mathcal{E} \mathcal{L}_{m_{0}}\mathcal{N}_{2}^{'}\mathcal{N}_{4}- \chi a\mathcal{E} \mathcal{L}_{m_{0}}\mathcal{N}_{2}^{'}+\frac{2\chi c^{'}\mathcal{E}\mathcal{L}_{m_{0}}}{r^{3}}+\frac{2\chi \bar{\mathcal{L}_{m}}}{r^{3}}
-\frac{6\chi c\mathcal{E}\mathcal{L}_{m_{0}}}{r^{4}}+2b^{'}\mathcal{E}\right.\\\nonumber&\left. \times \chi \mathcal{L}_{m_{0}}\mathcal{N}_{6}+2\chi \bar{\mathcal{L}_{m}\mathcal{N}_{1}^{'}\mathcal{N}_{6}} +6 \chi b \mathcal{E} \mathcal{L}_{m_{0}}\mathcal{N}_{8}\mathcal{N}_{1}^{'}\right\}P_{xx_{0}}
+\left\{a^{'}\chi \mathcal{E}\mathcal{L}_{m_{0}}\mathcal{N}_{1}+a\mathcal{E}\chi \mathcal{L}_{m_{0}}\mathcal{N}_{2}^{'}\mathcal{N}_{3}\right.\\\nonumber&\left.+\mathcal{N}_{2}\mathcal{N}_{2}^{'}\mathcal{N}_{3}\chi \bar{\mathcal{L}_{m}}
-2b\mathcal{E} \chi \mathcal{L}_{m_{0}}\mathcal{N}_{2}^{'}-4 \chi \mathcal{N}_{2}^{'}\mathcal{N}_{3}a\mathcal{E}\mathcal{L}_{m_{0}}\right\}\mu_{0}-\chi \mathcal{N}_{1}
\mathcal{N}_{2}^{'}\mathcal{L}_{m_{0}}\mathcal{K}-\left\{\frac{2\chi c^{'}\mathcal{E}\mathcal{L}_{m_{0}}\mathcal{N}_{4}}{r}\right.\\\nonumber&\left.-\frac{2\chi c\mathcal{E}\mathcal{L}_{m_{0}}\mathcal{N}_{4}}{r^{2}}-\frac{4\chi b\mathcal{E}\mathcal{L}_{m_{0}}\mathcal{N}_{6}}{r}+2\chi \bar{\mathcal{L}}_{m}\mathcal{N}_{4}\right\}P_{yy_{0}}+\Gamma \eta_{2} \mathcal{K}+4\pi i^{'}\mathbb{E}\mathcal{E}\mathcal{N}_{4}
+\frac{2 \pi c^{'}\mathcal{E}\mathbb{E}_{0}^{2}\mathcal{N}_{4}}{r^{2}}\\\nonumber&-\frac{16 \pi b^{'}\mathcal{E}\mathbb{E}_{0}^{2}\mathcal{N}_{6}}{r}+\frac{12 \pi b\mathcal{E}\mathbb{E}_{0}^{2}\mathcal{N}_{8}\mathcal{N}_{1}^{'}}{r}-\frac{8 \pi i\mathcal{E}\mathbb{E}\mathcal{N}_{6}\mathcal{N}_{1}^{'}}{r}+\frac{4 \pi c\mathbb{E}_{0}^{2}\mathcal{E}\mathcal{N}_{6}\mathcal{N}_{1}^{'}}{r}-\frac{20 \pi^{2} f\mathbb{E}_{0}\mathcal{E}\mathcal{N}_{4}}{r}
-2\pi \\\label{29}& \times a^{'} \mathcal{E}\mathbb{E}_{0}^{2}\mathcal{N}_{2}
 -4 \pi i \mathbb{E}_{0}\mathcal{N}_{2}\mathcal{N}_{2}^{'}+4 \pi b \mathcal{E} \mathbb{E}_{0}^{2}\mathcal{N}_{4}\mathcal{N}_{2}^{'}+2\pi a^{'}\mathcal{E}\mathbb{E}_{0}^{2}\mathcal{N}_{2}-\mathcal{Y}_{2PN}=0.
\end{align}
We can describe the condition for $\Gamma$ as being unstable, which implies that extremely dense matter illustrate stability despite violating instability restriction set by considering the collapse equation in $pN$ regime.
\begin{align}\nonumber
\Gamma & < \left\{\chi \bar{\mathcal{L}}_{m}-\chi a^{'}\mathcal{E}\mathcal{L}_{m_{0}}\mathcal{N}_{2}+\chi \bar{\mathcal{L}}_{m}\mathcal{N}_{2}\mathcal{N}_{2}^{'}
-2b\mathcal{E} \chi \mathcal{L}_{m_{0}}\mathcal{N}_{2}^{'}\mathcal{N}_{4}-\chi a \mathcal{E} \mathcal{L}_{m_{0}}\mathcal{N}_{2}^{'}+\frac{2 \chi c^{'}\mathcal{E}\mathcal{L}_{m_{0}}}{r^{3}}+\right.\\\nonumber&\left.\frac{2 \chi \bar{\mathcal{L}}_{m}}{r^{3}}-\frac{6 \chi c\mathcal{E}\mathcal{L}_{m_{0}}}{r^{4}}+2b^{'}\mathcal{E}\chi \mathcal{L}_{m_{0}}\mathcal{N}_{6}+2 \chi \bar{\mathcal{L}}_{m}\mathcal{N}_{1}^{'}\mathcal{N}_{6}-6 \chi \mathcal{L}_{m_{0}}b\mathcal{E}\mathcal{N}_{1}^{'}\mathcal{N}_{8}\right\}P_{xx_{0}}+\left\{a^{'}\mathcal{E} \chi \mathcal{L}_{0}\right.\\\nonumber&\left. \times \mathcal{N}_{1}+a\mathcal{E} \chi \mathcal{L}_{0}\mathcal{N}_{2}^{'}\mathcal{N}_{3}+\chi \bar{\mathcal{L}}_{m}\mathcal{N}_{2}\mathcal{N}_{2}^{'}\mathcal{N}_{3}-2b\chi \mathcal{E} \mathcal{L}_{m_{0}}\mathcal{N}_{2}^{'}-4a\mathcal{E}\chi \mathcal{L}_{m_{0}}
\mathcal{N}_{2}^{'}\mathcal{N}_{3}\right\}\mu_{0}-\chi \mathcal{K}\mathcal{L}_{m_{0}}\\\nonumber& \times \mathcal{N}_{1}\mathcal{N}_{2}^{'}+\mathcal{Y}_{2pN}-\left\{
\frac{2 \chi c^{'}\mathcal{E}}{r} \mathcal{L}_{m_{0}}\mathcal{N}_{4}+\frac{2 \chi c\mathcal{E}\mathcal{L}_{m_{0}}\mathcal{N}_{4}}{r^{2}}
+2 \chi \bar{\mathcal{L}}_{m}\mathcal{N}_{4}-\frac{4 \chi c\mathcal{E}\mathcal{L}_{m_{0}}\mathcal{N}_{4}}{r}-\frac{4 \chi b\mathcal{E}\mathcal{L}_{m_{0}}\mathcal{N}_{6}}{r}\right\}P_{yy_{0}}\\\label{pnc}& \times \frac{\mathcal{F}_{4}}{-\eta _{1}^{'}\mathcal{K}^{'}+\eta _{1}\mathcal{K}\left(1+\mathcal{N}_{2}^{'}\mathcal{N}_{6}+\frac{2}{r^{3}}+\mathcal{N}_{6}\mathcal{N}_{1}^{'}\right)\chi \mathcal{L}_{m_{0}}-\eta_{2}\mathcal{K}}.
\end{align}
where, the extra curvature terms relating to considered $f(R, \mathcal{L}_{m})$ theory in $pN$ regime are specified as $\mathcal{Y}_{2pN}$  which is described in Appendix. The terms mentioned in Eq. \eqref{pnc} must be satisfied corresponding to unstable restricted spherical configuration particularly in $pN$ regime. The following constraints must be fulfilled for $\Gamma$ :\\
\begin{align}\nonumber
&\chi \bar{\mathcal{L}}_{m}+\chi \bar{\mathcal{L}}_{m}\mathcal{N}_{2}\mathcal{N}_{2}^{'}
+\frac{2 \chi c^{'}\mathcal{E}\mathcal{L}_{m_{0}}}{r^{3}} > \chi a^{'}\mathcal{E}\mathcal{L}_{m_{0}}\mathcal{N}_{2}+2b\mathcal{E} \chi \mathcal{L}_{m_{0}}\mathcal{N}_{2}^{'}\mathcal{N}_{4},
\\\nonumber&
\frac{2 \chi \bar{\mathcal{L}}_{m}}{r^{3}}+2b^{'}\mathcal{E}\chi \mathcal{L}_{m_{0}}\mathcal{N}_{6}+2 \chi \bar{\mathcal{L}}_{m}\mathcal{N}_{1}^{'}\mathcal{N}_{6} > \frac{6 \chi c\mathcal{E}\mathcal{L}_{m_{0}}}{r^{4}} +6 \chi \mathcal{L}_{m_{0}}b\mathcal{E}\mathcal{N}_{1}^{'}\mathcal{N}_{8}, \\\nonumber&  \quad   a^{'}\mathcal{E}\chi \mathcal{L}_{m_{0}}\mathcal{N}_{1}+a\mathcal{E}\chi \mathcal{L}_{m_{0}}\mathcal{N}_{2}^{'}\mathcal{N}_{3}+\chi \bar{\mathcal{L}_{m}}\mathcal{N}_{2} \mathcal{N}_{2}^{'}\mathcal{N}_{3} > 2b\chi \mathcal{E}\mathcal{L}_{m_{0}}\mathcal{N}_{2}^{'}+4a\mathcal{E}\chi \mathcal{L}_{m_{0}}\mathcal{N}_{2}^{'}\mathcal{N}_{3}, \\\nonumber&   \quad \mathcal{Y}_{2pN} > \chi \mathcal{K}\mathcal{L}_{m_{0}},
\quad \frac{2 \chi c^{'}\mathcal{E}}{r}+\frac{2\chi c\mathcal{E}\mathcal{L}_{m_{0}}\mathcal{N}_{4}}{r^{2}}+2\chi \bar{\mathcal{L}}_{m}\mathcal{N}_{4} > \frac{2 \chi c\mathcal{E}\mathcal{L}_{m_{0}\mathcal{N}_{4}}}{r}+\frac{4b\chi \mathcal{E}\mathcal{L}_{m_{0}}\mathcal{N}_{6}}{r}, \\\nonumber& \quad \eta_{1}\mathcal{K}\chi \mathcal{L}_{m_{0}}> 0,
\quad \mathcal{F}_{4} > 0,
\quad \eta_{1}\mathcal{K}\left(1+\mathcal{N}_{2}^{'}\mathcal{N}_{6}+\frac{2}{r^{3}}
+\mathcal{N}_{6}\mathcal{N}_{1}^{'}\right)>\eta_{1}^{'}\mathcal{K}+\eta_{2}\mathcal{K}.
\end{align}
To simplify and comprehend our computations, we include various and repeating expressions listed below:
\begin{align}\nonumber
&\eta_{1}=\frac{P_{xx_{0}}}{\rho_{0}+P_{xx_{0}}}, \quad \eta_{2}=\frac{P_{yy_{0}}}{\rho_{0}+P_{yy_{0}}}, \quad \mathcal{N}_{1}=1+\frac{m_{0}}{r_{1}} \quad \mathcal{N}_{2}=1-\frac{m_{0}}{r_{1}} \quad \mathcal{N}_{3}=1+\frac{2m_{0}}{r_{1}} \\\nonumber&
\quad \mathcal{N}_{4}=1-\frac{2m_{0}}{r_{1}}, \quad \mathcal{N}_{5}=1+\frac{3m_{0}}{r_{1}}, \quad \mathcal{N}_{6}=1-\frac{3m_{0}}{r_{1}}.
\\\nonumber
\mathcal{F}_{4}&=4\pi i^{'}\mathbb{E}\mathcal{E}\mathcal{N}_{4}
+\frac{2 \pi c^{'}\mathcal{E}\mathbb{E}_{0}^{2}\mathcal{N}_{4}}{r^{2}}-\frac{16 \pi b^{'}\mathcal{E}\mathbb{E}_{0}^{2}\mathcal{N}_{6}}{r}+\frac{12 \pi b\mathcal{E}\mathbb{E}_{0}^{2}\mathcal{N}_{8}\mathcal{N}_{1}^{'}}{r}-\frac{8 \pi i\mathbb{E}\mathcal{E}\mathcal{N}_{6}\mathcal{N}_{1}^{'}}{r}+\frac{4 \pi c\mathbb{E}_{0}^{2}\mathcal{E}\mathcal{N}_{6}\mathcal{N}_{1}^{'}}{r}\\\nonumber&-\frac{20 \pi^{2} i\mathbb{E}_{0}\mathcal{E}\mathcal{N}_{4}}{r}-2\pi a^{'}\mathcal{E}\mathbb{E}_{0}^{2}\mathcal{N}_{2}-4 \pi i \mathbb{E}_{0}\mathcal{N}_{2}\mathcal{N}_{2}^{'}+4 \pi b \mathcal{E} \mathbb{E}_{0}^{2}\mathcal{N}_{4}\mathcal{N}_{2}^{'}+2\pi a^{'}\mathcal{E}\mathbb{E}_{0}^{2}\mathcal{N}_{2}.
\end{align}

\section{Final comments}

The study of stability associated with the compact structures within the context of modified gravity is an interesting subject for research. In this manuscript, specific limitations for evaluating the stability of spherically symmetric charged fluid are established after the consideration of $f(R,\mathcal{L}_{m})$ gravity. We determined the modified equations of motion and analyzed the conservation laws following the constrained form of Bianchi identities depending on the energy-momentum tensor. The perturbation technique is employed for the revised equations of motion along with the dynamical equations resulting in their representation as static and non-static components. We contemplated that the matter is in stable state, but it experiences deviations with passage of time. The collapse equation is derived by using the perturbed form of revised field equations that plays a crucial role in establishing the instability restraints for N and $pN$ approximations in the form of stiffness parameter $\Gamma$. It is concluded that the spherically symmetric fluid dispersion exhibits unstable dynamics until it fulfills Eq. \eqref{28} and \eqref{pnc} corresponding to N and $pN$ domains. While, the adiabatic index along with the model-based and charged terms having correspondence with dynamical anisotropy and energy density significantly contribute in establishing these constraints. Our analysis concluded that the repellent nature of dark source terms corresponding to $f(R, \mathcal{L}_{m})$ gravity determines the stable configuration of matter. Furthermore, the absence of non-diagnol terms signifies the disappearance of gravitating radiations resulting in no transfer of high energy \cite{herrera2014dissipative}. Our study assesses the complications associated with the gravitational collapse of compact structures and evaluates the stable and unstable dynamics of highly dense charged matter after considering the formalism of  $f(R, \mathcal{L}_{m})$ gravity.

A compact distribution of charged matter after consuming the internal energy experiences the gravitational collapse as the result of dominance of gravitational force over the outward pressure. In our case study, metric functions suggest the more complexity of the system that adds the interest in our work. The perturbation technique is considered to simplify the  non-linear modified differential equations, and the restraints for the dynamical instability are evaluated providing the thorough understanding of the gravitational interactions that govern the accelerated expansion of our universe. Our study determines the dynamics of collapsed spherically symmetric fluids by using the conservation laws and the Harrison-Wheeler EOS and establishes the instability restraints for $N$ and $pN$ domains, highlighting the importance of $\Gamma$ in determining the conduct of collapse. We conclude that the hydrostatic equilibrium is important in understanding the stability of highly dense matter and the intrinsically related equilibrium systems emerged from the interaction between fundamental parameters and the matter distribution by the revised equations of motion. This illustrates the significance of $\Gamma$ in defining restraints for gravitational collapse, which validates Chandrasekhar's analysis of stability. The mathematical form of $\Gamma$ has significant relevance in determining the mechanical features of dense charged matter and governs the stability and evolution of compact structures.
The additional terms from the considered gravity model $f(R, \mathcal{L}_{m})=R+\zeta R^{2}+\chi \mathcal{L}_{m}^{2}$, results the modified form of Ricci scalar $R$ and matter Lagrangian $\mathcal{L}_{m}$ and these modifications introduce extra curvature terms that are not contemplated in GR. Specifically, the terms $\zeta R^{2}$ and $\chi \mathcal{L}_{m}^{2}$ imply significant  effects on fluid dynamics, ultimately revising the instability restraints. These effects are described as correction terms which seems to be the additional forces that disturbs the balance between gravity and pressure.

In GR, the stability of compact objects is primarily established by the interaction between attractive gravitational force and pressure fluctuations, while in the context of  $f(R, \mathcal{L}_{m})$ gravity the consideration of dark source terms in addition to the matter Lagrangian determine the instability restraints. The inclusion of dark source terms modifies the limiting value of $\Gamma$ establishing a new restraint corresponding to the matter instability. The modified field equations associated with the $f(R, \mathcal{L}_{m})$ gravity lead to additional forces that can be of attraction or repulsion depending on the gravity model. Anti-gravitational effects can impede the earlier stage of collapse or may conclude different equilibrium configurations as compared to GR in N and $pN$ domains. The terms corresponding to $f(R, \mathcal{L}_{m})$ theory prescribe cosmic implications to characterize the mechanisms of late-time cosmological evolution particularly the accelerated expansion that is not thoroughly described in GR. The revised collapse equation in $f(R, \mathcal{L}_{m})$ theory modifies the needed time for hydrostatic equilibrium and the density dispersion, changes in pressure along with the energy flux are interrelated in a different way in considered gravity, featuring novel attributes such as regular instabilities and delayed collapse. The obtained findings consider the static components of the fluid dispersion and are compatible with \cite{chandrasekhar1964dynamical,herrera2012dynamical}.
The mathematical framework of $f(R, \mathcal{L}_{m})$ gravity offers physical features that minimize the stability limitations. Furthermore, the relations for $\Gamma$ are derived in Eqs. \eqref{28} and \eqref{pnc} in our case study, the adiabatic index should satisfy these restraints for ensuring the dynamical  instability, that can be interpreted as the considered system will be in stable state if certain inequalities relating to N and $pN$ doamins are avoided. Additionally, the dynamical instability relies on the constraints related to the hydrostatic equilibrium which are derived from the revised equations of motion.

The correction terms provide stability for the system that may collapse in N and $pN$ eras in the case of GR and maintain the stability of modified parameters for cosmic radiations. Moreover, the contemplated gravity model along with the higher-order metric derivatives have significant relevance in establishing the stability constraints. We studied the physical instability of anisotropic matter relating to the N and $pN$ domains in $f(R, \mathcal{L}_{m})$ gravity along with the stability conditions derived in Eqs. \eqref{28} and \eqref{pnc}, in which the dispersion of energy density, anisotropic pressure and correction parameter impacts the stability. The adiabatic index should satisfy the derived restraints in order to establish the hydrostatic equilibrium, demonstrating the importance of the interaction between extra curvature terms and characteristics of charged matter in assessing the stability of highly dense matter. Consequently, it can be asserted that the additional curvature effects, revised stability restraints, extra forces are the attributes associated with the considered self-gravitational fluid in $f(R, \mathcal{L}_{m})$ theory. The instability conditions related to the spherically symmetric charged matter are derived after ensuring that these inequalities are addressed in GR and we evaluated how the increased curvature affects the stability restrictions corresponding to the considered gravity model. The field equations corresponding to $f(R, \mathcal{L}_{m})$ theory are used to establish distinct equilibrium states which illustrates the gravitational interaction with charged matter. These modifications provide a more thorough framework for evaluating the evolution of celestial objects and explain the phenomenon that relates to gravity changes beyond GR.

\subsection*{Acknowledgements:}

The authors extend their appreciation to the Deanship of Research and Graduate Studies at King Khalid University for funding this work through large Research Groups Program under grant number RGP 2/289/47.

\section*{Appendix}

\renewcommand{\theequation}{A\arabic{equation}}
\setcounter{equation}{0}

The assumption of $f(R, \mathcal{L}_{m})$ theory leads to extra curvature terms in modified field equations and their values are specified as dark source elements represented by $\varphi_{ij}^{'s}$ where $ij=00,01,11, 22$ having the subsequent form:
\begin{align}\nonumber
\varphi_{00}&=-\frac{D^{2}}{E^{2}}f^{''}_{R}-\frac{D^{2}F^{'}}{E^{2}F}f_{R}^{'}+\frac{\dot{E}}{E}\dot{f_{R}}+\frac{D^{2}E^{'}}{E^{3}}f_{R}^{'}
-\frac{D^{2}F^{'}}{E^{2}F}f_{R}^{'},
\\\nonumber
\varphi_{01}&=-\dot{f_{R}^{'}}+\frac{\dot{D}}{D}\dot{f_{R}}+\frac{\dot{E}}{E}f_{R}^{'}
\\\nonumber
\varphi_{11}&=-\frac{E^{2}}{D^{2}}\ddot{f_{R}}+\frac{\dot{D}E^{2}}{D^{3}}\dot{f_{R}}+\frac{D^{'}}{D}f_{R}^{'}+f_{R}^{''}-\frac{\dot{E}E}{D^{2}}
\dot{f_{R}}-\frac{E^{'}}{E}f_{R}^{'}-\frac{E^{2}\dot{F}}{D^{2}F}\dot{f_{R}}+\frac{F^{'}}{F}f_{R}^{'}+\frac{E^{2}\dot{F}}{D^{2}F}\dot{f_{R}}
+\frac{\dot{F}}{F}f_{R}^{'}+\frac{E \dot{E}}{D^{2}}\\\nonumber& \times \dot{f_{R}}+\frac{E^{'}}{E}f_{R}^{'}-f_{R}^{''},
\\\nonumber
\varphi_{22}&=\frac{-F^{2}}{D^{2}}\ddot{f_{R}}+\frac{\dot{D}F^{2}}{D^{3}}\dot{f_{R}}+
\frac{\dot{F}F}{D^{2}}\dot{f_{R}}+\frac{D^{'}F^{2}}{D E^{2}}f_{R}^{'}+\frac{F^{2}}{E^{2}}f_{R}^{''}-\frac{\dot{E}F^{2}}{D^{2}E}\dot{f_{R}}-\frac{E^{'}F^{2}}{E^{3}}f_{R}^{'}
-\frac{\dot{F}F}{D^{2}}\dot{f_{R}}-\frac{F F^{'}}{E^{2}}f_{R}^{'}.
\end{align}
The static components of the modified equations of motion are described as:
\begin{align}\nonumber&
-\frac{2\zeta R_{0}D_{0}D_{0}^{''}}{E_{0}^{2}}-\frac{4\zeta R_{0}D_{0}D_{0}^{'}F_{0}^{'}}{E_{0}^{2}F_{0}}+\frac{2\zeta R_{0}D_{0}D_{0}^{'}E_{0}^{'}}{E_{0}^{3}}-\frac{R_{0}D_{0}^{2}}{2}-2 \pi D_{0}^{2}\mathbb{E}^{2}-\frac{\zeta R_{0}^{2}D_{0}^{2}}{2}-\frac{\chi \mathcal{L}_{m_{0}}^{2}D_{0}^{2}}{2}+\frac{2D_{0}^{2}F_{0}^{''}}{F_{0}E_{0}^{2}}\\\nonumber&+\frac{D_{0}^{'}F_{0}^{'2}}{F_{0}^{2}E_{0}^{2}}
-\frac{2D_{0}F_{0}^{'}E_{0}^{'}}{E_{0}^{3}}+\rho_{0}\chi \mathcal{L}_{m_{0}}+\chi \eta \mathcal{L}_{m_{0}}D_{0}^{2}-\chi \mathcal{L}_{m_{0}}D_{0}^{2},
\\\nonumber&
\chi \mathcal{L}_{m_{0}}F_{0}^{2}-2\zeta R_{0}-\chi \mathcal{L}_{m_{0}}F_{0}^{2}P_{yy_{0}}+\frac{2\zeta R_{0}F_{0}F_{0}^{'}}{E_{0}^{2}}+\frac{2\zeta R_{0}F_{0}^{'2}}{E_{0}^{2}}-\frac{2\zeta R_{0}F_{0}F_{0}^{'}E_{0}^{'}}{E_{0}^{3}}-2 \pi F_{0}^{2}\mathbb{E}_{0}^{2}+\frac{R_{0}F_{0}^{2}}{2}+\frac{\zeta R_{0}}{2}\\\nonumber&\times F_{0}^{2}+\frac{2 \zeta R_{0}D_{0}^{'}F_{0}^{2}}{D_{0}E_{0}^{2}}+\frac{2 \zeta R_{0}^{''}F_{0}^{2}}{E_{0}^{2}}-\frac{2 \zeta R_{0}^{'}F_{0}^{'}F_{0}}{E_{0}^{2}}-\frac{2 \zeta R_{0}^{'}E_{0}^{'}F_{0}^{2}}{E_{0}^{3}}.
\end{align}
The expression for $\mathcal{Y}_{1}$ mentioned in non-static component of non-conservation equation has following form
\begin{align}\nonumber
\mathcal{Y}_{1}&=\left\{-\frac{2e\zeta D_{0}^{''}}{D_{0}^{3}E_{0}^{2}}-\frac{2a^{''}\zeta R_{0}}{D_{0}^{3}E_{0}^{2}}+\frac{2a\zeta D_{0}^{''}R_{0}}{D_{0}^{4}E_{0}^{2}}-\frac{4e\zeta D_{0}^{'}F_{0}^{'}}{D_{0}^{3}E_{0}^{2}}-\frac{4a^{'}\zeta F_{0}^{'}R_{0}}{D_{0}^{3}E_{0}^{2}F_{0}}-\frac{8\zeta D_{0}^{'}R_{0}c^{'}}{D_{0}^{3}E_{0}^{2}F_{0}}-\frac{8c^{'}\zeta D_{0}^{'}R_{0}}{D_{0}^{3}E_{0}^{2}F_{0}}+\frac{4a\zeta D_{0}^{'} F_{0}^{'}R_{0}}{D_{0}^{4}E_{0}^{2}F_{0}}\right.\\\nonumber&
+\frac{8b\zeta D_{0}^{'} F_{0}^{'}R_{0}}{D_{0}^{3}E_{0}^{3}F_{0}}+\frac{6c\zeta D_{0}^{'} F_{0}^{'}R_{0}}{D_{0}^{3}E_{0}^{2}F_{0}^{2}}-\frac{2e\zeta D_{0}^{'} E_{0}^{'}}{D_{0}^{3}E_{0}^{3}}+\frac{2a^{'}\zeta E_{0}^{'} R_{0}}{D_{0}E_{0}^{3}}+\frac{2b^{'}\zeta D_{0}^{'} R_{0}}{D_{0}^{3}E_{0}^{3}}-\frac{2a\zeta D_{0}^{'} E_{0}^{'} R_{0}}{D_{4}E_{0}^{3}}-\frac{4b\zeta D_{0}^{'} E_{0}^{'} R_{0}}{D_{0}^{3}E_{0}^{4}}-\frac{e}{2}\\\nonumber&\times \frac{1}{D_{0}^{2}}-\frac{\zeta eR_{0}}{D_{0}^{2}}+\frac{aR_{0}}{D_{0}^{3}}+\frac{\zeta a R_{0}^{2}}{D_{0}^{3}}+\frac{\chi a\mathcal{L}_{m_{0}}^{2}}{D_{0}^{3}}
-\frac{4\zeta c^{'}R_{0}^{'}}{D_{0}^{2}E_{0}^{2}F_{0}}-\frac{4\zeta c^{''}R_{0}}{D_{0}^{2}E_{0}^{2}F_{0}}-\frac{12\zeta c^{'}E_{0}^{'}R_{0}}{D_{0}^{2}E_{0}^{3}F_{0}}-\frac{8\zeta c^{'}F_{0}^{'}R_{0}}{D_{0}^{2}E_{0}^{2}F_{0}^{2}}+\frac{4\zeta bF_{0}^{'}R_{0}^{'}}{F_{0}D_{0}^{2}E_{0}^{3}}\\\nonumber&+\frac{4b^{'}\zeta R_{0}F_{0}^{'}}{D_{0}^{2}E_{0}^{3}F_{0}}+\frac{4b\zeta R_{0}F_{0}^{''}}{D_{0}^{2}E_{0}^{3}F_{0}}-\frac{12b\zeta R_{0}F_{0}^{'}E_{0}^{'}}{D_{0}^{2}E_{0}^{4}F_{0}}+\frac{4c\zeta D_{0}^{'}}{D_{0}^{3}E_{0}^{2}F_{0}}-\frac{8c\zeta R_{0}D_{0}^{''}}{D_{0}^{3}E_{0}^{2}F_{0}}-\frac{4c\zeta D_{0}^{'}R_{0}}{D_{0}^{4}E_{0}^{2}F_{0}}-\frac{2c D_{0}^{'}F_{0}^{'}}{D_{0}^{3}E_{0}^{2}F_{0}^{2}}-\frac{a R_{0}}{D_{0}^{3}}-\frac{\zeta  }{D_{0}}\\\nonumber& \times \frac{aR_{0}^{2}}{D_{0}^{2}}
-\frac{a \chi \mathcal{L}_{m_{0}}^{2}}{D_{0}^{3}}-\frac{b R_{0}}{D_{0}^{2}E_{0}}-\frac{\zeta b R_{0}}{D_{0}^{2}E_{0}}-\frac{b \chi \mathcal{L}_{m_{0}}^{2}}{D_{0}^{2}E_{0}}-\frac{4 \zeta c D_{0}^{''} R_{0}}{D_{0}^{3}E_{0}^{2}F_{0}}-\frac{\zeta cR_{0}}{D_{0}^{2}F_{0}}-\frac{cR_{0}}{D_{0}^{2}F_{0}}-\frac{2 \chi c \mathcal{L}_{m_{0}}}{D_{0}^{2}F_{0}}+\frac{\chi b \mathcal{L}_{m_{0}}}{D_{0}^{2}E_{0}}-\frac{2 \zeta }{D_{0}^{2}}\\\nonumber&\left. \times \frac{c R_{0}F_{0}^{''}}{F_{0}^{2}E_{0}^{2}}-\frac{2 \zeta c R_{0}F_{0}^{'}E_{0}^{'}}{D_{0}^{2}E_{0}^{3}F_{0}^{2}}-\frac{2 \zeta c R_{0}}{D_{0}^{2}F_{0}^{3}}+\frac{ c R_{0}}{2D_{0}^{2}F_{0}}
+\frac{ c \zeta R_{0}}{2D_{0}^{2}F_{0}}+\frac{c \chi \mathcal{L}_{m_{0}}^{2}}{2D_{0}^{2}F_{0}}-\frac{2 \chi a \eta \mathcal{L}_{m_{0}}}{D_{0}^{3}}
-\frac{\chi b \eta E_{0} \mathcal{L}_{m_{0}}}{D_{0}^{2}E_{0}^{2}}\right\}\dot{\mathcal{\mathcal{E}}}+\frac{4 \zeta c R_{0}}{D_{0}^{4}}\\\nonumber& \times \dddot{\mathcal{\mathcal{E}}}
+\frac{2 \zeta b R_{0}\dddot{\mathcal{\mathcal{E}}}}{D_{0}^{4}E_{0}}+\frac{2a\mathcal{\mathcal{E}}}{D_{0}^{3}E_{0}^{2}}\varphi_{01}^{'}
+\frac{2b\mathcal{\mathcal{E}}}{D_{0}^{4}E_{0}^{2}}\varphi_{01}^{'}
-\frac{b^{'}\mathcal{\mathcal{E}}}{D_{0}^{2}E_{0}^{3}}+\frac{6a\mathcal{\mathcal{E}}D_{0}^{'}}{D_{0}^{4}E_{0}^{2}}
\varphi_{01}+\frac{4b\mathcal{\mathcal{E}}D_{0}^{'}}{D_{0}^{3}E_{0}^{3}}\varphi_{01}
-\frac{2\mathcal{\mathcal{E}}b^{'}}{D_{0}^{2}E_{0}^{3}}\varphi_{01}+\frac{2a\mathcal{\mathcal{E}}E_{0}^{'}}{D_{0}^{3}E_{0}^{3}}
\varphi_{01}
\\\nonumber& +\frac{4b}{D_{0}^{2}} \frac{\mathcal{\mathcal{E}}E_{0}^{'2}}{E_{0}^{4}}\varphi_{01}-\frac{a^{'}\mathcal{\mathcal{E}}}{D_{0}^{3}E_{0}^{2}}
\varphi_{01}+\frac{2b\mathcal{\mathcal{E}}D_{0}^{'2}}{D_{0}^{3}E_{0}^{3}}\varphi_{01}+\frac{3a\mathcal{\mathcal{E}}
D_{0}}{D_{0}^{4}E_{0}^{2}}\varphi_{01}+\frac{3b\mathcal{\mathcal{E}}E_{0}^{'}}{D_{0}^{2}
E_{0}^{4}}+\frac{2a\mathcal{\mathcal{E}}E_{0}^{'}}{D_{0}^{3}E_{0}^{3}}-\frac{2c^{'}\mathcal{\mathcal{E}}}{D_{0}^{2}
E_{0}^{2}F_{0}} \varphi_{01}+\frac{4bF_{0}^{'}\mathcal{\mathcal{E}}}{D_{0}^{2}
E_{0}^{3}F_{0}}\varphi_{01}\\\nonumber&+\frac{4aF_{0}^{'}\mathcal{\mathcal{E}}}{D_{0}^{3}E_{0}^{2}F_{0}}
\varphi_{01}+\frac{4cF_{0}^{'}\mathcal{\mathcal{E}}}{D_{0}^{2}
E_{0}^{2}F_{0}^{2}}\varphi_{01}-\frac{\chi \bar{\mathcal{L}_{m}}\mu_{0}}{D_{0}^{4}}-\frac{\chi \bar{\mathcal{L}_{m}}\eta}{D_{0}^{2}}-\frac{b\chi \mathcal{\mathcal{E}}\eta \mathcal{L}_{m_{0}}}{D_{0}^{2}E_{0}}+\frac{2b\chi \mathcal{\mathcal{E}} \mathcal{L}_{m_{0}}F_{0}}{D_{0}^{3}E_{0}}+\frac{2ab\chi \mathcal{\mathcal{E}}\eta \mathcal{L}_{m_{0}}}{D_{0}^{3}E_{0}}-\frac{2a\eta \mathcal{\mathcal{E}}}{D_{0}^{3}}\\\nonumber&-\frac{4a^{'}\chi \mathcal{\mathcal{E}} \mathcal{L}_{m_{0}}}{D_{0}^{4}E_{0}^{3}}+\frac{12b\mathcal{\mathcal{E}}\chi \mathcal{L}_{m_{0}}D_{0}^{'}}{D_{0}^{4}E_{0}^{4}}+\frac{16a\mathcal{\mathcal{E}}\chi \mathcal{L}_{m_{0}}D_{0}^{'}}{D_{0}^{5}E_{0}^{3}}-\frac{4a^{'}\mathcal{\mathcal{E}}\chi \mathcal{L}_{m_{0}}\eta}{D_{0}^{4}E_{0}^{3}}+\frac{4a^{'}s\mathcal{\mathcal{E}}\chi \mathcal{L}_{m_{0}}}{D_{0}^{4}E_{0}^{3}}-\frac{4\chi \eta D_{0}\bar{\mathcal{L}_{m}}}{D_{0}^{4}E_{0}^{3}}+\frac{4\chi s D_{0}\bar{\mathcal{L}_{m}}}{D_{0}^{4}E_{0}^{3}}\\\nonumber&+\frac{12\chi \eta b\mathcal{\mathcal{E}} D_{0}^{'}\mathcal{L}_{m_{0}}}{D_{0}^{4}E_{0}^{4}} -\frac{12\chi s b\mathcal{\mathcal{E}} D_{0}^{'}\mathcal{L}_{m_{0}}}{D_{0}^{4}E_{0}^{4}}-\frac{2\chi \eta b^{'}\mathcal{\mathcal{E}} \mathcal{L}_{m_{0}}}{D_{0}^{3}E_{0}^{2}}+\frac{2\chi s b^{'}\mathcal{\mathcal{E}} \mathcal{L}_{m_{0}}}{D_{0}^{3}E_{0}^{2}}-\frac{2 \eta \chi E_{0}^{'}\bar{\mathcal{L}_{m}}}{D_{0}^{3}E_{0}^{2}}+\frac{2 s \chi E_{0}^{'}\bar{\mathcal{L}_{m}}}{D_{0}^{3}E_{0}^{2}}.
\end{align}
The mathematical form of $\mathcal{Y}_{2}$ appearing in non-static component of non-conservation equation corresponding to $\nu=1$ is
\begin{align}\nonumber
\mathcal{Y}_{2}&=-\frac{4 \zeta R_{0}c^{'}\ddot{\mathcal{E}}}{D_{0}^{2}E_{0}^{2}F_{0}}+\frac{2b\mathcal{E}}{D_{0}^{2}E_{0}^{3}}\dot{\varphi_{01}}
+\frac{2a\mathcal{E}}{D_{0}^{3}E_{0}^{2}}\dot{\varphi_{01}}
-\frac{2a\dot{\mathcal{E}}}{D_{0}^{3}E_{0}^{2}}\varphi_{01}-\frac{2b\dot{\mathcal{E}}}{D_{0}^{2}E_{0}^{3}}\varphi_{01}+\frac{4\zeta c^{''}\mathcal{E} R_{0}^{'}}{E_{0}^{2}F_{0}}-\frac{8\zeta b\mathcal{E} R_{0}^{'}F_{0}^{''}}{E_{0}^{3}F_{0}}+\frac{4\zeta e\mathcal{E} F_{0}^{''}}{E_{0}^{4}F_{0}}
\\\nonumber&+\frac{4\zeta c^{'''}\mathcal{E} R_{0}}{E_{0}^{4}F_{0}}-\frac{16\zeta F_{0}^{'''} b\mathcal{E} R_{0}^{'}}{E_{0}^{5}F_{0}}-\frac{4\zeta F_{0}^{'''}c\mathcal{E} R_{0}}{E_{0}^{4}F_{0}^{2}}+\frac{2\zeta F_{0}^{''}c^{''}\mathcal{E} R_{0}}{E_{0}^{4}F_{0}^{2}}+\frac{2\zeta F_{0}^{''}c^{'}\mathcal{E} R_{0}}{E_{0}^{4}F_{0}^{2}}+\frac{2\zeta F_{0}^{'}F_{0}^{''}e\mathcal{E} R_{0}}{E_{0}^{4}F_{0}^{2}}-\frac{4\zeta c\mathcal{E} F_{0}^{'} }{E_{0}^{4}F_{0}}\\\nonumber& \times \frac{F_{0}^{''}R_{0}}{F_{0}^{2}}-\frac{8 \zeta b\mathcal{E} R_{0}F_{0}^{'}F_{0}^{''}}{E_{0}^{5}F_{0}^{2}}
-\frac{2\zeta b \ddot{\mathcal{E}}R_{0}^{'}}{D_{0}^{2}E_{0}^{3}}-\frac{2\zeta b^{'} \dddot{\mathcal{E}}R_{0}}{D_{0}^{2}E_{0}^{3}}+\frac{4\zeta b \ddot{\mathcal{E}}R_{0}D_{0}^{'}}{D_{0}^{3}E_{0}^{3}}+\frac{2\zeta b \ddot{\mathcal{E}}R_{0}E_{0}^{'}}{D_{0}^{2}E_{0}^{4}}+\frac{2\zeta a^{''} \mathcal{E}R_{0}^{'}}{D_{0}E_{0}^{4}}+\frac{2\zeta e^{'} \mathcal{E}D_{0}^{''}}{D_{0}E_{0}^{4}}\\\nonumber& -\frac{2\zeta a\mathcal{E} R_{0}^{'}D_{0}^{''}}{D_{0}^{2}E_{0}^{4}}-\frac{8\zeta b \mathcal{E}R_{0}^{'}D_{0}^{''}}{D_{0}E_{0}^{5}}+\frac{2\zeta a^{'''} \mathcal{E}R_{0}}{D_{0}E_{0}^{4}}+\frac{2\zeta D_{0}^{'''} e\mathcal{E}}{D_{0}E_{0}^{4}}-\frac{2\zeta R_{0} D_{0}^{'''} a\mathcal{E}}{D_{0}^{2}E_{0}^{4}}
-\frac{4\zeta R_{0} D_{0}^{'''} b\mathcal{E}}{D_{0}E_{0}^{5}}-\frac{2\zeta R_{0} D_{0}^{''} a^{'}\mathcal{E}}{D_{0}^{2}E_{0}^{4}}\\\nonumber&
 -\frac{2\zeta D_{0}^{'} D_{0}^{''} e\mathcal{E}}{D_{0}^{2}E_{0}^{4}}+\frac{4\zeta R_{0} D_{0}^{'} D_{0}^{''} b\mathcal{E}}{D_{0}^{2}E_{0}^{5}}+\frac{4\zeta R_{0} D_{0}^{'} D_{0}^{''} a\mathcal{E}}{D_{0}^{3}E_{0}^{5}}-\frac{8\zeta R_{0} D_{0}^{''} E_{0}^{'} a^{'}\mathcal{E}}{D_{0}^{2}E_{0}^{5}}-\frac{8\zeta R_{0} D_{0}^{'} E_{0}^{'} a^{''}\mathcal{E}}{D_{0}^{2}E_{0}^{5}} - \frac{8\zeta R_{0} D_{0}^{'} E_{0}^{'} a^{''}\mathcal{E}}{E_{0}^{5}D_{0}^{2}}\\\nonumber&-\frac{8\zeta  D_{0}^{'} D_{0}^{''}E_{0}^{'} e\mathcal{E}}{D_{0}^{2}E_{0}^{5}}-\frac{16\zeta R_{0} D_{0}^{'} D_{0}^{''} b\mathcal{E} E_{0}^{'}}{D_{0}^{2}E_{0}^{6}}-\frac{16\zeta R_{0} D_{0}^{'} D_{0}^{''} E_{0}^{'} a\mathcal{E}}{D_{0}^{3}E_{0}^{5}}+\frac{24\zeta R_{0} D_{0}^{'} D_{0}^{''} E_{0}^{'} b\mathcal{E}}{D_{0}^{2}E_{0}^{6}} - \frac{4\zeta R_{0}^{'} F_{0}^{'}  b_{0}^{'} \mathcal{E}}{F_{0}E_{0}^{5}}-\frac{4}{F_{0}}\\\nonumber& \times \frac{\zeta E_{0}^{'} F_{0}^{'}  e^{'}\mathcal{E}}{E_{0}^{5}}+\frac{4\zeta R_{0}^{'} E_{0}^{'} F_{0}^{'}  b\mathcal{E}}{F_{0}E_{0}^{6}}+\frac{4\zeta R_{0}^{'} E_{0}^{'} F_{0}^{'}  c\mathcal{E}}{F_{0}^{2}E_{0}^{5}}-\frac{4\zeta R_{0} F_{0}^{'} b^{''}\mathcal{E}}{F_{0}E_{0}^{5}}-\frac{4\zeta R_{0} E_{0}^{''} c^{'}\mathcal{E}}{F_{0}E_{0}^{5}}+\frac{4\zeta F_{0}^{'}e\mathcal{E} E_{0}^{'}}{F_{0}E_{0}^{5}}+\frac{20b\mathcal{E}\zeta R_{0}}{E_{0}F_{0}}\\\nonumber& \times \frac{ E_{0}^{''} F_{0}^{'}}{E_{0}^{5}}+\frac{4\zeta R_{0} E_{0}^{''} c\mathcal{E}F_{0}^{'}}{F_{0}^{2}E_{0}^{5}}+\frac{24\zeta R_{0} E_{0}^{'} b^{'}\mathcal{E}F_{0}^{'}}{F_{0}E_{0}^{6}}+\frac{12 \zeta R_{0}E_{0}^{'}c^{'}\mathcal{E}}{E_{0}^{6}F_{0}}+\frac{12 \zeta E_{0}^{'2}F_{0}^{'}e\mathcal{E}}{E_{0}^{6}F_{0}}-\frac{72 \zeta }{E_{0}^{7}}  \frac{b\mathcal{E}E_{0}^{'2}F_{0}^{'}R_{0}}{F_{0}}+\frac{6 \zeta R_{0}}{E_{0}^{5}}\\\nonumber&\times\frac{ F_{0}^{'2}b^{'}\mathcal{E}}{F_{0}^{2}}+\frac{12 \zeta R_{0} E_{0}^{'}F_{0}^{'}c^{'}\mathcal{E}}{E_{0}^{5}F_{0}^{2}}+\frac{6 \zeta E_{0}^{'}F_{0}^{'2}e\mathcal{E}}{E_{0}^{5}F_{0}^{2}}-\frac{30 \zeta R_{0} E_{0}^{'}F_{0}^{'2}b\mathcal{E}}{E_{0}^{6}F_{0}^{2}}-\frac{12 \zeta c\mathcal{E} R_{0} E_{0}^{'}F_{0}^{'2}}{F_{0}^{3}E_{0}^{5}}- \frac{2 \zeta E_{0}^{'}R_{0}^{'}}{b^{'}\mathcal{E}D_{0}E_{0}^{5}}
-\frac{2 \zeta b^{'}\mathcal{E}D_{0}^{'}}{D_{0}}\\\nonumber& \times \frac{R_{0}^{'}}{E_{0}^{5}}-\frac{2 \zeta e^{'}\mathcal{E}E_{0}^{'}D_{0}^{'}}{D_{0}E_{0}^{5}}+\frac{2 \zeta a\mathcal{E}E_{0}^{'}R_{0}^{'}D_{0}^{'}}{D_{0}^{2}E_{0}^{5}}+\frac{10 \zeta b\mathcal{E}E_{0}^{'}R_{0}^{'}D_{0}^{'}}{D_{0}E_{0}^{6}}
+\frac{2 \zeta a^{''}\mathcal{E}E_{0}^{''}R_{0}}{D_{0}E_{0}^{5}}+\frac{2 \zeta b^{'}}{D_{0}} \frac{\mathcal{E}D_{0}^{''}R_{0}}{E_{0}^{5}}+\frac{2 \zeta e\mathcal{E}E_{0}^{'}D_{0}^{''}}{D_{0}E_{0}^{5}}\\\nonumber&-\frac{2 \zeta a\mathcal{E}E_{0}^{'}R_{0}D_{0}^{''}}{D_{0}^{2}E_{0}^{5}}-\frac{10 \zeta b\mathcal{E}E_{0}^{'}R_{0}D_{0}^{''}}{D_{0}E_{0}^{6}}-\frac{2 \zeta R_{0} a^{'}\mathcal{E}E_{0}^{''}}{D_{0}E_{0}^{5}} -\frac{2 \zeta b^{''}\mathcal{E}D_{0}^{'}R_{0}}{D_{0}E_{0}^{5}}-\frac{2 \zeta b\mathcal{E}}{D_{0}} \frac{E_{0}^{''}D_{0}^{'}}{E_{0}^{5}}+\frac{2 \zeta a\mathcal{E}E_{0}^{''}D_{0}^{'}R_{0}}{D_{0}^{2}E_{0}^{5}}+\frac{10 }{D_{0}}\\\nonumber& \times \frac{\zeta b\mathcal{E}E_{0}^{''}D_{0}^{'}R_{0}}{E_{0}^{6}}+\frac{6 \zeta a\mathcal{E}E_{0}^{'2}R_{0}}{D_{0}E_{0}^{6}}+\frac{6 \zeta e\mathcal{E}E_{0}^{'2}D_{0}^{'2}}{D_{0}E_{0}^{6}}+\frac{12 \zeta b^{'}\mathcal{E}E_{0}^{'}R_{0}D_{0}^{'}}{D_{0}E_{0}^{6}}+\frac{18 \zeta }{D_{0}} \frac{b\mathcal{E}R_{0}E_{0}^{'2}D_{0}^{'}}{E_{0}^{7}}+\frac{6 \zeta a\mathcal{E}R_{0}E_{0}^{'2}D_{0}^{'}}{D_{0}^{2}E_{0}^{6}}\\\nonumber&+\frac{e^{'}\mathcal{E}}{2E_{0}^{2}}+\frac{\zeta e\mathcal{E}R_{0}^{'}}{E_{0}^{2}}+\frac{\zeta e^{'}\mathcal{E}R_{0}}{E_{0}^{2}}+\frac{\chi \bar{\mathcal{L}_{m}}\mathcal{L}_{m_{0}}^{'}}{E_{0}^{2}} +\frac{b \mathcal{E}R_{0}^{'}}{E_{0}^{3}}+\frac{2\zeta b\mathcal{E} R_{0}R_{0}^{'}}{E_{0}^{3}}-\frac{2 \chi}{E_{0}} \frac{ b\mathcal{E} \mathcal{L}_{m_{0}}\mathcal{L}_{m_{0}}^{'}}{E_{0}^{2}}-\frac{e\mathcal{E}E_{0}^{'}}{E_{0}^{2}}+\frac{2\zeta e\mathcal{E}R_{0}E_{0}^{'}}{E_{0}^{2}}\\\nonumber&-\frac{2 \chi \mathcal{L}_{m_{0}}\bar{\mathcal{L}_{m}}E_{0}^{'}}{E_{0}^{2}}-\frac{b^{'}\mathcal{E}R_{0}}{E_{0}^{2}}-\frac{b^{'}\mathcal{E}}{E_{0}} \frac{\zeta R_{0}^{2}}{E_{0}}-\frac{b^{'}\mathcal{E}\chi \mathcal{L}_{m_{0}^{2}}}{E_{0}^{2}}-\frac{3b\mathcal{E}R_{0}E_{0}^{'}}{E_{0}^{3}}-\frac{3 \zeta b\mathcal{E}R_{0}^{2}E_{0}^{'}}{E_{0}^{3}}+\frac{3 \chi b\mathcal{E}E_{0}^{'}\mathcal{L}_{m_{0}}^{2}}{E_{0}^{3}}-\frac{2 \chi b\mathcal{E}\mathcal{L}_{m_{0}}^{'}}{E_{0}^{3}}\\\nonumber&
-\frac{8 \zeta a^{'}\mathcal{E}R_{0}D_{0}^{'}F_{0}^{'}}{D_{0}^{2}E_{0}^{4}F_{0}}+\frac{4 \zeta c^{'}\mathcal{E}R_{0}D_{0}^{'}}{D_{0}^{2}E_{0}^{4}F_{0}}-\frac{4 \zeta e\mathcal{E}F_{0}^{'}D_{0}^{'}}{D_{0}^{2}E_{0}^{4}F_{0}}+\frac{16 \zeta b\mathcal{E}R_{0}}{D_{0}^{2}E_{0}^{5}} \frac{D_{0}^{'}F_{0}^{'}}{F_{0}}+\frac{8 \zeta R_{0} a\mathcal{E}D_{0}^{'}F_{0}^{'}}{D_{0}^{3}E_{0}^{3}F_{0}}+\frac{4 \zeta c\mathcal{E}R_{0}D_{0}^{'}F_{0}^{'}}{D_{0}^{2}E_{0}^{4}F_{0}^{2}}+\frac{a^{'}}{2}\\\nonumber& \times\frac{\mathcal{E} R_{0}}{E_{0}^{2}D_{0}}-\frac{e\mathcal{E} D_{0}^{'}}{2E_{0}^{2}D_{0}}-\frac{\zeta a^{'}\mathcal{E} R_{0}^{2}}{2E_{0}^{2}R_{0}}-\frac{\chi a^{'}\mathcal{E} \mathcal{L}_{m_{0}}^{2}}{2E_{0}^{2}D_{0}}
-\frac{\chi D_{0}^{'} \mathcal{L}_{m_{0}}\bar{\mathcal{L}_{m_{0}}}}{E_{0}^{2}D_{0}}+\frac{b\mathcal{E}R_{0}D_{0}^{'}}{E_{0}^{3}D_{0}}
+\frac{a\mathcal{E}R_{0}D_{0}^{'}}{2E_{0}^{2}D_{0}^{2}}
+\frac{\zeta b\mathcal{E}R_{0}^{2}D_{0}^{'}}{E_{0}^{3}D_{0}}+\frac{\zeta a\mathcal{E}}{D_{0}^{2}} \\\nonumber& \times \frac{R_{0}^{2}D_{0}^{'}}{2E_{0}^{2}}+\frac{\chi b\mathcal{E}D_{0}^{'}\mathcal{L}_{m_{0}}^{2}}{E_{0}^{3}D_{0}}+\frac{\chi a\mathcal{E}D_{0}^{'}\mathcal{L}_{m_{0}}^{2}}{2E_{0}^{2}D_{0}^{2}}-\frac{\chi a^{'}\mathcal{E}\mathcal{L}_{m_{0}}}{E_{0}^{2}D_{0}}-\frac{\chi D_{0}^{'}\bar{\mathcal{L}_{m}}}{E_{0}^{2}D_{0}}+\frac{\chi a\mathcal{E}\mathcal{L}_{m_{0}}D_{0}^{'}}{E_{0}^{2}D_{0}^{2}}+\frac{2\chi b\mathcal{E}\mathcal{L}_{m_{0}}D_{0}^{'}}{E_{0}^{3}}+\frac{a^{'}}{E_{0}^{2}} \frac{\mathcal{E}}{D_{0}^{3}}\\\nonumber&-\frac{3a\mathcal{E}D_{0}^{'}}{E_{0}^{2}D_{0}^{4}}
-\frac{2b\mathcal{E}D_{0}^{'}}{E_{0}^{3}D_{0}^{3}}+\frac{6\zeta c^{'}\mathcal{E}F_{0}^{'2}R_{0}}{E_{0}^{4}F_{0}^{3}}+\frac{2\zeta e\mathcal{E}F_{0}^{'3}}{E_{0}^{4}F_{0}^{3}}+\frac{6\zeta c\mathcal{E}F_{0}^{'3}R_{0}}{E_{0}^{4}F_{0}^{4}}-\frac{8\zeta b\mathcal{E}F_{0}^{'}R_{0}}{E_{0}^{5}F_{0}^{3}}-\frac{2\zeta c^{'}\mathcal{E}R_{0}}{E_{0}^{4}F_{0}^{3}}-\frac{2\zeta e\mathcal{E}F_{0}^{'}}{E_{0}^{4}F_{0}^{3}}+\frac{8}{E_{0}^{5}}\\\nonumber& \times \frac{\zeta b\mathcal{E}F_{0}^{'}R_{0}}{F_{0}^{3}}+\frac{6\zeta c\mathcal{E}F_{0}^{'}R_{0}}{E_{0}^{4}F_{0}^{4}}-\frac{2\zeta a^{'}\mathcal{E}F_{0}^{'2}R_{0}}{D_{0}E_{0}^{2}F_{0}^{2}}-\frac{4\zeta c^{'}\mathcal{E}F_{0}^{'}R_{0}D_{0}^{'}}{D_{0}E_{0}^{2}F_{0}^{2}}
-\frac{2\zeta e\mathcal{E}F_{0}^{'2}D_{0}^{'}}{D_{0}E_{0}^{2}F_{0}^{2}}+\frac{a\mathcal{E}F_{0}^{'2}}{D_{0}^{2}E_{0}^{2}}
 \frac{D_{0}^{'}}{F_{0}^{2}}+\frac{2 b\mathcal{E}F_{0}^{'2}D_{0}^{'}}{D_{0}E_{0}^{3}F_{0}^{2}}+2\\\nonumber& \times \frac{c\mathcal{E}D_{0}^{'}F_{0}^{'}}{D_{0}E_{0}^{2}F_{0}^{3}}+\frac{2\zeta c^{'}\mathcal{E}R_{0}}{E_{0}^{2}F_{0}^{3}}-\frac{2\zeta e\mathcal{E}F_{0}^{'}}{E_{0}^{2}F_{0}^{3}}-\frac{4\zeta b\mathcal{E}R_{0}F_{0}^{'}}{F_{0}^{3}E_{0}^{3}}-\frac{6\zeta c\mathcal{E}R_{0}}{F_{0}^{4}} \frac{F_{0}^{'}}{E_{0}^{2}}+\frac{e\mathcal{E}F_{0}^{'}}{2E_{0}^{2}F_{0}}+\frac{2\zeta e\mathcal{E}R_{0}F_{0}^{'}}{2F_{0}E_{0}^{2}}+\frac{\chi \mathcal{L}_{m_{0}}\bar{\mathcal{L}_{m}}F_{0}^{'}}{E_{0}^{2}F_{0}}\\\nonumber&+\frac{c^{'}\mathcal{E}R_{0}}{E_{0}^{2}F_{0}}
+\frac{\zeta c^{'}\mathcal{E}R_{0}^{2}}{2E_{0}^{2}F_{0}}+\frac{\chi c^{'}\mathcal{E}\mathcal{L}_{m_{0}}^{2}}{2E_{0}^{2}F_{0}}-\frac{b\mathcal{E}R_{0}F_{0}^{'}}{E_{0}^{3}F_{0}}
-\frac{c\mathcal{E}R_{0}F_{0}^{'}}{2E_{0}^{2}F_{0}^{2}}+\frac{c^{'}\mathcal{E}\chi \mathcal{L}_{m_{0}}}{E_{0}^{2}F_{0}}+\frac{\chi \bar{\mathcal{L}_{m}}F_{0}^{'}}{E_{0}^{2}F_{0}}-\frac{2 \chi b\mathcal{E}\chi \mathcal{L}_{m_{0}}F_{0}^{'}}{E_{0}^{3}F_{0}}-\frac{c\mathcal{E}\chi}{E_{0}^{2}}\\\nonumber \times &\frac{ \mathcal{L}_{m_{0}}F_{0}^{'}}{F_{0}^{2}}-\frac{c^{'}\mathcal{E}}{E_{0}^{2}F_{0}^{3}}\varphi_{22}+\frac{2b\mathcal{E}F_{0}^{'}}{E_{0}^{3}F_{0}^{3}}\varphi_{22}
+\frac{3c\mathcal{E}F_{0}^{'}}{E_{0}^{2}F_{0}^{4}}\varphi_{22}-\frac{2b\dot{\mathcal{E}}}{D_{0}^{2}E_{0}^{3}}\varphi_{01}
-\frac{2}{D_{0}^{2}}
  \frac{c\dot{\mathcal{E}}}{E_{0}^{2}F_{0}}\varphi_{01}+\frac{4 \zeta c^{''}\mathcal{E}R_{0}D_{0}^{'}}{D_{0}E_{0}^{4}F_{0}}+\frac{4 \zeta a^{'}\mathcal{E}R_{0}}{D_{0}E_{0}^{4}}\\\nonumber& \times \frac{F_{0}^{''}}{F_{0}}+\frac{4 \zeta e\mathcal{E}F_{0}^{''}D_{0}^{'}}{D_{0}E_{0}^{4}F_{0}}-\frac{4 \zeta F_{0}^{''}a\mathcal{E}R_{0}D_{0}^{'}}{D_{0}^{2}E_{0}^{4}F_{0}}-\frac{16 \zeta b\mathcal{E}R_{0}D_{0}^{'}F_{0}^{''}}{D_{0}E_{0}^{5}F_{0}}+\frac{4 \zeta }{D_{0}} \frac{b^{'}\mathcal{E}R_{0}D_{0}^{'}F_{0}^{'}}{E_{0}^{5}F_{0}}+\frac{e\mathcal{E}D_{0}^{'}}{2D_{0}E_{0}^{2}}
+\frac{\chi \mathcal{L}_{m_{0}\bar{\mathcal{L}_{m}}}D_{0}^{'}}{D_{0}E_{0}^{2}}\\\nonumber&+\frac{\zeta a^{'}\mathcal{E}R_{0}^{2}}{2D_{0}E_{0}^{2}}-\frac{ a\mathcal{E}R_{0}D_{0}^{'}}{2D_{0}^{2}E_{0}^{2}}-\frac{\chi \mathcal{L}_{m_{0}}^{2}a\mathcal{E}D_{0}^{'}}{2D_{0}^{2}E_{0}^{2}}-\frac{ b\mathcal{E}R_{0}D_{0}^{'}}{D_{0}E_{0}^{3}}-\frac{\chi \mathcal{L}_{m_{0}}^{2}b\mathcal{E}D_{0}^{'}}{D_{0}E_{0}^{3}} -\frac{\chi \mathcal{L}_{m_{0}}b\mathcal{E}D_{0}^{'}}{D_{0}E_{0}^{3}}+\frac{\chi D_{0}^{'}\bar{\mathcal{L}_{m}}}{D_{0}E_{0}^{2}}+\frac{e\mathcal{E}E_{0}^{'}}{E_{0}^{3}}+\frac{2}{E_{0}}\\\nonumber& \times \frac{\chi E_{0}^{'}\mathcal{L}_{m_{0}}\bar{\mathcal{L}_{m}}}{E_{0}^{2}}+\frac{b^{'}\mathcal{E}R_{0}}{E_{0}^{3}}+\frac{\chi b^{'}\mathcal{E}\mathcal{L}_{m_{0}}^{2}}{E_{0}^{3}}-\frac{3\chi b\mathcal{E}\mathcal{L}_{m_{0}}^{2}E_{0}^{'}}{E_{0}^{4}}+\frac{2 \chi b^{'}\mathcal{E}\mathcal{L}_{m_{0}}}{E_{0}^{3}} +\frac{2b^{'}\mathcal{E}}{E_{0}^{5}}\varphi_{11}
+\frac{4 \zeta a^{''}\mathcal{E} R_{0}F_{0}^{'}}{D_{0}E_{0}^{4}F_{0}}-\frac{16 \zeta b\mathcal{E} R_{0}}{D_{0}}\\\nonumber& \times \frac{F_{0}^{'}D_{0}^{''}}{E_{0}^{5}F_{0}}
+\frac{8 \zeta b^{'}\mathcal{E} R_{0}F_{0}^{'2}}{E_{0}^{5}F_{0}}+\frac{8 \zeta e\mathcal{E} E_{0}^{'}F_{0}^{'2}}{E_{0}^{5}F_{0}}-\frac{16 \zeta c\mathcal{E} R_{0}F_{0}^{'2}E_{0}^{'}}{E_{0}^{5}F_{0}^{3}}-\frac{4 \zeta a^{'}}{D_{0}}
 \frac{\mathcal{E} R_{0}F_{0}^{'}E_{0}^{'}}{E_{0}^{5}F_{0}}-\frac{4 \zeta e\mathcal{E} D_{0}^{'}F_{0}^{'}E_{0}^{'}}{D_{0}E_{0}^{5}F_{0}}
+\frac{4 \zeta a\mathcal{E} R_{0} }{D_{0}^{2}F_{0}}\\\nonumber&\times \frac{D_{0}^{'}F_{0}^{'}E_{0}^{'}}{E_{0}^{5}}-\frac{a^{'}\mathcal{E} \chi \mathcal{L}_{m_{0}}}{D_{0}E_{0}^{2}}-\eta \chi \bar{\mathcal{L}_{m}} -\frac{a\mathcal{E} \eta \chi \mathcal{L}_{m_{0}}D_{0}^{'}}{D_{0}^{2}E_{0}^{2}}-\frac{ \eta \chi \bar{\mathcal{L}_{m}}D_{0}^{'}}{D_{0}E_{0}^{2}}+\frac{2 \chi  \eta b\mathcal{E} \mathcal{L}_{m_{0}}D_{0}^{'}}{D_{0}E_{0}^{3}}+\frac{2 \chi  \eta a\mathcal{E} \mathcal{L}_{m_{0}}D_{0}^{'}}{D_{0}^{2}E_{0}^{3}}-\frac{\chi \eta a^{'}}{D_{0}}\\\nonumber& \times \frac{\mathcal{L}_{m_{0}}\mathcal{E}}{E_{0}^{2}}-\frac{\chi \eta \bar{\mathcal{L}_{m}}D_{0}^{'}}{D_{0}E_{0}^{2}}+\frac{2\chi \eta b}{D_{0}} \frac{\mathcal{L}_{m_{0}}\eta D_{0}^{'}\mathcal{E}}{E_{0}^{3}}
+\frac{\chi \eta a\mathcal{E}\mathcal{L}_{m_{0}} D_{0}^{'}}{D_{0}^{2}E_{0}^{2}}-\frac{2\chi c^{'}\mathcal{E}\eta \mathcal{L}_{m_{0}}}{F_{0}^{3}}-\frac{2\chi F_{0}^{'}\eta \bar{\mathcal{L}_{m}}}{F_{0}^{3}}+\frac{6\chi c\mathcal{E}\eta \mathcal{L}_{m_{0}}F_{0}^{'}}{F_{0}^{4}}\\\nonumber& +\frac{6\chi E_{0}^{'}b\mathcal{E}\eta \mathcal{L}_{m_{0}}}{E_{0}^{4}}.
\end{align}

The coefficient values for Eq. \eqref{M10} have following representations
\begin{align}\nonumber
u&=\frac{2 \zeta b R_{0}}{D_{0}^{2}E_{0}}+\frac{2 \zeta b R_{0}}{\chi \mathcal{L}_{m_{0}}D_{0}^{2}E_{0}}-\frac{2c}{\chi D_{0}^{2}F_{0}},
\\\nonumber
v&=-\frac{4 \zeta b F_{0}^{'}R_{0}}{D_{0}E_{0}^{2}F_{0}}-\frac{4 \zeta c D_{0}^{'}R_{0}}{D_{0}F_{0}^{2}E_{0}}-\frac{4 \zeta c^{'} R_{0}}{D_{0}E_{0}F_{0}},
\\\nonumber
w&=\frac{a \varphi_{01}}{D_{0}^{2}E_{0}}+\frac{b \varphi_{01}}{E_{0}^{2}D_{0}}-\frac{4 \zeta c F_{0}^{''}}{E_{0}^{2}F_{0}}
-\frac{4 \zeta c^{''} R_{0} }{E_{0}^{2}F_{0}}-\frac{2 \zeta a D_{0}^{''} R_{0} }{D_{0}^{2}E_{0}^{2}}-\frac{4 \zeta b D_{0}^{''} R_{0} }{D_{0}E_{0}^{3}}-\frac{2 \zeta e D_{0}^{''}  }{D_{0}E_{0}^{2}}-\frac{2 \zeta a^{''}  R_{0} }{D_{0}E_{0}^{2}}
+\frac{4 \zeta c^{'} E_{0}^{'} R_{0} }{F_{0}E_{0}^{3}}+\frac{4  }{F_{0}}\\\nonumber& \times \frac{\zeta b^{'} F_{0}^{'} R_{0}}{E_{0}^{3}}
+\frac{4 \zeta c^{'}R_{0}E_{0}^{'}}{E_{0}^{3}F_{0}}-\frac{1}{\chi E_{0}^{2}\mathcal{L}_{m_{0}}}\left\{4 \zeta e F_{0}^{''}
+4 \zeta c^{''} R_{0}-\frac{4 \zeta c R_{0}F_{0}^{''}}{F_{0}^{2}}+\frac{2 \zeta e R_{0}A_{0}^{''}}{D_{0}}
+\frac{2 \zeta  R_{0}a^{''}}{D_{0}}-\frac{2 \zeta a R_{0}D_{0}^{''}}{D_{0}^{2}}\right.\\\nonumber&\left.
+\frac{2 \zeta b R_{0}E_{0}^{'}F_{0}^{'}}{E_{0}^{2}F_{0}}+\frac{2 \zeta c R_{0}E_{0}^{'}F_{0}^{'}}{F_{0}^{2}E_{0}}
+\frac{2 \zeta c^{'} R_{0}E_{0}^{'}}{E_{0}F_{0}}+\frac{2 \zeta b^{'} R_{0}F_{0}^{'}}{E_{0}F_{0}}
+\frac{2 \zeta e E_{0}^{'}F_{0}^{'}}{E_{0}F_{0}}+\frac{2 \zeta b R_{0}E_{0}^{'}D_{0}^{'}}{E_{0}^{2}D_{0}}
+\frac{2 \zeta a R_{0}E_{0}^{'}D_{0}^{'}}{D_{0}^{2}E_{0}}+\frac{2 \zeta}{E_{0}}\right.\\\nonumber& \times \left.\frac{  R_{0}a^{'}E_{0}^{'}}{D_{0}}+\frac{2 \zeta b^{'} R_{0}D_{0}^{'}}{D_{0}E_{0}}+\frac{eE_{0}^{2}}{2}+\zeta e R_{0}E_{0}^{2}
+bE_{0}R_{0}+4 \pi b E_{0}\mathbb{E}_{0}^{2}+\zeta b E_{0}R_{0}^{2}+b \chi E_{0}\mathcal{L}_{m_{0}}^{2}+2b \chi E_{0}\mathcal{L}_{m_{0}}+\varphi_{11}\right.\\\nonumber&\left.-b \eta \chi E_{0}^{2}\mathcal{L}_{m_{0}}+\frac{2 c D_{0}^{'}F_{0}^{'}}{D_{0}F_{0}^{2}}+\frac{2 a D_{0}^{'}F_{0}^{'}}{F_{0}D_{0}^{2}}-\frac{2 c^{'} F_{0}^{'}}{F_{0}^{2}}+\frac{2 c F_{0}^{'}}{F_{0}^{3}}
+\frac{2 b E_{0}}{F_{0}^{2}}-\frac{2 c E_{0}^{2}}{F_{0}^{3}}\right\}.
\end{align}
The non-static components of non-conservation equations incorporate several lengthy terms $\mathcal{Y}_{i}$. The alternative gravity within the context of N domain result extra curvature term $\mathcal{Y}_{2N}$ having the subsequent form:
The value of $\mathcal{Y}_{2N}$ is as follows
\begin{align}\nonumber
\mathcal{Y}_{2N}&=\frac{-4\zeta c^{'}R_{0}\ddot{\mathcal{E}}}{r}+2b\mathcal{E} \dot{\varphi_{01}}+2a\mathcal{E} \dot{\varphi_{01}}-2a\dot{\mathcal{E}} \varphi_{01}-2b\dot{\mathcal{E}} \varphi_{01}+\frac{4\zeta c^{''}\mathcal{E}R_{0}^{'}}{r}+\frac{4\zeta c^{'''}\mathcal{E}R_{0}}{r}-2\zeta b \ddot{\mathcal{E}}R_{0}^{'}-2\zeta b^{'} \ddot{\mathcal{E}}R_{0} \\\nonumber& +2\zeta a^{''}\mathcal{E}R_{0}^{'}+2\zeta a^{'''} \mathcal{E}R_{0}-\frac{4 \zeta b^{'}R_{0}^{'}\mathcal{E}}{r}-\frac{4 \zeta b^{''}R_{0}\mathcal{E}}{r}+\frac{6 \zeta b^{'}R_{0}\mathcal{E}}{r^{2}}+\frac{e^{'}\mathcal{E}}{2}+\zeta e\mathcal{E} R_{0}^{'}+\zeta e^{'}\mathcal{E} R_{0}+\chi \bar{\mathcal{L}_{m}}\mathcal{L}_{m_{0}}^{'}\\\nonumber&+b \mathcal{E}R_{0}^{'}+2 \zeta b \mathcal{E}R_{0}R_{0}^{'}-2\chi \mathcal{L}_{m_{0}}\mathcal{L}_{m_{0}^{'}}b\mathcal{E}-\zeta b^{'}\mathcal{E}R_{0}^{2}-b^{'}\mathcal{E}R_{0}-\chi b^{'}\mathcal{E}\mathcal{L}_{m_{0}}^{2}-2\chi b\mathcal{E}\mathcal{L}_{m_{0}}^{'}+\frac{a^{'}\mathcal{E}R_{0}}{2}-\frac{\zeta a^{'}\mathcal{E}R_{0}}{2}
\\\nonumber&-\frac{\chi a^{'}\mathcal{E}\mathcal{L}_{m_{0}}^{2}}{2}-\chi a^{'}e \mathcal{L}_{m_{0}}+a^{'}\mathcal{E}+\frac{6\zeta c^{'}\mathcal{E}R_{0}}{r^{3}}+\frac{2\zeta e\mathcal{E}}{r^{3}}+\frac{6\zeta c\mathcal{E}R_{0}}{r^{4}}-\frac{8\zeta b\mathcal{E}R_{0}}{r^{3}}-\frac{2\zeta c^{'}\mathcal{E}R_{0}}{r^{3}}-\frac{2\zeta e\mathcal{E}}{r^{3}}
+\frac{8\zeta b}{r^{2}}\\\nonumber& \times \frac{\mathcal{E}R_{0}}{r}-\frac{2\zeta a^{'}\mathcal{E}R_{0}}{r^{2}}+\frac{2 \zeta R_{0}c^{'}\mathcal{E}}{r^{3}}
+\frac{2 \zeta e\mathcal{E}}{r^{3}}-\frac{4 \zeta R_{0}b\mathcal{E}}{r^{3}}-\frac{6 \zeta R_{0}c\mathcal{E}}{r^{4}}+\frac{e\mathcal{E}}{2r}+\frac{\zeta e\mathcal{E}R_{0}}{r}
+\frac{\chi \mathcal{L}_{m_{0}}\bar{\mathcal{L}_{m}}}{r}+\frac{c^{'}\mathcal{E}R_{0}}{r}\\\nonumber&+\frac{\zeta c^{'}\mathcal{E}R_{0}}{2r}+\frac{c^{'}\chi \mathcal{E}\mathcal{L}_{m_{0}}^{2}}{2r}-\frac{b\mathcal{E}R_{0}}{r}-\frac{c\mathcal{E}R_{0}}{2r^{2}}+\frac{\chi c^{'}\mathcal{E}\mathcal{L}_{m_{0}}}{r}+\frac{\chi \bar{\mathcal{L}_{m}}}{r}
-\frac{2\chi b\mathcal{E}\mathcal{L}_{m_{0}}}{r}-\frac{\chi c\mathcal{E}\mathcal{L}_{m_{0}}}{r^{2}}-\frac{c^{'}\mathcal{E}}{r^{3}}\varphi_{22}+\frac{2b}{r^{3}}\\\nonumber& \times \mathcal{E}\varphi_{22}+\frac{3c\mathcal{E}}{r^{4}}\varphi_{22}-2b\dot{\mathcal{E}}\varphi_{01}
-\frac{2c\dot{\mathcal{E}}}{r}\varphi_{01}+\frac{\zeta a^{'}\mathcal{E}R_{0}^{2}}{2}+b^{'}\mathcal{E}R_{0}
+\chi b^{'}\mathcal{E}\mathcal{L}_{m_{0}}^{2}+2b^{'}\mathcal{E}\varphi_{11}+\frac{4 \zeta a^{''}\mathcal{E}R_{0}}{r}+\frac{8 \zeta b^{'}}{r}\\\nonumber& \times \mathcal{E} R_{0}
-\eta \chi \bar{\mathcal{L}_{m}}-a^{'}\chi \mathcal{L}_{m_{0}}\mathcal{E}-a^{'}\eta \chi \mathcal{L}_{m_{0}}\mathcal{E}-\frac{2\chi c^{'}\mathcal{E}\mathcal{L}_{m_{0}}}{r^{3}}-\frac{2\chi \eta \bar{\mathcal{L}_{m}}}{r^{3}}+\frac{6\chi \eta c\mathcal{E}\mathcal{L}_{m_{0}}}{r^{4}}
\end{align}
The additional curvature terms relating to $f(R, \mathcal{L}_{m})$ gravity in non-static component of the second non-conservation equation in $pN$ regime are specified as $Y_{2pN}$ having the value:
\begin{align}\nonumber
\mathcal{Y}_{2pN}&=-\frac{4 \zeta c^{'}R_{0}\ddot{\mathcal{E}}}{r}+\frac{8 \zeta c\mathcal{N}_{1}R_{0}\ddot{\mathcal{E}}}{r}-\frac{12 \zeta c\dot{\mathcal{E}}\mathcal{N}_{3}\mathcal{N}_{2}^{'}R_{0}}{r}+4b\mathcal{E}\mathcal{N}_{2}\dot{\varphi_{01}}+4a\mathcal{E}\mathcal{N}_{1}\dot{\varphi_{01}}
-4a\dot{\mathcal{E}}\mathcal{N}_{1}\varphi_{01}-4b\dot{\mathcal{E}}\mathcal{N}_{2} \\\nonumber& \times \varphi_{01}+\frac{4\zeta c^{''}\mathcal{E}R_{0}^{'}\mathcal{N}_{3}}{r}+\frac{4\zeta c^{'''}\mathcal{E}R_{0}\mathcal{N}_{8}}{r}-2 \zeta b\mathcal{N}_{2}R_{0}^{'}\ddot{\mathcal{E}}
-2 \zeta b^{'}\mathcal{N}_{2}R_{0}\ddot{\mathcal{E}}+4\zeta b\mathcal{N}_{2}^{'}R_{0}\ddot{\mathcal{E}}+2\zeta b\mathcal{N}_{4}\mathcal{N}_{1}^{'}R_{0} \ddot{\mathcal{E}}\\\nonumber& +2\zeta a^{''} \mathcal{N}_{6}R_{0}^{'}\mathcal{E}+2\zeta D_{0}^{''}e^{'}\mathcal{N}_{6}\mathcal{E}-2\zeta R_{0}^{'}a\mathcal{E}\mathcal{N}_{4}\mathcal{N}_{2}^{''}-8\zeta R_{0}^{'}b\mathcal{E}\mathcal{N}_{8}\mathcal{N}_{2}^{''}+2\zeta R_{0}a^{'''}\mathcal{E}\mathcal{N}_{6}+2\zeta R_{0}a^{'}\mathcal{E} \mathcal{N}_{4}\\\nonumber& \times \mathcal{N}_{2}^{''}-2  \zeta  a^{''}\mathcal{E}R_{0}\mathcal{N}_{4}\mathcal{N}_{2}^{'}-2\zeta e\mathcal{E}\mathcal{N}_{4}\mathcal{N}_{2}^{'}\mathcal{N}_{2}^{''}+4\zeta b \mathcal{E}R_{0}\mathcal{N}_{6}\mathcal{N}_{2}^{'}\mathcal{N}_{2}^{''}
4\zeta a \mathcal{E}R_{0}\mathcal{N}_{4}\mathcal{N}_{2}^{'}\mathcal{N}_{2}^{''}-8\zeta b a^{'}R_{0}\mathcal{N}_{6}\\\nonumber&-8\zeta a^{'''} \mathcal{E}R_{0}\mathcal{N}_{6}\mathcal{N}_{1}^{'}-8\zeta a^{''} \mathcal{E} R_{0}\mathcal{N}_{6}-8\zeta e \mathcal{E}\mathcal{N}_{6}\mathcal{N}_{1}^{'}\mathcal{N}_{2}^{''}-16\zeta b \mathcal{E}R_{0}\mathcal{N}_{8}\mathcal{N}_{2}^{'}\mathcal{N}_{2}^{''}\mathcal{N}_{1}^{'}+16\zeta a \mathcal{E}R_{0} \mathcal{N}_{4}\mathcal{N}_{1}^{'}\\\nonumber& \times \mathcal{N}_{2}^{''}\mathcal{N}_{2}^{'}+24\zeta b \mathcal{E}R_{0}\mathcal{N}_{8}\mathcal{N}_{1}^{'}\mathcal{N}_{2}^{''}\mathcal{N}_{2}^{'}-\frac{4\zeta b^{'}\mathcal{E}R_{0}^{'}\mathcal{N}_{10}}{r}
+\frac{4\zeta c^{'}\mathcal{E}R_{0}^{'}\mathcal{N}_{1}^{'}\mathcal{N}_{9}}{r}-\frac{4\zeta e^{'}\mathcal{E}\mathcal{N}_{1}^{'}\mathcal{N}_{9}}{r}+ \frac{4\zeta b\mathcal{E}R_{0}^{'}}{r}\\\nonumber& \times \mathcal{N}_{1}^{'}\mathcal{N}_{12} +\frac{4}{r} \frac{\zeta c\mathcal{E}R_{0}^{'}\mathcal{N}_{1}^{'}\mathcal{N}_{10}}{r}-\frac{72\zeta b\mathcal{E}R_{0}\mathcal{N}_{1}^{'2}\mathcal{N}_{14}}{r}
+\frac{4\zeta b^{'}\mathcal{E}R_{0}\mathcal{N}_{10}}{r^{2}}+\frac{12\zeta c^{'}\mathcal{E}R_{0}\mathcal{N}_{1}^{'}\mathcal{N}_{10}}{r^{2}}
+\frac{6 }{5}\frac{\mathcal{N}_{1}^{'}\zeta e\mathcal{E}}{r^{2}}\\\nonumber&-\frac{30\zeta b\mathcal{E}R_{0}}{r} \frac{\mathcal{N}_{1}^{'}\mathcal{N}_{12}}{r}-\frac{12\zeta c\mathcal{E}R_{0}\mathcal{N}_{1}^{'}\mathcal{N}_{10}}{r^{3}}-2 \zeta b^{'}\mathcal{E} R_{0}^{'}\mathcal{N}_{1}^{'}\mathcal{N}_{8}-3 \zeta b^{'}\mathcal{E} R_{0}^{'}\mathcal{N}_{2}^{'}\mathcal{N}_{8}+\frac{6 e\mathcal{E}\zeta \mathcal{N}_{1}^{'}}{5r^{2}} + \frac{30 b\mathcal{E}}{r}\\\nonumber& \times \frac{\zeta R_{0} \mathcal{N}_{1}^{'}\mathcal{N}_{12}}{r}-\frac{12 c\mathcal{E}\zeta R_{0} \mathcal{N}_{1}^{'}\mathcal{N}_{10}}{r^{3}}+10 \zeta b\mathcal{E} R_{0}^{'}\mathcal{N}_{1}^{'}\mathcal{N}_{2}^{'}\mathcal{N}_{10}
+2 \zeta a^{''}\mathcal{E} R_{0}\mathcal{N}_{1}^{'}\mathcal{N}_{8}+2 \zeta b^{'}\mathcal{E} R_{0}\mathcal{N}_{2}^{''}\mathcal{N}_{8}+\\\nonumber&2 \zeta e\mathcal{E} \mathcal{N}_{2}^{''}\mathcal{N}_{1}^{'}\mathcal{N}_{8}-2 \zeta a\mathcal{E} R_{0} \mathcal{N}_{2}^{''}\mathcal{N}_{1}^{'}\mathcal{N}_{8}-10 \zeta b\mathcal{E} R_{0} \mathcal{N}_{2}^{''}\mathcal{N}_{1}^{'}\mathcal{N}_{10}
-2 \zeta a^{'}\mathcal{E} R_{0} \mathcal{N}_{1}^{''}\mathcal{N}_{8}-2 \zeta b^{''}\mathcal{E} R_{0} \mathcal{N}_{2}^{'}\mathcal{N}_{8}
\\\nonumber&-2 \zeta e\mathcal{E} R_{0} \mathcal{N}_{1}^{''}\mathcal{N}_{2}^{'} \mathcal{N}_{8}+2 \zeta a\mathcal{E} R_{0} \mathcal{N}_{1}^{''}\mathcal{N}_{2}^{'}\mathcal{N}_{6}+6 \zeta a\mathcal{E} R_{0} \mathcal{N}_{1}^{'2}\mathcal{N}_{10}+6 \zeta e\mathcal{E} \mathcal{N}_{1}^{'2}\mathcal{N}_{2}^{'2}\mathcal{N}_{10}+12 \zeta R_{0} b^{'}\mathcal{E} \mathcal{N}_{1}^{'}\\\nonumber& \times \mathcal{N}_{2}^{'}\mathcal{N}_{10}+18 \zeta R_{0}  b\mathcal{E} \mathcal{N}_{1}^{'2}\mathcal{N}_{2}^{'}\mathcal{N}_{12}
+6 \zeta R_{0} a\mathcal{E} \mathcal{N}_{1}^{'2}\mathcal{N}_{2}^{'}\mathcal{N}_{8}+\frac{e^{'}\mathcal{E}\mathcal{N}_{4}}{2}+\chi \mathcal{N}_{4}\bar{\mathcal{L}_{m}}\mathcal{L}_{m_{0}}^{'}+\zeta e\mathcal{E} R_{0}^{'}\mathcal{N}_{4}+\zeta \\\nonumber& \times e^{'}\mathcal{E} R_{0}\mathcal{N}_{4}+ b\mathcal{E} R_{0}^{'} \mathcal{N}_{6}+2 \zeta R_{0}R_{0}^{'}b\mathcal{E}\mathcal{N}_{6}-2 \chi b \mathcal{E} \mathcal{L}_{m_{0}}\mathcal{L}_{m_{0}}^{'}\mathcal{N}_{6}-e\mathcal{E}\mathcal{N}_{1}^{'}\mathcal{N}_{4}-2\zeta e\mathcal{E} R_{0}\mathcal{N}_{1}^{'}\mathcal{N}_{4}
-\chi b^{'}\mathcal{E}\\\nonumber& \times \mathcal{L}_{m_{0}}^{2}\mathcal{N}_{4}+4 \zeta c^{'}\mathcal{E} R_{0}  \frac{\mathcal{N}_{2}^{'}\mathcal{N}_{4}}{r}
-\frac{4\zeta R_{0}\mathcal{N}_{2}^{'}\mathcal{N}_{4}}{r}-\frac{4\zeta e\mathcal{E} \mathcal{N}_{2}^{'}\mathcal{N}_{4}}{r}+\frac{16\zeta b\mathcal{E} R_{0} \mathcal{N}_{2}^{'}\mathcal{N}_{6}}{r}+\frac{8\zeta a\mathcal{E} R_{0} \mathcal{N}_{2}^{'}\mathcal{N}_{2}}{r}\\\nonumber&+\frac{4\zeta c\mathcal{E} R_{0} \mathcal{N}_{2}^{'}\mathcal{N}_{4}}{r^{2}} +  \frac{ a^{'}\mathcal{E}R_{0} \mathcal{N}_{2}}{2}-\frac{e\mathcal{E} \mathcal{N}_{2}\mathcal{N}_{2}^{'}}{2}-\frac{\zeta e^{'}\mathcal{E} R_{0}^{2}\mathcal{N}_{4}}{2R_{0}}-\frac{\chi a^{'}\mathcal{E} \mathcal{L}_{m_{0}}^{2}\mathcal{N}_{2}}{2}-\chi \mathcal{L}_{m_{0}}\bar{\mathcal{L}_{m}}\mathcal{N}_{2}\mathcal{N}_{2}^{'}+b\mathcal{E} \\\nonumber& \times R_{0}\mathcal{N}_{4}\mathcal{N}_{2}^{'}+\frac{a\mathcal{E}R_{0}\mathcal{N}_{2}^{'}}{2}
+ \zeta  R_{0}^{2}b\mathcal{E}\mathcal{N}_{2}^{'}\mathcal{N}_{4}+\frac{\zeta R_{0}^{2}a\mathcal{E}\mathcal{N}_{2}^{'}}{2}+b\mathcal{E}\chi \mathcal{L}_{m_{0}}^{2}\mathcal{N}_{4}\mathcal{N}_{2}^{'}+\frac{a\mathcal{E}\chi \mathcal{L}_{m_{0}}^{2}\mathcal{N}_{2}^{'}}{2}-\chi a^{'}\mathcal{E}\mathcal{L}_{m_{0}}\\\nonumber& \times \mathcal{N}_{2}-\chi \bar{\mathcal{L}_{m}}\mathcal{N}_{2}\mathcal{N}_{2}^{'}+a\mathcal{E} \chi  \mathcal{L}_{m_{0}}\mathcal{N}_{2}^{'}+2 \chi b \mathcal{E} \mathcal{L}_{m_{0}}\mathcal{N}_{2}^{'}\mathcal{N}_{6}+a^{'}\mathcal{E}\mathcal{N}_{1}-3a\mathcal{E} \mathcal{N}_{3}\mathcal{N}_{2}^{'}-2b\mathcal{E}\mathcal{N}_{2}^{'}+\frac{6 \zeta c^{'}\mathcal{E}R_{0}}{r}\\\nonumber& \times \frac{\mathcal{N}_{8}}{r^{2}}+\frac{2 \zeta e\mathcal{E}\mathcal{N}_{8}}{r^{3}}+\frac{6 \zeta R_{0} c}{r} \frac{\mathcal{E}\mathcal{N}_{8}}{r^{3}} -\frac{8 \zeta R_{0} b\mathcal{E}\mathcal{N}_{10}}{r^{3}}-\frac{2 \zeta R_{0} c^{'}\mathcal{E}\mathcal{N}_{8}}{r^{3}}-\frac{8 \zeta e\mathcal{E}\mathcal{N}_{8}}{r^{3}}+\frac{8 \zeta R_{0} b\mathcal{E}\mathcal{N}_{10}}{r^{3}}-\frac{2 \zeta R_{0} }{r}\\\nonumber& \times \frac{a^{'}\mathcal{E}\mathcal{N}_{2}}{r}-\frac{4 \zeta R_{0} c^{'}\mathcal{E}\mathcal{N}_{2}\mathcal{N}_{2}^{'}}{r^{2}}+\frac{a\mathcal{E}\mathcal{N}_{2}^{'}}{r^{2}}
+\frac{2b\mathcal{E}\mathcal{N}_{2}^{'}\mathcal{N}_{4}}{r^{2}}
+\frac{2c\mathcal{E}\mathcal{N}_{2}^{'}\mathcal{N}_{2}}{r^{3}}+\frac{2\zeta c^{'}\mathcal{E}R_{0}\mathcal{N}_{4}}{r^{3}}-\frac{4\zeta b\mathcal{E}R_{0}\mathcal{N}_{6}}{r^{3}}+\\\nonumber&\frac{e\mathcal{E}\mathcal{N}_{4}}{2r}+\frac{\chi \mathcal{L}_{m_{0}}\bar{\mathcal{L}}_{m}\mathcal{N}_{4}}{r}+\frac{c^{'}\chi \mathcal{E}}{2}  \frac{\mathcal{L}_{m_{0}}^{2}\mathcal{N}_{4}}{r}+\frac{\chi c^{'}\mathcal{E}\mathcal{L}_{m_{0}}\mathcal{N}_{4}}{r}+\frac{a\mathcal{E}\mathcal{N}_{2}^{'}\mathcal{N}_{2}}{2}+\chi \mathcal{L}_{m_{0}}\bar{\mathcal{L}_{m}}\mathcal{N}_{2}\mathcal{N}_{2}^{'}-\frac{a\mathcal{E}R_{0}\mathcal{N}_{1}^{'}}{2}-\\\nonumber&\frac{\chi \mathcal{L}_{m_{0}}^{2}a\mathcal{E}\mathcal{N}_{2}^{'}}{2}-b\mathcal{E}R_{0}\mathcal{N}_{2}\mathcal{N}_{2}^{'}-\chi b\mathcal{E} \mathcal{L}_{m_{0}}^{2}\mathcal{N}_{2} \mathcal{N}_{2}^{'}+\chi \bar{\mathcal{L}}_{m}\mathcal{N}_{2}\mathcal{N}_{2}^{'}-\chi b \mathcal{E}\mathcal{L}_{m_{0}}\mathcal{N}_{2}^{'}\mathcal{N}_{4}
+e \mathcal{E}\mathcal{N}_{1}^{'}\mathcal{N}_{6}+2\chi \mathcal{L}_{m_{0}}\\\nonumber& \times \bar{\mathcal{L}}_{m}\mathcal{N}_{1}^{'}\mathcal{N}_{6}-3 \chi b\mathcal{E}\mathcal{L}_{m_{0}}^{2}\mathcal{N}_{1}^{'}\mathcal{N}_{8}+2b^{'}\chi \mathcal{E}\mathcal{L}_{m_{0}}\mathcal{N}_{6}+\frac{4 \zeta a^{''}\mathcal{E} R_{0}\mathcal{N}_{6}}{r}-\frac{16 \zeta b\mathcal{E} R_{0}\mathcal{N}_{2}^{''}\mathcal{N}_{8}}{r}+\frac{8 \zeta b^{'}\mathcal{E} R_{0}\mathcal{N}_{10}}{r}\\\nonumber&+\frac{8 \zeta e\mathcal{E} \mathcal{N}_{1}^{'}\mathcal{N}_{10}}{r} -\chi \eta \bar{\mathcal{L}}_{m}\mathcal{N}_{2} \mathcal{N}_{2}^{'}-\eta \chi \bar{\mathcal{L}}_{m}+4 \eta b\mathcal{E}  \chi \mathcal{L}_{m_{0}}\mathcal{N}_{4}\mathcal{N}_{2}^{'}+6 \eta b\mathcal{E}  \chi \mathcal{L}_{m_{0}}\mathcal{N}_{8}\mathcal{N}_{1}^{'}+2 \eta a\mathcal{E} \chi \mathcal{L}_{m_{0}}\mathcal{N}_{2}^{'}.
\end{align}


\vspace{0.5cm}


\begin{thebibliography}{10}

\bibitem{chandrasekhar1957introduction}
S.~Chandrasekhar, vol.~2.
\newblock Courier Corporation, 1957.

\bibitem{chandrasekhar1964dynamical}
S.~Chandrasekhar and R.~F. Tooper {\em Astrophy. J.}, vol.~139, p.~1396, 1964.

\bibitem{herrera2009expansion}
L.~Herrera, G.~Le~Denmat, and N.~O. Santos {\em Phys. Rev. D}, vol.~79,
  p.~087505, 2009.

\bibitem{herrera2010collapsing}
L.~Herrera and N.~O. Santos {\em Gen. Relativ. Gravit.}, vol.~42, p.~2383,
  2010.

\bibitem{herrera2018tilted}
L.~Herrera, A.~Di~Prisco, and J.~Carot {\em Phys. Rev. D}, vol.~97, p.~124003,
  2018.

\bibitem{herrera2022non}
L.~Herrera, A.~Di~Prisco, and J.~Ospino {\em Universe}, vol.~8, p.~296, 2022.

\bibitem{herrera2023expansion}
L.~Herrera, A.~Di~Prisco, and J.~Ospino {\em Symmetry}, vol.~15, p.~754, 2023.

\bibitem{sharif2014dynamical}
M.~Sharif and Z.~Yousaf {\em Astrophys. Space Sci.}, vol.~354, p.~471, 2014.

\bibitem{sharif2016instability}
M.~Sharif and R.~Manzoor {\em J. Exp. Theor. Phys.}, vol.~122, p.~849, 2016.

\bibitem{yousaf2018dynamical}
Z.~Yousaf, K.~Bamba, M.~Z. Bhatti, {\em et~al.} {\em Eur. Phys. J. A}, vol.~54,
  p.~122, 2018.

\bibitem{torres2005some}
R.~Torres {\em Class. Quantum Gravity}, vol.~22, p.~4335, 2005.

\bibitem{mitra2006gravitational}
A.~Mitra {\em Phys. Rev. D}, vol.~74, p.~024010, 2006.

\bibitem{ivanov2010importance}
B.~Ivanov {\em Int. J. Theor. Phys.}, vol.~49, p.~1236, 2010.

\bibitem{yousaf2016influence}
Z.~Yousaf, K.~Bamba, and M.~Z. Bhatti {\em Phys. Rev. D}, vol.~93, p.~064059,
  2016.

\bibitem{baffou2016exploring}
E.~Baffou, M.~Houndjo, and J.~Tosssa {\em Astrophys. Space Sci.}, vol.~361,
  p.~376, 2016.

\bibitem{wu2011stability}
P.~Wu and H.~Yu {\em Phys. Lett B}, vol.~703, p.~223, 2011.

\bibitem{izumi2014acausality}
K.~Izumi, J.~A. Gu, and Y.~C. Ong {\em Phys. Rev. D}, vol.~89, p.~084025, 2014.

\bibitem{herrera1989dynamical}
L.~Herrera, G.~Le~Denmat, and N.~Santos {\em Mon. Not. R. Astron. Soc.},
  vol.~237, p.~257, 1989.

\bibitem{chan1993dynamical}
R.~Chan, L.~Herrera, and N.~Santos {\em Mon. Not. R. Astron. Soc.}, vol.~265,
  p.~533, 1993.

\bibitem{herrera2012dynamical}
L.~Herrera, G.~L. Denmat, and N.~Santos {\em Gen. Relativ. Gravit.}, vol.~44,
  p.~1143, 2012.

\bibitem{bhatti2020stability}
M.~Z. Bhatti, Z.~Yousaf, and M.~Yousaf {\em Phys. Dark Universe}, vol.~28,
  p.~100501, 2020.

\bibitem{ur2024dynamically}
A.~Rehman, M.~Z. Bhatti, and Z.~Yousaf {\em Fortschr. der Phys.}, vol.~72,
  p.~2300247, 2024.

\bibitem{bekenstein1971hydrostatic}
J.~D. Bekenstein {\em Phys. Rev. D}, vol.~4, p.~2185, 1971.

\bibitem{esculpi2010conformal}
M.~Esculpi and E.~Aloma {\em Eur. Phys. J. C}, vol.~67, p.~521, 2010.

\bibitem{riess1998observational}
A.~G. Riess, A.~V. Filippenko, P.~Challis, {\em et~al.} {\em Astron. J.},
  vol.~116, p.~1009, 1998.

\bibitem{perlmutter1999astrophys}
S.~Perlmutter {\em et~al.} {\em Astron. J.}, vol.~116, p.~1009, 1999.

\bibitem{riess2007new}
A.~G. Riess, L.~G. Strolger, S.~Casertano, {\em et~al.} {\em Astrophys. J.},
  vol.~659, p.~98, 2007.

\bibitem{nojiri2007introduction}
S.~Nojiri and S.~D. Odintsov {\em Int. J. Geom. Methods Mod. Phys.}, vol.~4,
  p.~115, 2007.

\bibitem{nojiri2006new}
S.~Nojiri and S.~D. Odintsov {\em Phys. Lett. B}, vol.~639, p.~144, 2006.

\bibitem{bamba2012dark}
K.~Bamba, S.~Capozziello, S.~Nojiri, {\em et~al.} {\em Astrophys. Space Sci.},
  vol.~342, p.~155, 2012.

\bibitem{copeland2006dynamics}
E.~J. Copeland, M.~Sami, and S.~Tsujikawa {\em Int. J. Mod. Phys. D}, vol.~15,
  p.~1753, 2006.

\bibitem{nojiri2011unified}
S.~Nojiri and S.~D. Odintsov {\em Phys. Rep.}, vol.~505, p.~59, 2011.

\bibitem{bamba2015inflationary}
K.~Bamba and S.~D. Odintsov {\em Symmetry}, vol.~7, p.~220, 2015.

\bibitem{rehman2026complexity}
A.~Rehman {\em et~al.} {\em Nucl. Phys. B}, p.~117434, 2026.

\bibitem{ditta2026structure}
A.~Ditta, M.~Yousaf, {\em et~al.} {\em J. Cosmol. Astropart. Phys.}, vol.~2026,
  p.~053, 2026.

\bibitem{yousaf2025anisotropic}
M.~Yousaf {\em Int. J. Geom. Methods Mod. Phys.}, p.~2650099, 2025.

\bibitem{nojiri2017modified}
S.~Nojiri, S.~Odintsov, and V.~Oikonomou {\em arXiv preprint arXiv:1705.11098},
  2017.

\bibitem{faraoni2005stability}
V.~Faraoni and S.~Nadeau {\em Phys. Rev. D}, vol.~72, p.~124005, 2005.

\bibitem{feng2024brief}
Y.~Feng {\em et~al.} {\em Chin. J. Phys.}, vol.~90, p.~372, 2024.

\bibitem{sotiriou2010f}
T.~P. Sotiriou and V.~Faraoni {\em Rev. Mod. Phys.}, vol.~82, p.~451, 2010.

\bibitem{de2010f}
A.~De~Felice and S.~Tsujikawa {\em Living Rev. Relativ.}, vol.~13, p.~161,
  2010.

\bibitem{olmo2007limit}
G.~J. Olmo {\em Phys. Rev. D}, vol.~75, p.~023511, 2007.

\bibitem{capozziello2011hydrostatic}
S.~Capozziello, M.~De~Laurentis, S.~Odintsov, {\em et~al.} {\em Phys. Rev. D},
  vol.~83, p.~064004, 2011.

\bibitem{sharif2016charged}
M.~Sharif and Z.~Yousaf {\em Int. J. Theor. Phys.}, vol.~55, p.~470, 2016.

\bibitem{pan2018astronomical}
S.~Pan, A.~Mukherjee, and N.~Banerjee {\em Mon. Not. R. Astron. Soc.},
  vol.~477, p.~1189, 2018.

\bibitem{mustafa2020bardeen}
G.~Mustafa, M.~F. Shamir, and M.~Ahmad {\em Phys. Dark Universe}, vol.~30,
  p.~100652, 2020.

\bibitem{bertolami2007extra}
O.~Bertolami, C.~G. Boehmer, T.~Harko, {\em et~al.} {\em Phys. Rev. D},
  vol.~75, p.~104016, 2007.

\bibitem{harko2010f}
T.~Harko and F.~S. Lobo {\em Eur. Phys. J. C}, vol.~70, p.~373, 2010.

\bibitem{rehman2025orthogonal}
A.~Rehman, T.~Naseer, N.~Alessa, {\em et~al.} {\em Nucl. Phys. B}, vol.~1015,
  p.~116897, 2025.

\bibitem{wang2012energy}
J.~Wang and K.~Liao {\em Class. Quantum Gravity}, vol.~29, p.~215016, 2012.

\bibitem{jaybhaye2022cosmology}
L.~V. Jaybhaye, R.~Solanki, {\em et~al.} {\em Phys. Lett. B}, vol.~831,
  p.~137148, 2022.

\bibitem{naseer2023constructing}
T.~Naseer, M.~Sharif, A.~Fatima, {\em et~al.} {\em Chin. J. Phys.}, vol.~86,
  p.~350, 2023.

\bibitem{jaybhaye2023baryogenesis}
L.~V. Jaybhaye, S.~Bhattacharjee, and P.~Sahoo {\em Phys. Dark Universe},
  vol.~40, p.~101223, 2023.

\bibitem{thirukkanesh2008charged}
S.~Thirukkanesh and S.~Maharaj {\em Class. Quantum Gravity}, vol.~25,
  p.~235001, 2008.

\bibitem{nayak1989bianchi}
B.~Nayak and B.~Sahoo {\em Gen. Relativ. Gravit.}, vol.~21, p.~211, 1989.

\bibitem{maurya2017relativistic}
S.~Maurya {\em Eur. Phys. J. A}, vol.~53, p.~89, 2017.

\bibitem{yousaf2023cylindrical}
M.~Yousaf, M.~Z. Bhatti, and Z.~Yousaf {\em Nucl. Phys. B}, vol.~995,
  p.~116328, 2023.

\bibitem{jaybhaye2022constraints}
L.~V. Jaybhaye, S.~Mandal, and P.~Sahoo {\em Int. J. Geom. Methods Mod. Phys.},
  vol.~19, p.~2250050, 2022.

\bibitem{devi2024constraining}
Y.~K. Devi, S.~Narawade, and B.~Mishra {\em Phys. Dark Universe}, vol.~46,
  p.~101640, 2024.

\bibitem{semelin2001self}
B.~Semelin, N.~Sanchez, and H.~De~Vega {\em Phys. Rev. D}, vol.~63, p.~084005,
  2001.

\bibitem{yousaf2017role}
Z.~Yousaf, K.~Bamba, and M.~Z. Bhatti {\em Phys. Rev. D}, vol.~95, p.~024024,
  2017.

\bibitem{herrera2011physical}
L.~Herrera {\em Int. J. Mod. Phys. D}, vol.~20, p.~1689, 2011.

\bibitem{bhatti2026dynamics}
M.~Z. Bhatti, A.~Adeel, and M.~Yousaf {\em Int. J. Geom. Methods Mod. Phys.},
  vol.~23, p.~2550209, 2026.

\bibitem{herrera2010cavity}
L.~Herrera, G.~Le~Denmat, and N.~O. Santos {\em Class. Quantum Gravity},
  vol.~27, p.~135017, 2010.

\bibitem{harrison1965gravitation}
B.~K. Harrison, K.~S. Thorne, {\em et~al.} {\em Gravitation Theory and
  Gravitational Collapse}, 1965.

\bibitem{sharif2015instability}
M.~Sharif and Z.~Yousaf {\em Eur. Phys. J. C}, vol.~75, p.~194, 2015.

\bibitem{sharif2013dynamical}
M.~Sharif and Z.~Yousaf {\em Phys. Rev. D}, vol.~88, p.~024020, 2013.

\bibitem{herrera2014dissipative}
L.~Herrera, A.~Di~Prisco, J.~Ib{\'a}{\~n}ez, {\em et~al.} {\em Phys. Rev. D},
  vol.~89, p.~084034, 2014.

\end{thebibliography}
\end{document}